\documentclass[12pt]{article}

\usepackage{bm}
\usepackage{float}
\usepackage{relsize}
\usepackage{array}
\usepackage{fullpage}
\usepackage{amsfonts}
\usepackage{amsmath}
\usepackage{setspace}
\usepackage{easybmat}
\usepackage{slashed}
\usepackage{amsmath}
\usepackage{amssymb}
\usepackage{graphicx}
\usepackage{verbatim}
\usepackage{caption}
\usepackage{subcaption}
\usepackage{hyperref}

\allowdisplaybreaks
\def\timesbox{\hbox{$\scriptscriptstyle\times$}}
\def\ant{ {{\lower 1ex  \timesbox} \atop {\raise 1.5ex  \timesbox}}}
\def\f{\frac}
\def\pa{\partial}
\def\non{\nonumber\\}
\def\a{\alpha}
\def\b{\beta}

\def\d{\delta}
\def\e{\epsilon}
\def\g{\gamma}

\def\l{\lambda}
\def\lp{\lambda^{\prime}}
\def\m{\mu}
\def\n{\nu}
\def\o{\omega}

\def\s{\sigma}
\def\t{\tau}
\def\th{\theta}
\def\D{\Delta}

\def\G{\Gamma}

\newcommand{\mathsym}[1]{{}}
\newcommand{\unicode}[1]{{}}

\newcommand{\be}{\begin{equation}}
\newcommand{\ee}{\end{equation}}
\newcommand{\beqa}{\begin{eqnarray}}
\newcommand{\eeqa}{\end{eqnarray}}
\newcommand{\bsp}{\begin{split}}
\newcommand{\esp}{\end{split}}
\newcommand{\bgth}{\begin{gather}}
\newcommand{\egth}{\end{gather}}

\newcommand{\tr}{\hbox{tr}}

\begin{document}

\title{\textbf{Phases of theories with fermions in AdS}
\\
\author{Astha Kakkar$^{a}$\footnote{asthakakkar8@gmail.com} and 
Swarnendu Sarkar$^{a,b}$\footnote{swarnen@gmail.com}\\
$^a$\small{{\em Department of Physics and Astrophysics,
University of Delhi,}} \\ 
\small{{\em Delhi 110007, India}}\\
$^b$\small{{\em Department of Physics, Vidyasagar University,}}\\
\small{{\em Midnapore 721102, India}}\\
}}

\date{}

\maketitle

\abstract{We study the phases of Yukawa theories at weak coupling and the Gross-Neveu models in AdS spaces at zero and finite temperature. Following the method used in \cite{Kakkar:2022hub}, we first compute the one-loop partition functions, using the generalized eigenfunctions of the Laplacian on Euclidean AdS in the Poincar\'e coordinates. These functions satisfy desired periodicities under thermal identification. The method replicates results for partition functions known in the literature. We then study the phases of these field theories with fermions as regions in the corresponding parameter spaces at zero temperature. The phases and the corresponding phase boundaries are further identified as a function of the mass-squared of the scalar field and temperature for the Yukawa theories. While for the Gross-Neveu models, the changes in the phases as a function of the fermionic mass and the coupling constant at finite temperature are discussed. The Gross-Neveu-Yukawa model is studied for AdS$_4$. We also note certain deviations from phases of these theories in flat space.}

\newpage

\tableofcontents

\baselineskip=18pt

\section{Introduction}\label{introf}

Quantum field theories in Anti-de Sitter (AdS) spaces have been studied for quite some time \cite{Burgess:1984ti}-\cite{Aharony:2012jf}. In the recent years there has been renewed interest in this area mainly with the aim of obtaining information on non-perturbative flat-space S-Matrix. The idea involves taking flat-space limit, $L \rightarrow \infty$, ($L$ being the AdS length scale) of boundary correlation functions in AdS apace. This translates into taking appropriate limits in boundary conformal field theory where newly developed techniques in the conformal field theories can be used. See \cite{Kruczenski:2022lot} for a recent status and references therein. Another motivation has been to study the conformal field theories in flat-space with boundaries by considering theories at the critical point in AdS \cite{Carmi:2018qzm}-\cite{Giombi:2021cnr}.  

Our aim in this paper is however much more modest. We study phases of theories perturbatively in couplings and in $1/N$ ($N$ being the number of flavors) in AdS for a fixed $L$. The ultraviolet regime of these theories remains same as that in flat-space but the infrared behavior differs. The results in this case are thus expected to deviate from those of flat-space. An analysis along this line was done in \cite{Kakkar:2022hub} where phases of theories with only scalars at finite temperature was studied after computing effective potentials. Several features which differ form flat space were noted. Symmetry broken phase was found to exist at high temperatures for AdS$_3$ and AdS$_2$. In AdS$_2$ symmetry breaking was noted in \cite{Inami:1985dj} at zero temperature, evading the Coleman-Mermin-Wagner theorem \cite{Mermin:1966fe,Coleman:1973ci}. This was shown to persist at finite temperatures.   

In this paper we study the phases of theories with fermions concentrating on the Yukawa theories and the Gross-Neveu models. The essential ingredient utilized in studying these phases is the one-loop partition function for scalars and for fermions. These partitions functions have been obtained in the recent past using various methods, see for example \cite{Gibbons:2006ij}-\cite{Kraus:2020nga}. An alternate method was introduced in \cite{Kakkar:2022hub}. 

In the initial sections of the paper we derive expressions for fermion one-loop partition function at zero and finite temperatures using the method used in \cite{Kakkar:2022hub} for scalars. The trace (\ref{trf}) which is the primary object for our purpose, can be obtained by knowing the degeneracy of the eigenvalues. The spectrum in this case is continuous and ranges between $\pm \infty$. These degeneracies have been worked out quite some time back in \cite{Camporesi:1995fb}. We however take a different route to the computation of the trace. We use the eigenfunctions of the Laplacian operator for Euclidean AdS$_{d+1}$ in Poincar\'e coordinates which are worked out and studied in \cite{Henningson:1998cd}-\cite{Henneaux:1998ch}. Though the degeneracy is easily computed, in our method, one does not need to isolate it. The trace then is essentially the Green's function evaluated at coincident points. This method naturally generalizes to the case of thermal AdS$_{d+1}$. 

The finite temperature potentials that we study are obtained by the identification of the Euclidean time in global coordinates, $\t \rightarrow \t+\b$. This can be identified as the quotient space ${\mathbb H}^{d+1}/{\mathbb Z}$. For finite temperature we thus first find the generalized eigenfunctions that are periodic or anti-periodic for bosons and fermions respectively, under the action of the elements of $\mathbb{Z}$. A similar method has been used for studying field theories on quotient spaces, see for example \cite{Banach:1978dt} and for a more recent application for scalars see \cite{Miyagawa:2015sql,Sugishita:2016iel}. It can be shown that the trace obtained using such a method can be re-written in terms of infinite number of images of the Green's functions.

\begin{figure}[t] 
\begin{center} 
 \includegraphics[width=0.25\linewidth]{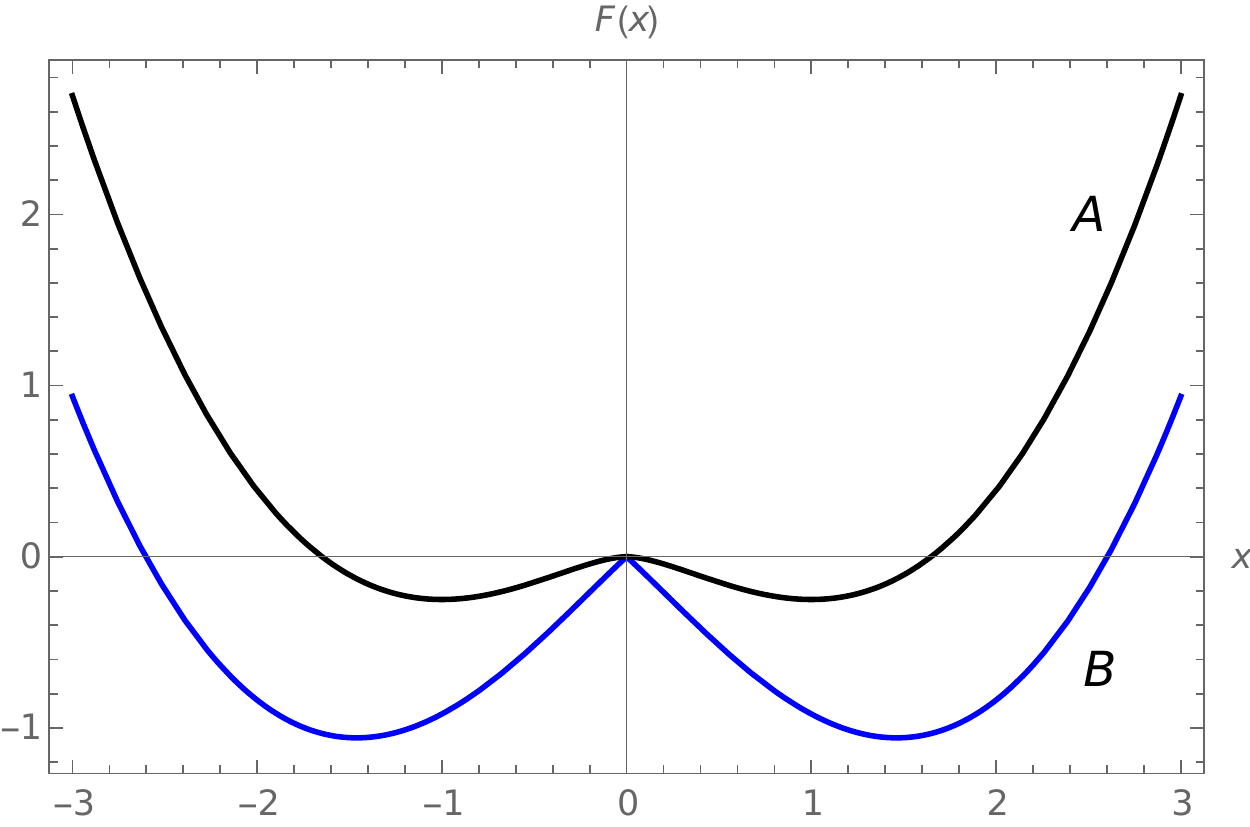} 
  \caption{General feature of potentials with cusp for theories with fermions in AdS is shown in blue. The corresponding feature in flat-space is smooth and is shown in black. For illustration we have plotted the function $F(x)=\int_0^x f(y)~ dy$ \it{vs} $x$. $A$: $f(x)=x\log (|x|)$ and $B$: $f(x)=x~\psi^{(0)}(|x|)$.}
  \label{cusp} 
  \end{center} 
\end{figure}

From the analysis of scalars in \cite{Kakkar:2022hub} it was seen that, though the zero temperature contribution to the trace is proportional to the divergent volume of AdS$_{d+1}$, the finite temperature piece is not. In writing down the effective potential we thus need to regularize this volume. The results of regularized volume which are used in this paper are reviewed in \cite{Kakkar:2022hub}. As mentioned before, the ultraviolet behavior of the theories is same as that of flat space and one needs to implement the usual renormalization. In our setup of the computation of the trace it is natural to use dimensional regularization. The ultraviolet renormalization then is done by imposing suitable renormalization conditions. A scheme similar to that of the minimal subtraction scheme is also discussed for Yukawa theories in appendix \ref{alternatef}. In the theories with fermions that we study, as compared to flat space effective potential, the \texttt{Log}'s are replaced by \texttt{PolyGamma} functions with absolute values of arguments. The potentials are thus not smooth and have a cusp. A simple such example is plotted in Figure \ref{cusp}. This feature of the potentials causes a difficulty in the analysis of the phases which are studied numerically. 

For Yukawa theories because of the lack of $\phi \rightarrow -\phi$ symmetry the potential is asymmetric about $\phi=0$. These theories have several parameters which poses an additional complication in the analysis. We first isolate the regions in the parameter spaces at zero temperature corresponding to the number of minima. The phases are then further identified as a function of the scalar mass-squared and temperature. In all the cases we find a phase boundary where two minima exchange dominance at zero temperature. At finite temperature such a transition is found for AdS$_2$ and AdS$_3$.  

We next study the Gross-Neveu model having $U(N)$ symmetry and four-fermion interaction, originally studied in two dimensions \cite{Gross:1974jv}.  In flat space the Gross-Neveu model has also been studied extensively in three dimensions. The theory though non-renormalizable perturbatively in the four-fermion coupling constant, has been shown to be renormalizable in the large $N$ expansion \cite{Rosenstein:1990nm}. In our analysis we consider this model in AdS$_2$ and in AdS$_3$ with tree-level fermion mass $m_f$. In the large $N$ limit we find that the discrete chiral symmetry that appears in the $m_f=0$ limit remains broken at zero and at all finite temperatures both for two and three dimensions. This is unlike in flat space where the discrete chiral symmetry is restored beyond a certain temperature \cite{Jacobs:1974ys}-\cite{Rosenstein:1988dj}. In four dimensions one needs to include the scalar self interactions for the four-fermion theory to be renormalizable \cite{Zinn-Justin:1991ksq}. In the large $N$ limit with proper re-scaling of the parameters by $N$ the potential reduces to that of the Yukawa model in AdS$_4$.

This paper is organized as follows. We begin section \ref{partf} with the basic setup. In section \ref{euclidf} we re-derive a known result for the fermion trace (\ref{trf}) for fermions in Euclidean AdS$_{d+1}$. The corresponding expression for thermal  AdS$_3$ is computed in section \ref{thermalf}. This result is then generalized for thermal AdS$_{d+1}$ in section \ref{thermaldf}.  Section \ref{phasesf} is devoted to the study of phases of theories with fermions. Yukawa theories in various dimensions are analysed in section \ref{yukawa}. In section \ref{gnmodel} we study the Gross-Neveu models in two and three dimensions and the Gross-Neveu-Yukawa model in four dimensions. We end the paper with discussions and some future directions in section \ref{summaryf}. Some details of computation of the fermion trace (\ref{trf}) at zero temperature is given in appendix \ref{appaf}. Appendix \ref{Yads4app} contains the expression for potential for AdS$_4$ using renormalized perturbation theory. An alternate renormalization scheme for Yukawa theories is considered in appendix \ref{alternateads2Y}. 

\section{One-loop partition function}\label{partf}

We first highlight the general setup taking the example of Yukawa theory with scalar self-interaction

\begin{equation}
S=-\int d^{d+1}x \sqrt{g}\left[\bar{\psi}(\slashed{D}+m_f+g\phi)\psi+\f{1}{2}(\partial_\mu \phi)^2+V(\phi)\right].
\end{equation}

The Dirac operator $\slashed{D}$ is given by

\beqa\label{cder}
\slashed{D}\psi=\g^MD_M\psi=\e^M_a\G^a\left(\pa_M+\f{1}{8}\o_M^{bc}[\G_b,\G_c]\right)\psi
\eeqa

$\e^M_a$ is the vielbein and $\o_M^{bc}$ the spin connection. The Euclidean Gamma matrices satisfy $\{\G_b,\G_c\}=2\d_{bc}$. 
Next, writing $\phi=\phi^{}_{cl}+\eta$ and integrating the quadratic action over $\eta$ and $\psi$ gives the one-loop partition function,

\begin{eqnarray}
Z^{(1)}&=&\det[\slashed{D}+M_f(\phi_{cl})]\det[-\square^{}_{E}+V''(\phi^{}_{cl})]^{-1/2} \exp(-{\cal V}^{}_{d+1} V(\phi^{}_{cl}))\non 
&=&\exp\left(\mbox{tr}\log[\slashed{D}+M_f(\phi_{cl})]\right)\exp\left(-\frac{1}{2}\mbox{tr} \log[-\square^{}_{E}+V''(\phi^{}_{cl})]\right)\exp\left(-{\cal V}_{d+1} V(\phi^{}_{cl})\right)
\end{eqnarray}

where $ M_f(\phi_{cl})=m_f+g \phi_{cl}$, ${\cal V}^{}_{d+1}$ is the volume of $d+1$ dimensional Euclidean space and
$\square^{}_{E}=\partial^{}_{\mu}[\sqrt{g}g^{\mu\nu} \partial^{}_{\nu}]/\sqrt{g}$. The effective potential in terms of the one-loop partition functions for bosons, $Z^{(1)}_b$ and fermions, $Z^{(1)}_f$ is given by,

\beqa
\label{c1e1f}
V^{}_{eff}(\phi^{}_{cl})&=&-\frac{1}{{\cal V}^{}_{d+1}}\log Z_f^{(1)}-\frac{1}{{\cal V}^{}_{d+1}}\log Z_b^{(1)}+V(\phi_{cl})\non
&=&-\f{1}{{\cal V}_{d+1}}\mbox{tr}\log[\slashed{D}+M_f(\phi_{cl})]+\frac{1}{2{\cal V}^{}_{d+1}}\mbox{tr} \log[-\square^{}_{E}+V''(\phi^{}_{cl})]+V(\phi^{}_{cl}).
\eeqa

The $\log$ of the trace can be obtained (up to a constant) by integrating the the following: 

\beqa\label{trf}
\frac{1}{2{\cal V}^{}_{d+1}}\mbox{tr}\left[ \frac{1}{-\square^{}_{E}+V''(\phi^{}_{cl})}\right]~~~~~~~\mbox{and}~~~~~~~\frac{1}{{\cal V}^{}_{d+1}}\mbox{tr}\left[ \frac{1}{\slashed{D}+M_f(\phi_{cl})}\right].
\eeqa

For theories with only scalars the traces were computed and phases were studied in \cite{Kakkar:2022hub}. In the following few sections we compute the trace (\ref{trf}) for the fermions at zero and finite temperature. We use the method similar to that of scalars in \cite{Kakkar:2022hub} for computation of the trace for the fermions.

\subsection{Zero temperature: Euclidean AdS$_{d+1}$}\label{euclidf}

In this section we compute the trace (\ref{trf}) for fermions at zero temperature. We begin by reviewing the solution to the following eigenvalue equation

\beqa\label{diraceq}
\slashed{D}\psi= i\l\psi.
\eeqa

We shall be using metric of AdS$_{d+1}$ in Poincar\'e coordinates 

\beqa
ds^2=\f{L^2}{y^2}\left(dy^2+\eta_{\mu\nu}dx^{\mu}dx^{\nu}\right).
\eeqa

In these coordinates, $\e^M_a=y\d^M_a$ and $\o^{bc}_M=(\d^a_y\d^b_M-\d^a_M\d^b_y)/y$. Thus $\slashed{D}=y\G^a\pa_a-\frac{d}{2}\G_y$. Writing the Dirac fermion $\psi$ as 
\beqa
\psi=\left(\begin{array}{c}a_+\psi_{+}\\ a_-\psi_{-}\end{array}\right)
\eeqa

where $\G_y\psi_{\pm}=\pm\psi_{\pm}$. When $d+1$ is even $\psi_{\pm}$ are of $2^{\frac{d-1}{2}}$ components and $a_{\pm}$ are $2^{\frac{d-1}{2}}\times2^{\frac{d-1}{2}}$ constant matrices, to be determined later. We have thus the following first order differential equations 
\beqa
a_+y\pa_y\psi_+-a_+\frac{d}{2}\psi_+ + a_- y\G_i\pa_i\psi_-&=&i\l \psi_+a_+\non
-a_-y\pa_y\psi_-+a_-\frac{d}{2}\psi_- + a_+ y\G_i\pa_i\psi_+&=&i\l \psi_-a_-
\eeqa

The above equations lead to a decoupled set of second order differential equations which on re-scaling $\psi_{\pm}=y^{\f{d+1}{2}} \tilde{\psi}_{\pm}$ and further writing $\tilde{\psi}_{\pm}(\vec{x},y)=e^{i\vec{k}.\vec{x}}\tilde{\psi}_{\pm}(y)$ have the following form
\beqa\label{difff}
y^2\pa_y^2\tilde{\psi}_{\pm}+y\pa_y\tilde{\psi}_{\pm}-\left[k^2y^2+(i\l\mp\frac{1}{2})^2\right]\tilde{\psi}_{\pm}=0.
\eeqa  

The solutions to (\ref{difff}) are given by modified Bessel functions  $\tilde{\psi}_{\pm}(ky)=(ky)^{\frac{d+1}{2}}K_{i\l\mp \frac{1}{2}}(ky)$. The solution must satisfy the Dirac equation (\ref{diraceq}), this gives
\beqa
i \G_i k_i a_+ = k a_-
\eeqa

leading to the following form of the eigenfunctions \cite{Mueck:1998iz}

\beqa\label{eigenf}
\psi_{\vec{k},\l}(\vec{x},y)=\left(\begin{array}{c}\tilde{\psi}_{+}(ky)\\ \frac{i\G_ik_i}{k}\tilde{\psi}_{-}(ky)\end{array}\right)e^{i\vec{k}.\vec{x}}. 
\eeqa

For each pair of $\psi_{\pm}$, the eigenfunctions are normalized as 

\beqa\label{normf}
&&\int d^{d+1}x\sqrt{g}~\psi^{\dagger}_{\vec{k}^{\prime},\lp}(\vec{x},y)\psi_{\vec{k},\l}(\vec{x},y)\non
&=&\int d^{d+1}x ~e^{i(\vec{k}-\vec{k}^{\prime}).\vec{x}}(k^{\prime}k)^{\f{d+1}{2}}\left[K_{i\l-\frac{1}{2}}(ky)K_{-i\lp-\frac{1}{2}}(k^{\prime}y)+ K_{i\l+\frac{1}{2}}(ky)K_{-i\lp+\frac{1}{2}}(k^{\prime}y)\right]\non
&=&(2\pi)^d\d^{d}(\vec{k}-\vec{k}^{\prime})k^d\d(\l-\lp)/\m(\l)
\eeqa
where
\beqa\label{measuremu1}
\m(\l)=\f{1}{\pi \G\left(\frac{1}{2}+i\l\right)\G\left(\frac{1}{2}-i\l\right)}=\f{1}{\pi^2}\cosh(\pi\l)
\eeqa

as shown in appendix \ref{measuref}. We can now compute the required trace (\ref{trf}) as follows.

\beqa%\label{trzerotf}
\tr\left[\frac{1}{\slashed{D}+M_f}\right]&=&\int d^{d+1}x\sqrt{g}\int \f{d^dk}{(2\pi)^d}\f{1}{k^d}\int_{-\infty}^{\infty}\f{d\l ~\m(\l)}{i\l+M_f}\psi^{\dagger}_{\vec{k},\l}(\vec{x},y)\psi_{\vec{k},\l}(\vec{x},y)\label{trzerotf1}\\
&=& 2^{\frac{d-1}{2}}M_f \int d^{d+1}x\sqrt{g}\int \f{d^dk}{(2\pi)^d}\f{1}{k^d}\int_{-\infty}^{\infty}\f{d\l ~\m(\l)}{\l^2+M_f^2}\times\non
&\times& (ky)^{d+1}\left[K_{i\l-\frac{1}{2}}(ky)K_{-i\l-\frac{1}{2}}(ky)+K_{i\l+\frac{1}{2}}(ky)K_{-i\l+\frac{1}{2}}(ky)\right]\label{trzerotf2}\\
&=& \frac{{\cal V}_{d+1}2^{\frac{d+1}{2}}M_f}{(4\pi)^{(d+1)/2}\G\left(\frac{d+1}{2}\right)}\int_{-\infty}^{\infty}\f{d\l }{\l^2+M_f^2}\frac{\G\left(\frac{d+1}{2}+i\l\right)\G\left(\frac{d+1}{2}-i\l\right)}{\G\left(\frac{1}{2}+i\l\right)\G\left(\frac{1}{2}-i\l\right)}\label{trzerotf3}\\
&=& \mbox{sgn}(M_f)\frac{{\cal V}_{d+1}2^{\frac{d+1}{2}}}{(4\pi)^{(d+1)/2}}\frac{\G\left(\frac{d+1}{2}+|M_f|\right)\G\left(\frac{1}{2}-\f{d}{2}\right)}{\G\left(\frac{1}{2}-\f{d}{2}+|M_f|\right)}\label{trzerotf4}
\eeqa

The factor of $2^{\frac{d-1}{2}}$ in the second line (\ref{trzerotf2}) comes from the number of components of $\psi_{\pm}(ky)$ for odd $d$. For even $d$ this factor needs to be replaced by $2^{\frac{d-2}{2}}$. The third line (\ref{trzerotf3}) is obtained after performing the integral over $k$ and putting in the measure $\m(\l)$ from (\ref{measuremu1}). Equation (\ref{trzerotf3}) contains the degeneracy for the eigenvalues $\pm i\l$ computed in \cite{Camporesi:1995fb}. Finally in the last line (\ref{trzerotf4}) we have performed the integral over $\l$ by closing the contour in the upper half complex $\l$ plane (see \ref{contourf} for details) recovering the result in \cite{Giombi:2021cnr}. An alternate computation of the trace reversing the order of $k$ and $\l$ integrals is discussed in appendix \ref{alternatef}.

\subsection{Finite temperature: Thermal AdS$_3$}\label{thermalf}

In this section we consider thermal AdS$_{d+1}$ which is the quotient space $\mathbb{H}^{d+1}/\mathbb{Z}$. Having got the eigenfunctions on Euclidean AdS$_{d+1}$ we can now proceed with the analysis for thermal AdS$_{d+1}$ following \cite{Kakkar:2022hub}. Define 

\beqa\label{qeigenf}
\Psi_{\vec{k},\l}(x)=\f{1}{\cal N}\sum_{n=-\infty}^{\infty}{\cal R}(\g^n)~\psi_{\vec{k},\l}(\g^n x)
\eeqa

where $\g^{n}\in \mathbb{Z}$. ${\cal N}$ is a normalization constant which regularizes the sum \footnote{${\cal N}$ is equal to $|\mathbb{Z}|$ which being infinite will be assumed to be regularized using some scheme.}. $\psi_{\vec{k},\l}(\g^n x)$ are eigenfunctions on $\mathbb{H}^{d+1}$ (\ref{eigenf}). ${\cal R}(\g^n)$ is a one dimensional representation of the group $\mathbb{Z}$ and can be written as ${\cal R}(\g^n)=e^{2\pi ina}$ with $a=1$ for bosons and $a=\f{1}{2}$ for fermions.  $\Psi_{\vec{k},\l}(x)$ by construction are eigenfunctions on the quotient space. The expression (\ref{qeigenf}) is a generalization of the expression in \cite{Kakkar:2022hub}. For fermions we thus have 
$\Psi_{\vec{k},\l}(\g x)=-\Psi_{\vec{k},\l}(x)$.

Let us now first specialize to the case of thermal $AdS_3$ which is $\mathbb{H}^3/\mathbb{Z}$ with the metric 
\beqa
ds^2=\f{L^2}{y^2}\left(dy^2+dzd\bar{z}\right)
\eeqa

The action of $\g^{n}$ on the coordinates as,

\beqa
\g^n(y,z)=(e^{-n\beta}y, e^{2\pi in\tau}z) ~~~~\mbox{where}~~~~\tau=\frac{1}{2\pi}(\theta+i\beta)
\eeqa

In terms of real coordinates $z=x_1+ix_2$, the action of the group element on the real coordinates $\vec{x}=(x_1,x_2)$ can be written as, 

\beqa\label{realtf}
\g^n \vec{x}=\left(e^{-n\beta}(x_1 \cos n\theta-x_2\sin n\theta),e^{-n\beta}(x_1 \sin n\theta+x_2\cos n\theta)\right)
\eeqa

In three dimensions $\psi_{\vec{k},\l}(\vec{x},y)$ is two dimensional. The normalization is worked out from $\psi_{\vec{k},\l}(\vec{x},y)$ as follows,
\beqa\label{normtf}
&&\int d^{3}x\sqrt{g}~\psi^{\dagger}_{\vec{k},\l}(\g^n x)\psi_{\vec{k}^{\prime},\l^{\prime}}(\g^{n^{\prime}}x)\\\nonumber
&=& \int d^{3}x\sqrt{g}~(ke^{-n\beta}y)^{3/2}(k^{\prime}e^{-n^{\prime}\beta}y)^{3/2}e^{-i\vec{k}.(\g^n \vec{x})}e^{i\vec{k}^{\prime}.(\g^{n^{\prime}} \vec{x})}\times\non
&\times&\left[K_{-i\l-\f{1}{2}}(k e^{-n\beta}y)K_{i\lp-\f{1}{2}}(k^{\prime} e^{-n^{\prime}\beta}y)+K_{-i\l+\f{1}{2}}(k e^{-n\beta}y)K_{i\lp+\f{1}{2}}(k^{\prime} e^{-n^{\prime}\beta}y)\right]
\label{normtf1}\\\nonumber
&=&\int dy~(ke^{-n\beta})^{3/2}(k^{\prime}e^{-n^{\prime}\beta})^{3/2}(2\pi)^2\d^2(\g^n \vec{k}-\g^{n^{\prime}} \vec{k}^{\prime})\times\non
&\times&\left[K_{-i\l-\f{1}{2}}(k e^{-n\beta}y)K_{i\lp-\f{1}{2}}(k^{\prime} e^{-n^{\prime}\beta}y)+K_{-i\l+\f{1}{2}}(k e^{-n\beta}y)K_{i\lp+\f{1}{2}}(k^{\prime} e^{-n^{\prime}\beta}y)\right]\label{normtf2}\\
&=&(ke^{-(n-n^{\prime})\b})^2(2\pi)^2\d^2(\g^{n-n^{\prime}} \vec{k}-\vec{k}^{\prime})\d(\l-\l^{\prime})/\m(\l).\label{normtf3}
\eeqa

In the above equation (\ref{normtf1}), the integration over $\vec{x}$ leads to the delta function which is then rewritten using the identity
\beqa\label{di1f}
\d^2(\g^n \vec{k}-\g^{n^{\prime}} \vec{k}^{\prime})=e^{2n^{\prime}\b}\d^2(\g^{(n-n^{\prime})} \vec{k}-\vec{k}^{\prime})
\eeqa

The action of $\g^{n}$ on $\vec{k}$ is defined as,

\[
\g^n \vec{k}=\left(e^{-n\beta}(k_1 \cos n\theta+k_2\sin n\theta),e^{-n\beta}(k_2 \cos n\theta-k_1\sin n\theta)\right)
\]

so that $|\g^n \vec{k}|=e^{-n\b}k$. In equation (\ref{normtf2}) the $y$-integral is performed as shown in 
\ref{measuref}.

The trace (\ref{trf}) for thermal AdS$_3$ can now be computed noting the fact that  $\psi_{\vec{k},\l}(\g^n x)$ for $n\ne 0$ are also eigenfunctions of the Dirac operator with eigenvalues $\pm i\l$. The computation of trace for fermion is similar to that of scalars with some modifications, the details of which are as follows.

\beqa
&&\tr\left[\frac{1}{\slashed{D}+M_f}\right]=\frac{1}{{\cal N}^2}\sum_{n,n^{\prime}} \int d^{3}x\sqrt{g}\int \f{d^2k}{(2\pi)^2}\f{1}{k^2}\int_{-\infty}^{\infty}\f{d\l ~\m(\l)}{i\l+M_f}\Psi^{\dagger}_{\vec{k},\l}(\vec{x},y)\Psi_{\vec{k},\l}(\vec{x},y)\\
&=&\frac{1}{{\cal N}^2}\sum_{n,n^{\prime}}e^{-i\pi(n-n^{\prime})} \int d^{3}x\sqrt{g}~\int\frac{d^2k}{(2\pi)^2}\f{1}{k^2}\int_{-\infty}^{\infty} \f{d\l~\m(\l)M_f}{\l^2 +M_f^2}(ke^{-n\beta}y)^{3/2}(ke^{-n^{\prime}\beta}y)^{3/2}\times\label{line1f}\\
&\times&\left[K_{-i\l-\f{1}{2}}(k e^{-n\beta}y)K_{i\l-\f{1}{2}}(k e^{-n^{\prime}\beta}y)+K_{-i\l+\f{1}{2}}(k e^{-n\beta}y)K_{i\l+\f{1}{2}}(k e^{-n^{\prime}\beta}y)\right]e^{i\vec{k}.(\g^{n}\vec{x})}e^{-i\vec{k}.(\g^{n^{\prime}}\vec{x})}\non
&=&\frac{2}{{\cal N}}\sum_{n}(-1)^n\f{e^{-3n\beta/2}}{|1-e^{2\pi i n\t}|^2} \int_0^{\infty} \f{dy}{y}~\int_{-\infty}^{\infty} \f{d\l~\m(\l)M_f}{\l^2 +M_f^2}\times\non
&\times& \int\frac{d^2k}{(2\pi)^2} (ky)\left[K_{-i\l-\f{1}{2}}(k e^{-n\beta}y)K_{i\l-\f{1}{2}}(k y)+K_{-i\l+\f{1}{2}}(k e^{-n\beta}y)K_{i\l+\f{1}{2}}(k y)\right](2\pi)^2\d^2(\vec{k})\label{line2f}
\eeqa

We have made several simplifications in going from (\ref{line1f}) and (\ref{line2f}). The integral over  $\vec{x}$ which results in a delta function has been rewritten using

\beqa\label{di2f}
\d^2(\g^n\vec{k}-\g^{n^{\prime}}\vec{k})=\frac{e^{2n^{\prime}\b}}{|1-e^{2\pi i (n-n^{\prime})\t}|^2}\d^2(\vec{k})  
\eeqa

$y$ has been re-scaled $y\rightarrow e^{n^{\prime}\b}y$ within each term in the summation. The summation over $n^{\prime}$ cancels a factor of ${\cal N}$ as each term in the summation over $n^{\prime}$ gives the same series in $n$. The factor of $2$ in front in (\ref{line2f}) is as a result of the symmetry of the trace under $n\rightarrow -n$. We have also excluded the zero temperature $n=0$ term which has already been computed in (\ref{trzerotf4}).

The $k$ integral in (\ref{line2f}) gives the following

\beqa\label{k0limitf}
(ky)\left[K_{-i\l-\f{1}{2}}(k e^{-n\beta}y)K_{i\l-\f{1}{2}}(k y)+K_{-i\l+\f{1}{2}}(k e^{-n\beta}y)K_{i\l+\f{1}{2}}(k y)\right]\non
\xrightarrow[]{k \rightarrow 0}  \f{1}{2}\underbrace{\G\left({1}/{2}-i\l\right)\G\left({1}/{2}+i\l\right)}_{(\pi\m(\l))^{-1}}\left[\left(e^{-n\b}\right)^{-1/2-i\l}+\left(e^{-n\b}\right)^{-1/2+i\l}\right] 
\eeqa

We can now perform the $\lambda$ integral by closing the contour in the upper/lower half complex $\l$ plane. Both the terms in (\ref{k0limitf}) are related to each other by $\l\rightarrow -\l$. The measure $\m(\l)$ being an even function, they give the same contribution. 

Thus

\beqa
&&\tr\left[\frac{1}{\slashed{D}+M_f}\right]=\frac{2}{{\cal N}}\sum_{n}(-1)^n\mbox{sgn}(M_f)\f{e^{-n\beta(1+|M_f|)}}{|1-e^{2\pi i n\t}|^2} \int_0^{\infty} \f{dy}{y}
\eeqa

The divergence in the integral over $y$ is taken care of as follows: noting that the fundamental region along $y$ is 
$e^{-\b}\le y \le 1$,  

\beqa
\int_0^{\infty} \f{dy}{y}=\sum_{m=-\infty}^{\infty}\int_{e^{-(m+1)\b}}^{e^{-m\b}} \f{dy}{y}
={\cal N}\b~.
\eeqa

The finite temperature contribution to the one-loop partition function is then 

\beqa\label{finalp3f}
\log Z^{(1)}_{\t}&=&-\int_{M_f}^{\infty}\tr\left[\frac{1}{\slashed{D}+M}\right] dM\label{paVf}\\
&=&-2\sum_{n=1}^{\infty}\f{(-1)^{n}}{n}\f{e^{-n\beta(1+|M_f|)}}{|1-e^{2\pi i n\t}|^2}\label{rhsp3f}  
\eeqa

Denoting $q=e^{2\pi i \t}$, and $\D=|M_f|+1$ we can rewrite (\ref{rhsp3f}) as
\beqa\label{newformf}
\log Z^{(1)}_{\t}&=&-2\sum_{n=1}^{\infty}\sum_{l,l^{\prime}=0}^{\infty}\f{(-1)^n}{n}q^{n(l+\D/2)}\bar{q}^{n(l^{\prime}+\D/2)}\\
&=&2\sum_{l,l^{\prime}=0}^{\infty}\log\left(1+q^{(l +\D/2)}\bar{q}^{(l^{\prime} +\D/2)} \right)\\
&=& 2\sum_{l=0}^{\infty}(l+1)\log\left(1+e^{-\b(\D+l)}\right)~~~~~~~\mbox{for $\th=0$}.
\eeqa

$\D$ is the dimension of the operator in the boundary dual to the fermion in AdS$_3$. From the boundary CFT point of view the contribution to the partition function comes from the primary operator of dimension $\D$ and well as its descendants of dimensions $\D+l$.

\subsubsection{Thermal AdS$_{d+1}$}\label{thermaldf}

We now generalize the computation for $AdS^{}_{3}$ to that of higher dimensional spaces. We first consider the case for even $d$. The mertic is written as
 
\beqa
ds^2=\f{L^2}{y^2}\left(dy^2+\sum_{i=1}^{d/2} dz_id\bar{z_i}\right).
\eeqa

For thermal $AdS_{d+1}$ the action of $\g^n_i$ on the coordinates is,

\beqa\label{realtd}
\g^n_i(y,z_i)=(e^{-n\beta}y, e^{2\pi in\tau_i}z_i) ~~~~\mbox{where}~~~~\tau_i=\frac{1}{2\pi}(\theta_i+i\beta).
\eeqa

Next define real coordinates $z_i=x_{i1}+ix_{i2}$. The action on the real coordinates $\vec{x}_i=(x_{i1},x_{i2})$ can be written as in equation (\ref{realtf}) with $\theta$ replaced by $\theta_i$. In odd $(d+1)$ dimensions $\psi_{\vec{k},\l}(\vec{x},y)$ is $2^{(d-2)/2}$ dimensional. For each pair of the eigenfunctions, the normalization is worked out from $\psi_{\vec{k},\l}(\vec{x},y)$ as follows,
\beqa\label{normtfd}
&&\int d^{d+1}x\sqrt{g}~\psi^{\dagger}_{\vec{k},\l}(\g^n x)\psi_{\vec{k}^{\prime},\l^{\prime}}(\g^{n^{\prime}}x)\\\nonumber
&=& \int d^{d+1}x\sqrt{g}~(ke^{-n\beta}y)^{(d+1)/2}(k^{\prime}e^{-n^{\prime}\beta}y)^{(d+1)/2}e^{-i\vec{k_i}.(\g^{n}_{i} \vec{x_i})}e^{i\vec{k_i}^{\prime}.(\g^{n^{\prime}}_{i} \vec{x_i})}\non
&\times&\left[K_{-i\l-\f{1}{2}}(k e^{-n\beta}y)K_{i\lp-\f{1}{2}}(k^{\prime} e^{-n^{\prime}\beta}y)+K_{-i\l+\f{1}{2}}(k e^{-n\beta}y)K_{i\lp+\f{1}{2}}(k^{\prime} e^{-n^{\prime}\beta}y)\right]
\label{normtfd1}\\\nonumber
&=& \int dy~(ke^{-n\beta})^{1/2}(k^{\prime}e^{-n^{\prime}\beta})^{1/2}\left[K_{-i\l-\f{1}{2}}(k e^{-n\beta}y)K_{i\lp-\f{1}{2}}(k^{\prime} e^{-n^{\prime}\beta}y)+K_{-i\l+\f{1}{2}}(k e^{-n\beta}y)K_{i\lp+\f{1}{2}}(k^{\prime} e^{-n^{\prime}\beta}y)\right]\non
&\times& \prod^{d/2}_{i=1}  (ke^{-n\beta})(k^{\prime}e^{-n^{\prime}\beta}) (2\pi)^{2}\d^{2}(\g_i^n \vec{k}-\g_i^{n^{\prime}} \vec{k}^{\prime})\label{normtfd2}\\
&=& \d(\l-\l^{\prime})/\m(\l)\prod^{d/2}_{i=1}  (ke^{-(n-n^{\prime})\b})^2(2\pi)^2\d^2(\g_i^{n-n^{\prime}} \vec{k}-\vec{k}^{\prime}).\label{normtfd3}
\eeqa

In the above equation (\ref{normtfd1}), the integration over $\vec{x}$ leads to the delta function which is then rewritten using the identity (\ref{di1f}). In equation (\ref{normtfd2}) the $y$-integral is performed as shown in 
\ref{measuref}.

The trace can now be computed following the steps as in AdS$_{3}$.

\beqa
&&\tr\left[\frac{1}{\slashed{D}+M_f}\right]=\frac{1}{{\cal N}^2}\sum_{n,n^{\prime}} \int d^{d+1}x\sqrt{g}\int \f{d^dk}{(2\pi)^d}\f{1}{k^d}\int_{-\infty}^{\infty}\f{d\l ~\m(\l)}{i\l+M_f}\Psi^{\dagger}_{\vec{k},\l}(\vec{x},y)\Psi_{\vec{k},\l}(\vec{x},y)\\
&=&\frac{2^{\frac{(d-2)}{2}}}{{\cal N}^2}\sum_{n,n^{\prime}}  e^{-i\pi(n-n^{\prime})} \int d^{d+1}x\sqrt{g}~\int\frac{d^dk}{(2\pi)^d}\f{1}{k^d}\int_{-\infty}^{\infty} \f{d\l~\m(\l)M_f}{\l^2 +M_f^2}(ke^{-n\beta}y)^{(d+1)/2}(ke^{-n^{\prime}\beta}y)^{(d+1)/2}\non
&\times&\left[K_{-i\l-\f{1}{2}}(k e^{-n\beta}y)K_{i\l-\f{1}{2}}(k e^{-n^{\prime}\beta}y)+K_{-i\l+\f{1}{2}}(k e^{-n\beta}y)K_{i\l+\f{1}{2}}(k e^{-n^{\prime}\beta}y)\right]e^{i\vec{k}_i.(\g^{n^{\prime}}_{i}\vec{x}_i)}e^{-i\vec{k}_i.(\g^{n}_{i}\vec{x}_i)}\non\label{line1fd}\\
&=&\frac{1}{{\cal N}}\sum_{n}(-1)^n \int_0^{\infty} \f{dy}{y}~\int\frac{d^dk}{(2\pi)^d} (ky) e^{-n\beta/2}~\int_{-\infty}^{\infty} \f{d\l~\m(\l)M_f}{\l^2 +M_f^2}\non
&\times&\left[K_{-i\l-\f{1}{2}}(k e^{-n\beta}y)K_{i\l-\f{1}{2}}(k y)+K_{-i\l+\f{1}{2}}(k e^{-n\beta}y)K_{i\l+\f{1}{2}}(k y)\right]\non
&\times& \prod^{d/2}_{i=1}\f{2e^{-n\beta}}{|1-e^{2\pi i n\t_i}|^2}(2\pi)^2\d^2(\vec{k}_i)\label{line2fd}
\eeqa

We have again made simplifications in going from (\ref{line1fd}) and (\ref{line2fd}) similar to that of $AdS^{}_{3}$ and have used the identities (\ref{di1f}) and (\ref{di2f}). The $k$, $\l$ and $y$ integrals can be computed as before. This results in the following expression for the finite temperature contribution to the one-loop partition function

\beqa\label{finalp3fd}
\log Z^{(1)}_{\t}&=&-\int_{M_f}^{\infty}\tr\left[\frac{1}{\slashed{D}+M}\right] dM\label{paVfd}\\
&=&-\sum_{n=1}^{\infty}\f{(-1)^{n}}{n}e^{-n\beta |M_f|}\prod^{d/2}_{i=1}\f{2 e^{-n\beta}}{|1-e^{2\pi i n\t_i}|^2}\label{rhsp3fd}  
\eeqa

Denoting $q=e^{2\pi i \t}$, and $\D=|M_f|+d/2$ we can rewrite (\ref{rhsp3fd}) as
\beqa\label{newformfd}
\log Z^{(1)}_{\t}&=&-\sum_{n=1}^{\infty}\prod^{d/2}_{i=1}\sum_{l_i,l_i^{\prime}=0}^{\infty}2\f{(-1)^n}{n}q_i^{n(l_i+\D/d)}\bar{q}_i^{n(l_i^{\prime}+\D/d)}\\
&=&2^{d/2}\sum_{\substack{l_1,l^{\prime}_1\\...\\l_{d/2}l^{\prime}_{d/2}=0}}^{\infty}\log\left(1+q_1^{(l_1 +\D/d)}\bar{q}_1^{(l_1^{\prime} +\D/d)}\cdots q_{d/2}^{(l_{d/2} +\D/d)}\bar{q}_{d/2}^{(l_{d/2}^{\prime} +\D/d)} \right)\\
&=&2^{d/2}\sum_{\substack{l_1,l^{\prime}_1\\...\\l_{d/2}l^{\prime}_{d/2}=0}}^{\infty}\log\left(1+e^{-\b(l_1+l_1^{\prime}+\cdots+l_{d/2}+l_{d/2}^{\prime} +\D)}\right) ~~~~\mbox{for $\th_i=0$}\\
&=& 2^{d/2}\sum_{l=0}^{\infty}g(l,d)\log\left(1+e^{-\b(\Delta+l)}\right)~~~~\mbox{where~~ $g(l,d)=\f{(d+l-1)!}{l!(d-1)!}$}\label{degen}
\eeqa

The metric for odd $d$ is written as
 
\beqa
ds^2=\f{L^2}{y^2}\left(dy^2+\sum_{i=1}^{(d-1)/2} dz_id\bar{z_i}+dx^{2}_{d}\right).
\eeqa 

The action of $\g^{n}$ on the coordinates is,

\beqa
\g^n(y,z_i,x_d)=(e^{-n\beta}y, e^{2\pi in\tau_i}z_i,e^{-n\b}x_d) ~~~~\mbox{where}~~~~\tau_i=\frac{1}{2\pi}(\theta_i+i\beta).
\eeqa

For odd $d$  the fermion wave function has $2^{(d-1)/2}$ components. The normalization for a pair of fermions is as follows,
\beqa\label{normtfdd}
&&\int d^{d+1}x\sqrt{g}~\psi^{\dagger}_{\vec{k},\l}(\g^n x)\psi_{\vec{k}^{\prime},\l^{\prime}}(\g^{n^{\prime}}x)\\\nonumber
&=& (ke^{-(n-n^{\prime})\b})(2\pi)\d(\g^{n-n^{\prime}} \vec{k}-\vec{k}^{\prime})\prod^{(d-1)/2}_{i=1}  (ke^{-(n-n^{\prime})\b})^2(2\pi)^2\d^2(\g^{n-n^{\prime}} \vec{k}-\vec{k}^{\prime})\d(\l-\l^{\prime})/\m(\l).\label{normtfd3d}
\eeqa

The finite temperature contribution to the one-loop partition function following computations similar to the previous cases is

\beqa\label{finalp3fdd}
\log Z^{(1)}_{\t}&=&-\sum_{n=1}^{\infty}\f{(-1)^{n}}{n}\f{2 e^{-n\beta(|M_f|+1/2)}}{|1-e^{-n \b}|}\prod^{(d-1)/2}_{i=1}\f{2 e^{-n\beta}}{|1-e^{2\pi i n\t_i}|^2}\label{rhsp3fdd}\\
&=&2^{(d+1)/2}\sum_{\substack{l_1,l^{\prime}_1\\...\\l_d=0}}^{\infty}\log\left(1+q_1^{(l_1 +\D^{\prime}/(d-1))}\bar{q}_1^{(l_1^{\prime} +\D^{\prime}/(d-1))}\cdots q_{(d-1)/2}^{(l_{(d-1)/2} +\D^{\prime}/(d-1))}\bar{q}_{(d-1)/2}^{(l_{(d-1)/2}^{\prime} +\D^{\prime}/(d-1))}|q_d|^{(1/2+l_d)} \right)\non
&=&2^{(d+1)/2}\sum_{\substack{l_1,l^{\prime}_1\\...\\l_d=0}}^{\infty}\log\left(1+e^{-\b(l_1+l_1^{\prime}+\cdots+l_{(d-1)/2}+l_{(d-1)/2}^{\prime}+l_d +\D)}\right) ~~~~\mbox{for $\th_i=0$}\\
&=& 2^{(d+1)/2}\sum_{l=0}^{\infty}g(l,d)\log\left(1+e^{-\b(\Delta+l)}\right).   
\eeqa

where $\D^{\prime}=\D-1/2$ and $\D$ is given above equation (\ref{newformfd}). $g(l,d)$ is given in equation (\ref{degen})

\section{Study of theories with fermions}\label{phasesf}

In this section we study the phases of Yukawa theory for $d=1,2,3$ and the massive Gross-Neveu models for $d=1,2$. In AdS$_4$ we also discuss the Gross-Neveu-Yukawa model. The phases are studied by writing down the effective potential utilizing the results from the previous sections for the fermion trace and those of scalars from \cite{Kakkar:2022hub}. As already seen in the study of phases with only scalars, unlike the finite temperature piece, the zero temperature contribution to the trace is proportional to the divergent volume of Euclidean AdS space. This divergence needs to be regulated. In the cutoff scheme these renormalized volumes are ${\cal V}_2=-\b$, ${\cal V}_3=-\pi\b/2$, ${\cal V}_4=2\pi\b/3$ (see \cite{Kakkar:2022hub} and references therein). The renormalized volume for AdS$_3$ differs both in magnitude and in sign from the above in the the dimensional regularization scheme \cite{Diaz:2007an}. 
The results for the phases corresponding to the positive regularized volume of AdS$_3$ will thus differ from those presented here.        
Though this modification can be easily implemented as in \cite{Kakkar:2022hub}, we have however not used this in our analysis below. 

For the numerical computations in the following sections, we have set $n_{\tiny{\mbox{max}}}=10$ for the sum in the finite temperature contributions (\ref{finalp3f}),(\ref{finalp3fd}),(\ref{finalp3fdd}). In fact the series converges within the first two terms up to at least four decimal places because of the exponential damping.

\subsection{Yukawa theory}\label{yukawa}

The general form of the effective potential is given by

\beqa\label{generalveff}
V_{eff}=\f{1}{2}m_s^2\phi_{cl}^2+\f{\l_3}{3!}\phi_{cl}^3+\f{\l_4}{4!}\phi_{cl}^4 +\l_1\phi_{cl} &-&\frac{1}{2{\cal V}^{}_{d+1}}\int_{M_s^2}^{\infty}\mbox{tr}\left[ \frac{1}{-\square^{}_{E}+M_s^2}\right]dM_s^2\\
&-&\frac{1}{{\cal V}^{}_{d+1}}\int_0^{M_f}\mbox{tr}\left[ \frac{1}{\slashed{D}+M_f}\right]dM_f+ \mbox{counterterms}\nonumber
\eeqa

\beqa\label{defmsmf}
\mbox{where}~~~~M^2_s=m_s^2+\l_3 \phi_{cl}+\f{\l_4}{2}\phi_{cl}^2~~~~\mbox{and}~~~~M_f=m_f+g\phi_{cl}.
\eeqa

In the absence of the $\phi \rightarrow -\phi$ symmetry, the linear and cubic terms in $\phi$ need to be included.
The zero temperature contribution to the traces (\ref{trzerotf4}) have poles corresponding to UV divergence for odd $d$, which in our case is $d=1,3$. For these values of $d$, the counterterms cancel the UV divergences coming from the trace. $m_s$, $\l_3$, $\l_4$, $\l_1$, $g$ and $m_f$ appearing in (\ref{generalveff}) are thus renormalized parameters. To be more specific, at one loop level, only bare values of $m_s$, $\l_3$, $\l_4$ and $\l_1$, get renormalized by the counterterms in (\ref{generalveff}). Bare values of $m_f$ and $g$ would be assumed to be renormalized at one-loop level by counterterms corresponding to fermion propagator and the Yukawa vertex. 

In the following sections we use a scheme in which counterterms are evaluated using suitable renormalization conditions. This we do only for the divergent $n$-point scalar terms. For example one and two-point for $d=1$ and one to four-point for $d=3$. Correspondingly the scheme renormalizes $m_s$ and $\l_1$ for $d=1$ and $m_s$, $\l_3$, $\l_4$ and $\l_1$ for $d=3$.  The linear term in $\phi_{cl}$ can be removed by setting a renormalization condition so that the tadpole vanishes. Thus the renormalized one-point coupling $\l_1$ will be set to zero. An alternate renormalization scheme that is similar to the Minimal Subtraction scheme is discussed for $d=1,3$ in appendix \ref{alternateads2Y}. 

\subsubsection{AdS$_3$}

The zero temperature contribution from the trace (\ref{trzerotf4}) has no pole for $d=2$ and is given by 
\beqa
\f{1}{{\cal V}_3}\mbox{tr}\frac{1}{\slashed{D}+M_f}=-\f{1}{2\pi}\mbox{sgn}(M_f)\left(M_f^2-\f{1}{4}\right)
\eeqa

Including the contribution from the scalar determinant, the effective potential at zero temperature  is
\beqa\label{YAdS30texp}
V^0_{eff}=\f{1}{2}m_s^2\phi_{cl}^2+\f{\l_3}{3!}\phi_{cl}^3+\f{\l_4}{4!}\phi_{cl}^4+\l_1\phi_{cl}-\f{\nu^3}{12\pi}+\f{1}{2\pi}\left(\f{|M_f|^3}{3}-\f{|M_f|}{4}\right)
\eeqa

where, $\n=\sqrt{(d/2)^2+M_s^2}$.

\begin{figure}[H] 
\begin{center} 
  \begin{minipage}{0.28\textwidth}%   
   \begin{subfigure}[b]{0.98\linewidth}
    \centering
    \includegraphics[width=0.98\linewidth]{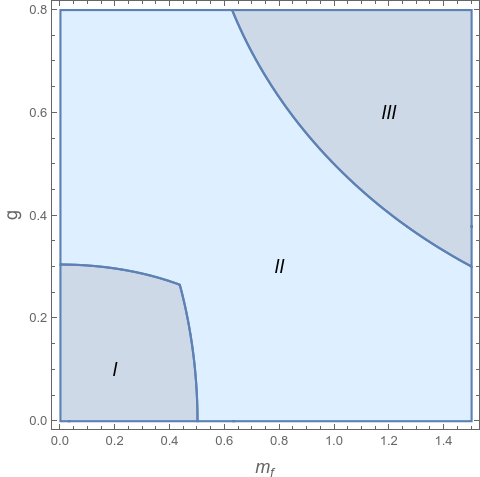} 
    \caption{} 
    \label{YAdS30tmfgr} 
    \vspace{1ex}
  \end{subfigure}
  \end{minipage}%%%
  \begin{minipage}{0.28\textwidth}%   
   \begin{subfigure}[b]{0.98\linewidth}
    \centering
    \includegraphics[width=0.98\linewidth]{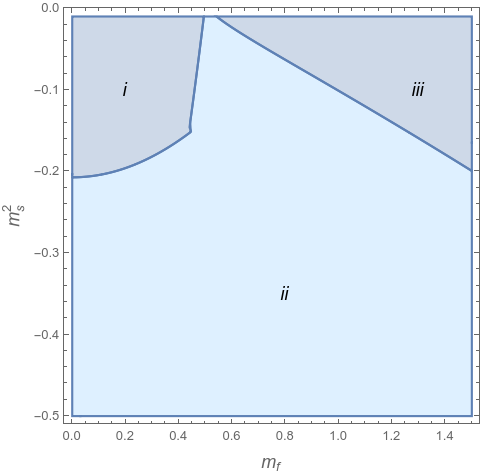} 
    \caption{} 
    \label{YAdS30tmfmsr} 
    \vspace{1ex}
  \end{subfigure}
  \end{minipage}%%%
  \begin{minipage}{0.44\textwidth}%
  \begin{subfigure}[b]{0.5\linewidth}
    \centering
    \includegraphics[width=0.95\linewidth]{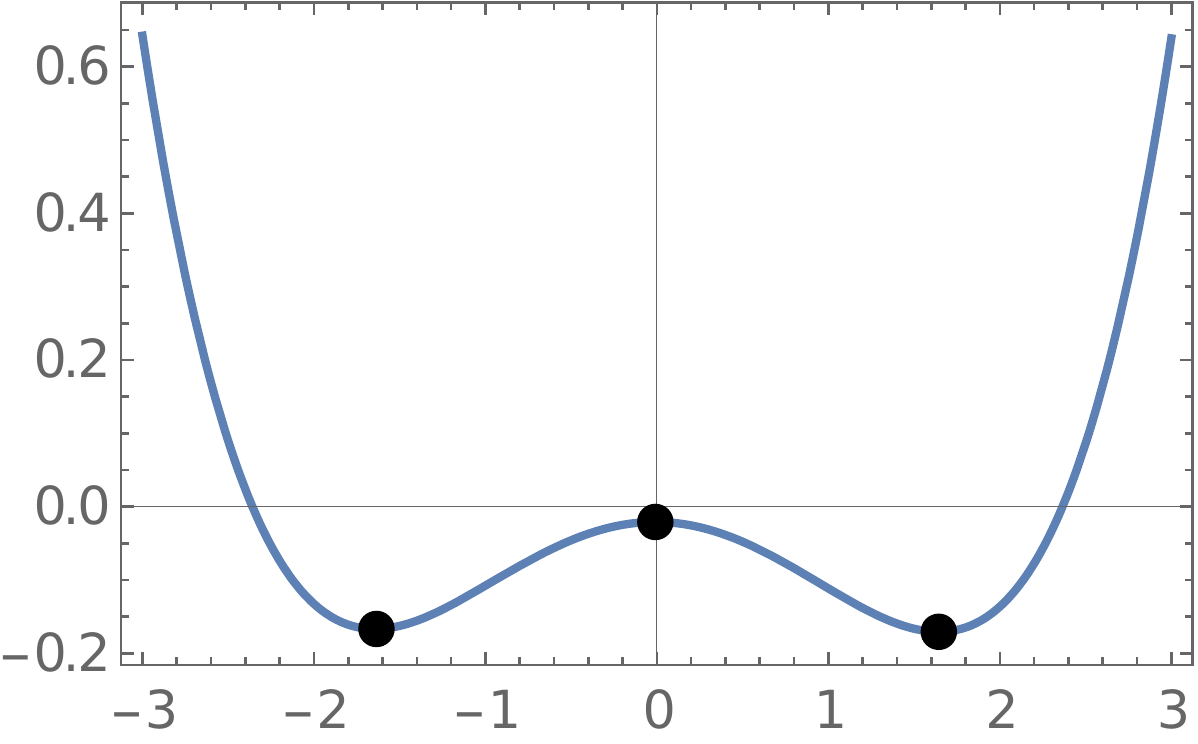} 
    \caption*{$I$: $m_f=0.05$, $g=0.05$} 
    %\label{a} 
    \vspace{1ex}
  \end{subfigure}%
  %\hspace{-10em}
  \begin{subfigure}[b]{0.5\linewidth}
    \centering
    \includegraphics[width=0.95\linewidth]{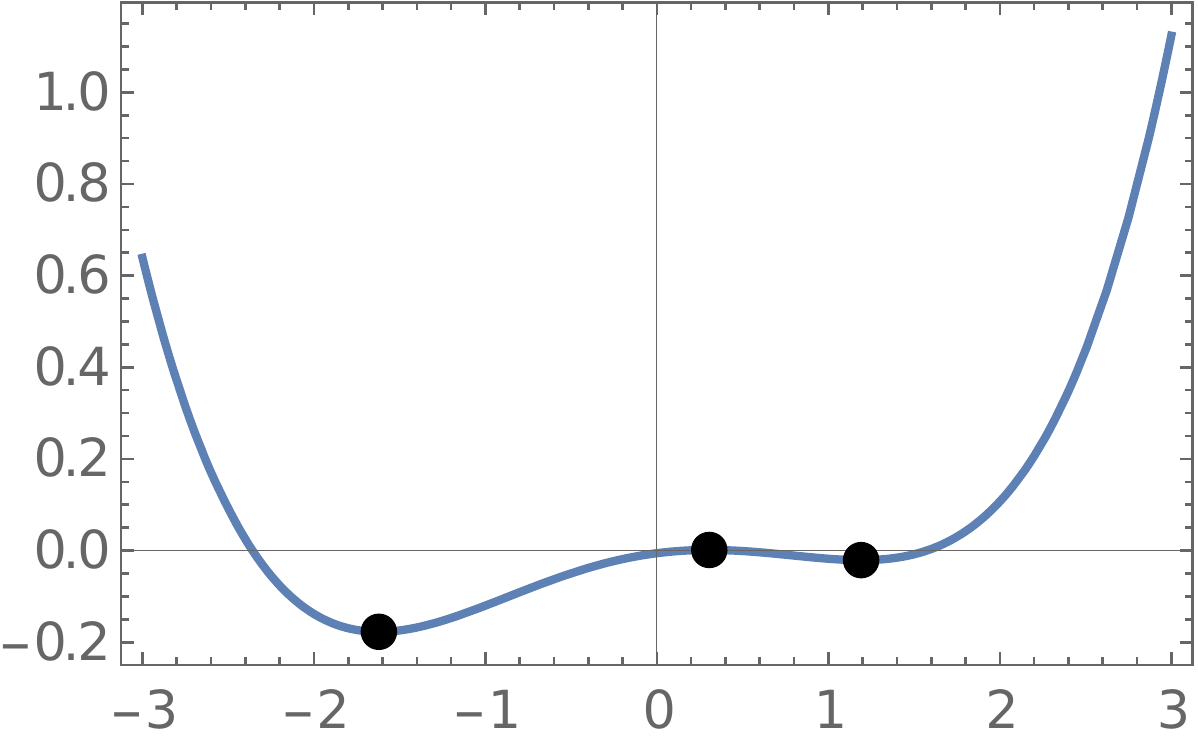} 
    \caption*{$II$: $m_f=1$, $g=0.4$} 
    %\label{b} 
    \vspace{1ex}
  \end{subfigure} 
  \begin{subfigure}[b]{0.5\linewidth}
    \centering
    \includegraphics[width=0.95\linewidth]{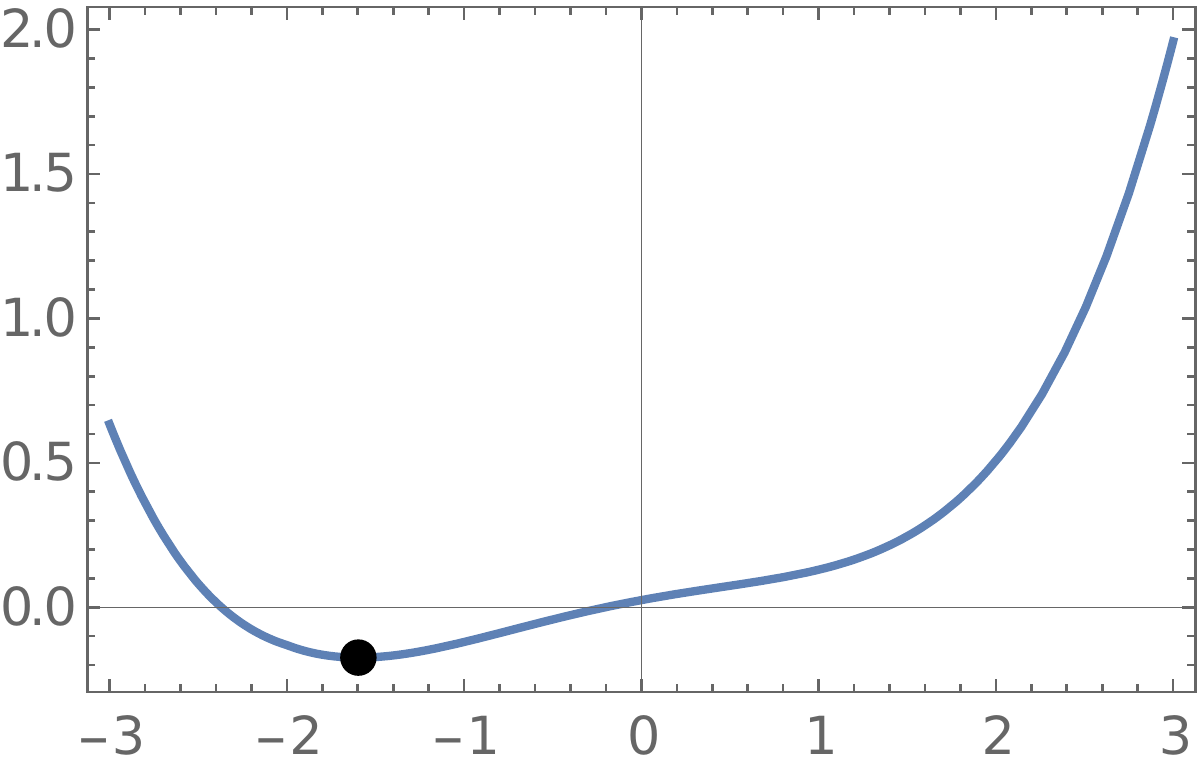} 
    \caption*{$III$: $m_f=1.2$, $g=0.6$} 
    %\label{c} 
  \end{subfigure}%%
  %\hspace{-5em}
  \begin{subfigure}[b]{0.5\linewidth}
    \centering
    \includegraphics[width=0.95\linewidth]{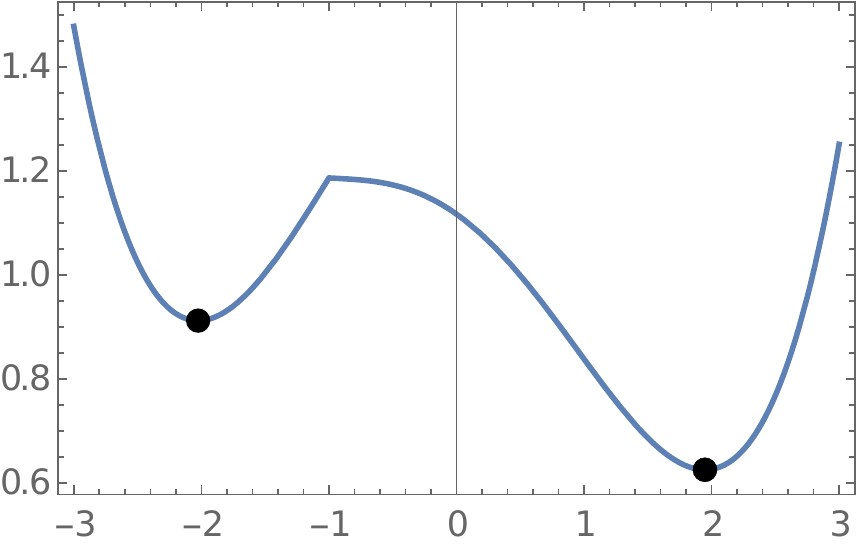} 
   \caption*{$B$: $\b=1, m^2_s=-0.2$} 
    %\label{d} 
  \end{subfigure} 
 % \label{fig7} 
\end{minipage}
  \caption{Regions in the $m_f$-$g$ plane for $m^2_s=-0.2$ (a), and $m_f$-$m_s^2$ plane for $g=0.3$ (b). Various representative potentials (extreme right) at zero temperature. Black dots on the potential plots are the extrema. $B$ is potential plot at finite temperature for region $B$ in the finite temperature phase plot Figure \ref{YAdS3tI}.}
  \label{YAdS30tmfg} 
  \end{center} 
\end{figure}

Figure \ref{YAdS30tmfg} shows the regions in $m_f$-$g$, $m_f$-$m_s^2$ spaces and potential plots for a fixed value of $m^2_s=-0.2$ at zero temperature. In the numerical plots we have set $\l_1=\l_3=0$ and $\l_4=0.5$. Some essential features of the phases are discussed below.\\
\noindent
1. For a fixed value of $m^2_s$, there are in general three regions. Figure \ref{YAdS30tmfgr} shows the ($m_f$, $g$) space plot for $m_s^2=-0.2$. In regions $I$ and $II$ there are two minima. Region $III$ has only one minimum. The more negative $m^2_s$ is, the $II$ region expands and the top right region diminishes in size where there is only one minimum. Figure \ref{YAdS30tmfmsr} shows similar regions in the $m_f$-$m_s^2$ plane for a fixed value of $g=0.3$.\\
\noindent
2. There is a first-order transition line, the boundary separating regions $I$ and $II$ where the both the minima have same height. The sharp edge of the boundary is due to the presence of absolute values (cusp) in the potential (\ref{YAdS30texp}) which lies between the left minimum and the maximum. At higher values of $g$ and $m_f$, the right minimum ceases to exist in region $III$.\\
\noindent  
3. For a fixed value of $m_f$ and $g$ there is also a transition at some critical value of $m^2_s$  (similar to that of symmetric potential) when the two minima start appearing. Whether the left or the right is lower depends on the values of $g$ and $m_f$.\\
\noindent 

%4.  There is no single $m_f$ value which gives both regions $I$ and $II$. Thus region-plots in $g$-$m^2_s$ plane give regions $A$-$C$ %or $B$-$C$. This can be seen from the Figure \ref{YAdS30to}. \\

\begin{figure}[H] 
\begin{center} 
  \begin{minipage}{1\textwidth}%   
   \begin{subfigure}[b]{0.3\linewidth}
    \centering
    \includegraphics[width=1\linewidth]{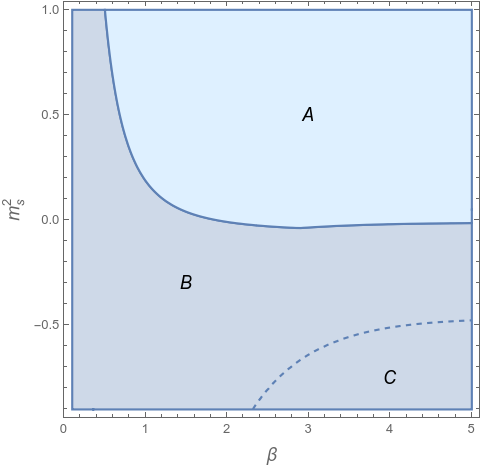} 
    \caption{} 
    \label{YAdS3tI} 
    \vspace{1ex}
  \end{subfigure}
  %\end{minipage}%%%
 %\begin{minipage}{0.5\textwidth}%    
   \begin{subfigure}[b]{0.3\linewidth}
    \centering
    \includegraphics[width=1\linewidth]{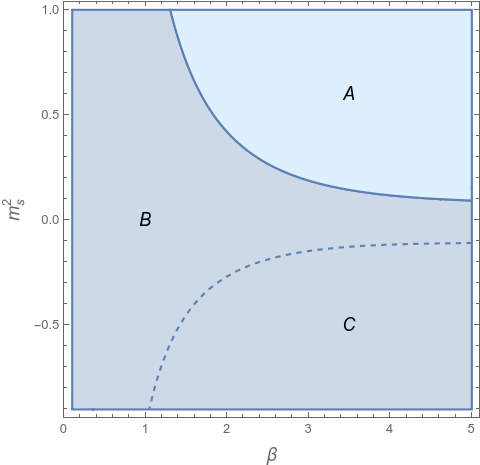} 
    \caption{} 
    \label{YAdS3tII} 
    \vspace{1ex}
  \end{subfigure}
%\end{minipage}
%\begin{minipage}{0.5\textwidth}%   
   \begin{subfigure}[b]{0.3\linewidth}
    \centering
    \includegraphics[width=1\linewidth]{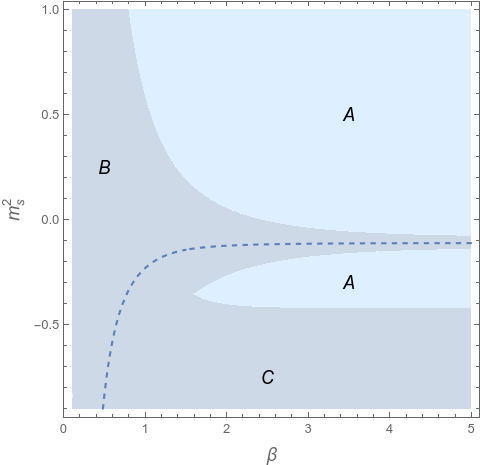} 
    \caption{} 
    \label{YAdS3tIII} 
    \vspace{1ex}
  \end{subfigure}
  \end{minipage}%%%
  \caption{Phase plots in the $m^2_s-\b$ plane for (a) $m_f=0.2$, $g=0.2$ (b) $m_f=0.1$, $g=0.4$ (c) $m_f=1$, $g=0.8$ which are points in region $I$, $II$ and $III$ of the zero temperature plot (Figure \ref{YAdS30tmfgr}) respectively.}
  \label{YAdS3t} 
  \end{center} 
\end{figure}

Inserting the finite temperature pieces,
\beqa
V_{eff}=V^0_{eff}+ \f{2}{\pi\b}\sum^{\infty}_{n=1}\frac{1}{n}\frac{e^{-n\beta\left(1+\sqrt{1+M_s^2}\right)}}{|1-e^{-n\beta}|^2}-\f{4}{\pi\b}\sum^{\infty}_{n=1}\frac{(-1)^n}{n}\frac{e^{-n\beta\left(1+|M_f|\right)}}{|1-e^{-n\beta}|^2}
\eeqa

The phase plots in the  $m^2_s-\b$ plane for various values of ($m_f$, $g$) are shown in Figure \ref{YAdS3t}. In all the plots, region $A$ has one minimum. Regions $B$ and $C$ have two minima which exchange dominance across the dashed lines. At the boundary of $A$ a new minimum appears from either of the two branches of the potential satisfying $m_f+g\phi_{cl} >0$ or $m_f+g\phi_{cl}<0$. The regions $B$ and $C$ which have two minima are the union of all the regions coming from either of the branches. The asymptotic zero temperature $m_s^2$ values for the phase boundaries are given in Table  \ref{YAdS30tv}.

\begin{table}[h]
\begin{center}
\begin{tabular}{|c|c|c|}
\hline
Figure & $m_{sA}^{2*}$ & $m_{sB}^{2*}$\\
\hline
\ref{YAdS3tI} &-0.014&-0.465\\
\hline
\ref{YAdS3tII} &0.079&-0.106\\
\hline
\ref{YAdS3tIII} &-0.082, -0.133, -0.421&-0.109\\
\hline
\end{tabular}
\end{center}
\caption{Zero temperature values of $m_s^2$ for phase boundaries.  $m_{sA}^{2*}$ is the zero temperature value for the boundary separating $A$ and $B$ in Figures \ref{YAdS3tI}, \ref{YAdS3tII} and boundary separating $A$-$B$ and $A$-$C$ in Figure \ref{YAdS3tIII}. $m_{sB}^{2*}$ that of the dashed boundary separating $B$ and $C$ in all phase plots of Figure \ref{YAdS3t}.}
\label{YAdS30tv}
\end{table}

At finite temperature, starting from a ($m_f$, $g$) point with $m_s^2=-0.2$ in the regions $I$, $II$ and $III$ of the zero-temperature phase plot (for example Figure  \ref{YAdS30tmfgr}) we have the following.\\
\noindent
1. At high temperatures the right minimum is always the global minimum.  A point in region $I$ (in Figure \ref{YAdS30tmfgr}) thus always has lower right minimum for all temperatures. This point lies in region $B$ in Figure \ref{YAdS3tI}. Points lying in region $C$ undergoes transition across the dashed line as $\b$ is decreased. \\ 
\noindent
2. A point in region $II$ (in Figure \ref{YAdS30tmfgr}) lies in region $C$ in Figure \ref{YAdS3tII}. The potential undergoes change of dominance of the minima across the dashed line shown in Figure \ref{YAdS3tII}.\\
\noindent
3. If one starts from a point in region $III$ (in Figure \ref{YAdS30tmfgr}), this point lies in region $A$ of Figure \ref{YAdS3tIII}. At finite temperature the right minimum appears which is followed by a first-order phase transition beyond which it becomes the lowest minimum as temperature is increased further.

\subsubsection{AdS$_2$}

The zero temperature traces are UV divergent and can be regularized using dimensional regularization. Expanding about $d=1$ with $\e=1-d$, the zero temperature traces for the scalar from \cite{Kakkar:2022hub} and the fermions (from equation (\ref{trzerotf4})) are
\beqa\label{expads2s}
\f{\m^{-\e}}{{2\cal V}_{d+1}}\mbox{tr} \frac{1}{-\square_E+M^{2}_s}=\frac{1}{4 \pi \e}+\frac{1}{8\pi}\left[-2 \psi^{(0)}\left(\n+\frac{1}{2}\right)-\gamma +\log (4\pi/\m^2 )\right]+{\cal O}\left(d-1\right)
\eeqa
\beqa\label{expads2f}
\f{\m^{-\e}}{{\cal V}_{d+1}}\mbox{tr} \frac{1}{\slashed{D}+M_f}=4 M_f\left[\frac{1}{4 \pi \e}-\frac{1}{8\pi}\left(\psi^{(0)}\left(|M_f|\right)+ \psi^{(0)}\left(|M_f|+1\right)+\gamma -\log (2\pi /\m^2)\right)\right]+{\cal O}\left(d-1\right)\non
\eeqa
     
In terms of Feynman diagrams, the UV divergences arise from the two-point and the one-point diagrams. We thus use the following structure of the counterterms
\[\phi_{cl}~\d\l_1+\f{1}{2}\phi_{cl}^2~\d m^2_s
\]

with renormalization conditions
\beqa\label{rcads2f1}
\left.\f{\pa V^0_{eff}}{\pa \phi_{cl}}\right|_{\phi_{cl}=0}=0~~~~~~~~~~\left.\f{\pa^2 V^0_{eff}}{\pa \phi_{cl}^2}\right|_{\phi_{cl}=0}=m_s^2.
\eeqa

These give the following zero temperature effective potential
\beqa\label{vztads2f}
V^0_{eff}&=&\f{1}{2}m_s^2\phi_{cl}^2+\f{\l_3}{3!}\phi_{cl}^3+\f{\l_4}{4!}\phi_{cl}^4\non 
&-&\f{1}{4\pi}\left[\int^{M_s^2}_{m_s^2}\psi^{(0)}\left(\nu(\phi_{cl})+1/2\right)dM_s^2-\int_{m_f}^{M_f}2 M_f\left(\psi^{(0)}\left(|M_f|\right)+\psi^{(0)}\left(1+|M_f|\right)\right)dM_f\right]\non
&+&\f{1}{8\pi}\phi_{cl}^2\left[\l_4\psi^{(0)}\left(\sqrt{1/4+m_s^2}+1/2\right)
-2g^2\left(\psi^{(0)}\left(m_f\right)+\psi^{(0)}\left(1+m_f\right)\right)\right.\non
&-&\left.2g^2 m_f\left(\psi^{(1)}\left(m_f\right)+\psi^{(1)}\left(1+m_f\right)\right)
+\f{1}{2}\l_3^2\f{\psi^{(1)}\left(\sqrt{1/4+m_s^2}+1/2\right)}{\sqrt{1/4+m_s^2}}\right]\non
&+&\f{1}{8\pi}\phi_{cl}\left[2\l_3\psi^{(0)}\left(\sqrt{1/4+m_s^2}+1/2\right)-4gm_f\left(\psi^{(0)}\left(m_f\right)+\psi^{(0)}\left(1+m_f\right)\right)\right]
\eeqa

We have set the lower limits of the integrals in the above expression as $m^2_s$ and $m_f$ for the scalar and the fermion traces respectively. This sets $V_{eff}^0$ at $\phi_{cl}=0$ equal to zero.

\begin{figure}[H] 
\begin{center} 
  \begin{minipage}{0.28\textwidth}%   
   \begin{subfigure}[b]{0.98\linewidth}
    \centering
    \includegraphics[width=0.98\linewidth]{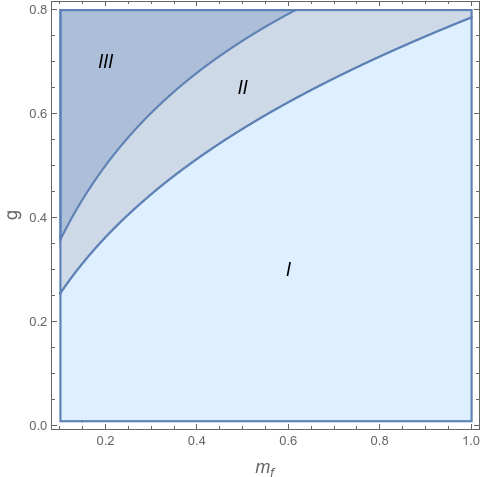} 
    \caption{} 
    \label{YAdS20tmfgr} 
    \vspace{1ex}
  \end{subfigure}
  \end{minipage}%%%
  \begin{minipage}{0.28\textwidth}%   
  \begin{subfigure}[b]{0.98\linewidth}
    \centering
    \includegraphics[width=0.98\linewidth]{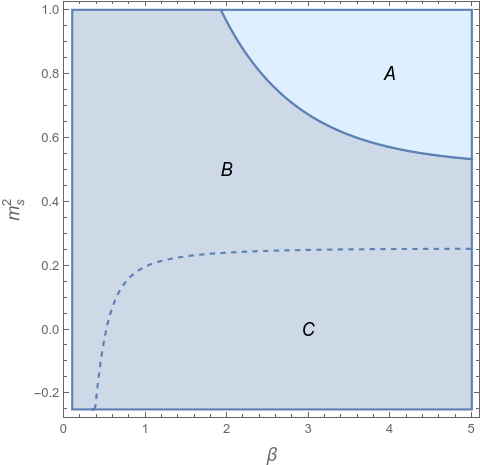} 
   \caption{} 
    \label{YAdS2tbms1r} 
  \end{subfigure}   
  \end{minipage}%%%
  \begin{minipage}{0.44\textwidth}%
  \begin{subfigure}[b]{0.5\linewidth}
    \centering
    \includegraphics[width=0.95\linewidth]{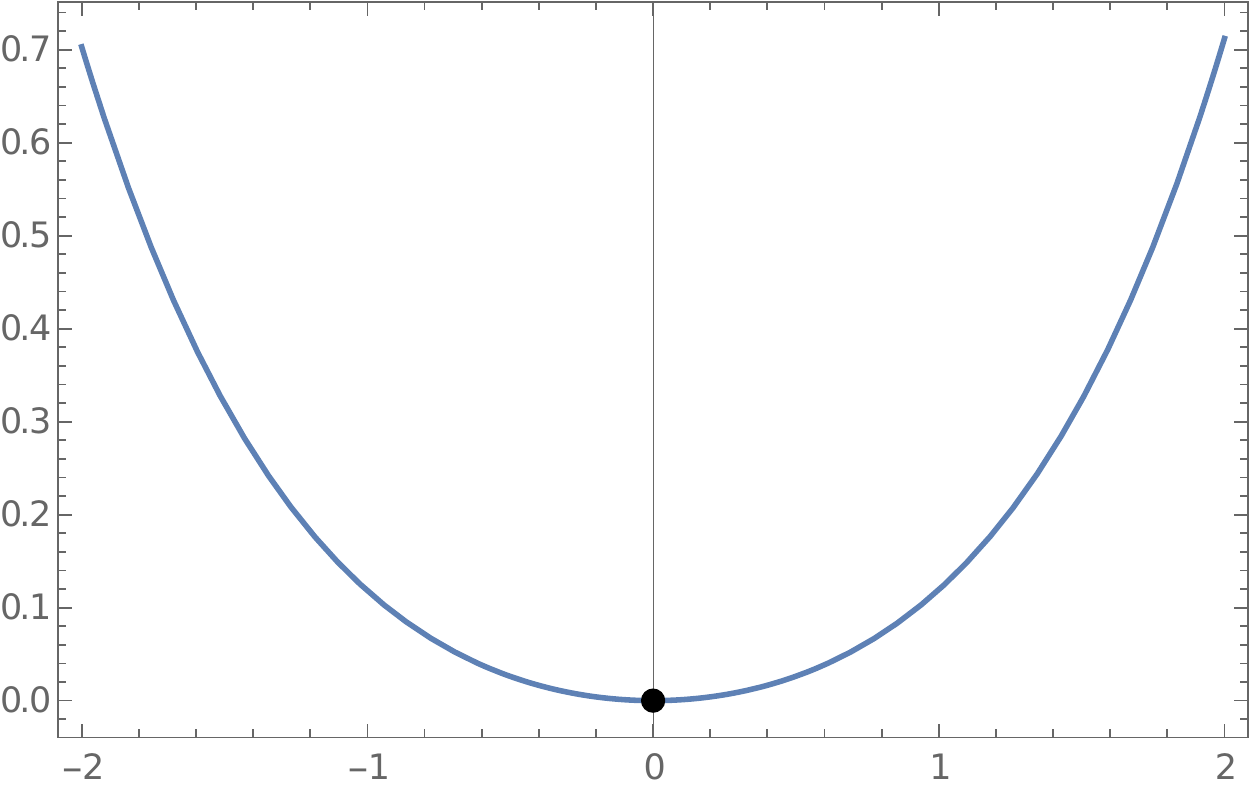} 
    \caption*{$I$: $m_f=0.6$, $g=0.2$} 
    %\label{a} 
    %\vspace{1ex}
  \end{subfigure}%
  %\hspace{-10em}
  \begin{subfigure}[b]{0.5\linewidth}
    \centering
    \includegraphics[width=0.95\linewidth]{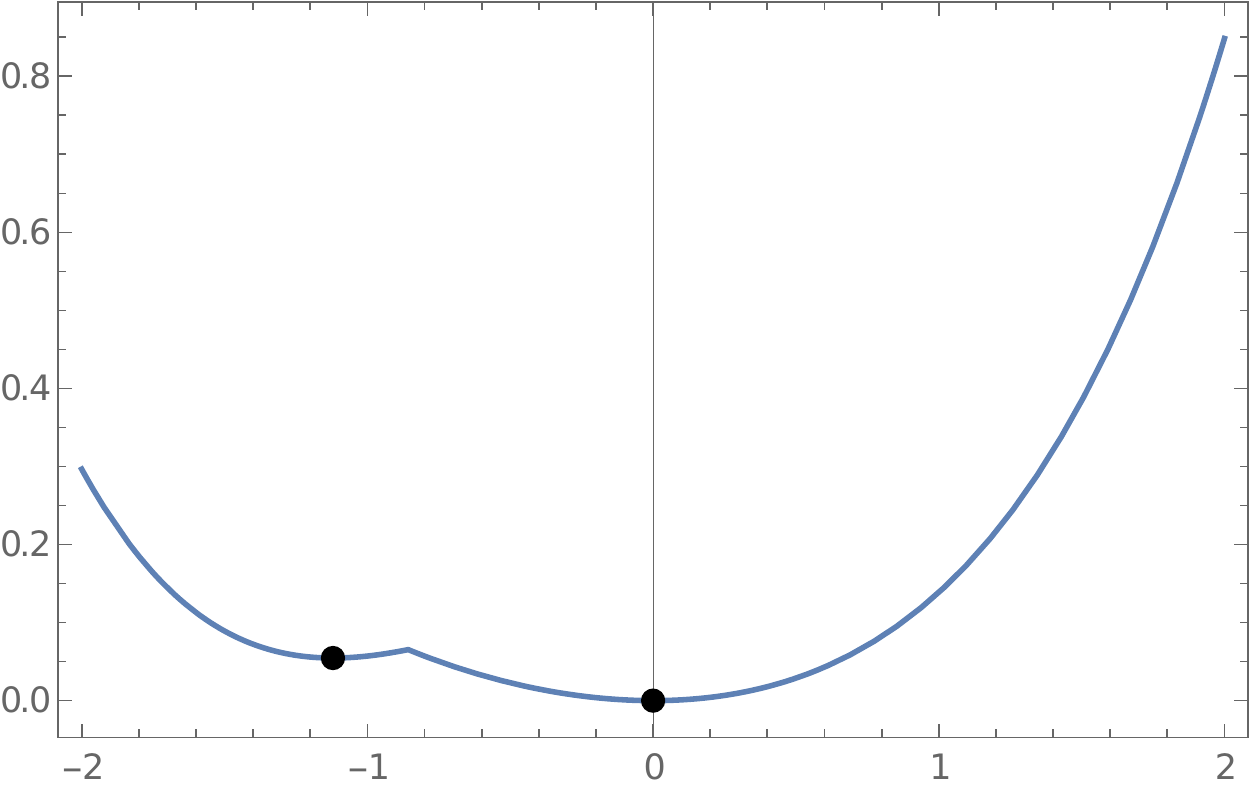} 
    \caption*{$II$: $m_f=0.6$, $g=0.7$} 
    %\label{b} 
   % \vspace{1ex}
  \end{subfigure} 
  \begin{subfigure}[b]{0.48\linewidth}
    \centering
    \includegraphics[width=0.95\linewidth]{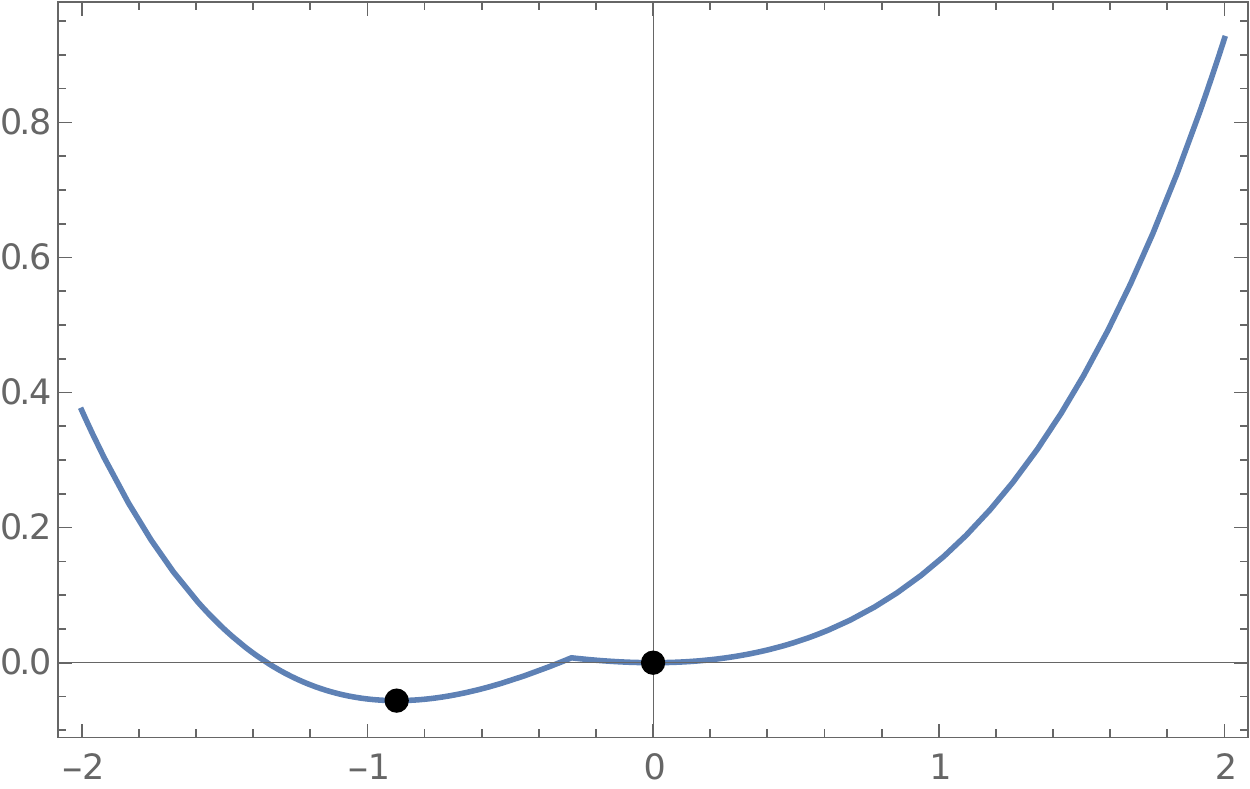} 
    \caption*{$III$: $m_f=0.2$, $g=0.7$} 
    %\label{b} 
    %\vspace{1ex}
  \end{subfigure} 
  \begin{subfigure}[b]{0.5\linewidth}
    \centering
    \includegraphics[width=0.95\linewidth]{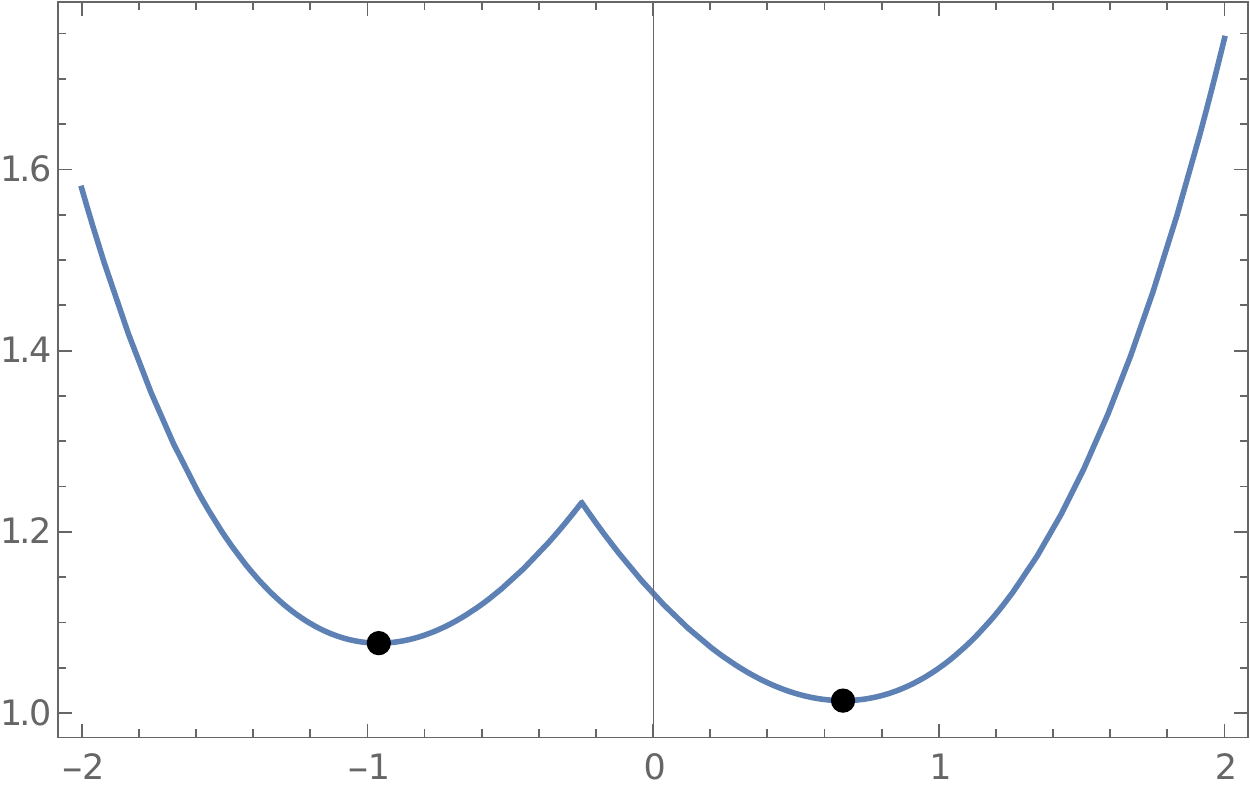} 
    \caption*{\small{$B$: $\b=1$, $m_s^2=0.4$}} 
    %\label{c} 
  \end{subfigure}%%
  %\hspace{-5em}
 % \label{fig7} 
\end{minipage}
  \caption{(a): Phases at zero temperature for $m_s^2=0.2$ in the $m_f$-$g$ plane  and representative potentials $I$, $II$ and $III$ (extreme right). (b): Regions in the $\b$-$m_s^2$ plane for $m_f=0.1$, $g=0.4$ a point in region $III$ in (a). A  plot of the potential at finite temperature corresponding to region $B$ in (b) is shown in the right. Black dots on the potential plots are the extrema. }
  \label{YAdS2tbms1} 
  \end{center} 
\end{figure}

The finite temperature effective potential using (\ref{rhsp3fdd}) and including the expression for scalars is 
\beqa
V_{eff}&=&V^0_{eff}+ \f{1}{\b}\sum^{\infty}_{n=1}\frac{1}{n}\frac{e^{-n\beta\left(1/2+\sqrt{1/4+M_s^2}\right)}}{|1-e^{-n\beta}|}-\f{2}{\b}\sum^{\infty}_{n=1}\frac{(-1)^n}{n}\frac{e^{-n\beta\left(1/2+|M_f|\right)}}{|1-e^{-n\beta}|}~.
\eeqa

\noindent
The phases of this theory are summarized as follows. In the plots we have set $\l_3=0$ and $\l_4=0.5$. 
\\
\noindent
1. The renormalization conditions (\ref{rcads2f1}) imply that at zero temperature the extremum at $\phi_{cl}=0$ is either a maximum or a minimum  which is determined by the sign of the scalar mass-squared, $m_s^2$. When $m_s^2<0$, there are two minima. The left minimum is lower than the right for all values of $m_f$ and $g$. When $m_s^2>0$, there is either a single minimum or two minima depending on the values of $m_f$ and $g$ and $m_s^2$, see Figure \ref{YAdS20tmfgr}. Region $I$ has one minimum. Two minima exchange dominance across the boundary separating regions $II$ and $III$. 

\begin{figure}[h] 
\begin{center} 
  \begin{minipage}{0.28\textwidth}%   
   \begin{subfigure}[b]{0.98\linewidth}
    \centering
    \includegraphics[width=0.98\linewidth]{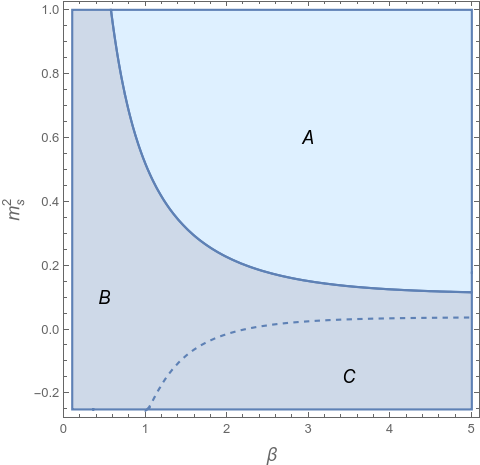} 
    \caption{} 
    \label{YAdS2tbmsr} 
    \vspace{1ex}
  \end{subfigure}
  \end{minipage}%%%
  \begin{minipage}{0.28\textwidth}%   
  \begin{subfigure}[b]{0.98\linewidth}
    \centering
    \includegraphics[width=0.98\linewidth]{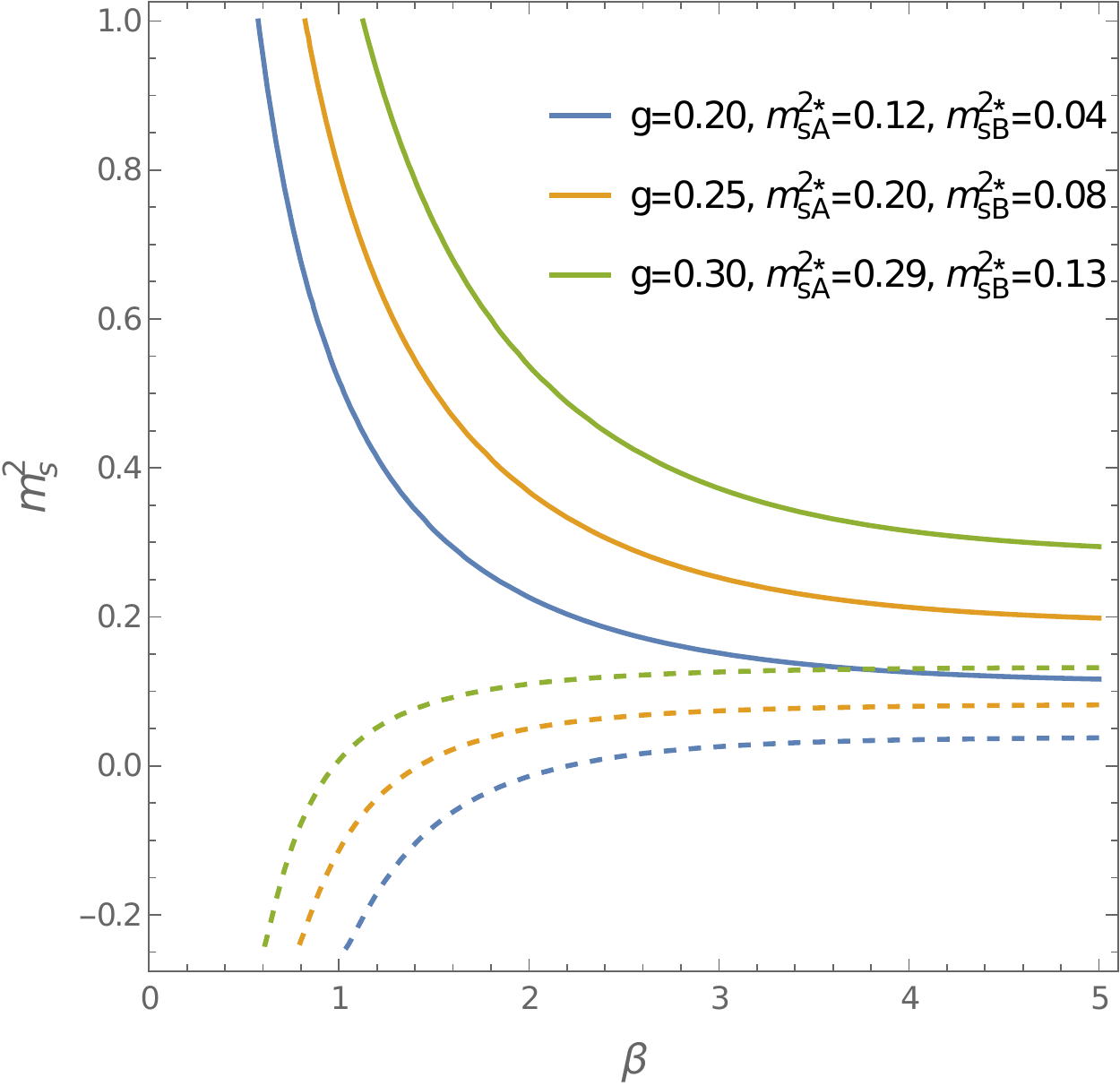} 
   \caption{} 
    \label{YAdS2tbmsc} 
  \end{subfigure}   
  \end{minipage}%%%
  \begin{minipage}{0.44\textwidth}%
  \begin{subfigure}[b]{0.5\linewidth}
    \centering
    \includegraphics[width=0.95\linewidth]{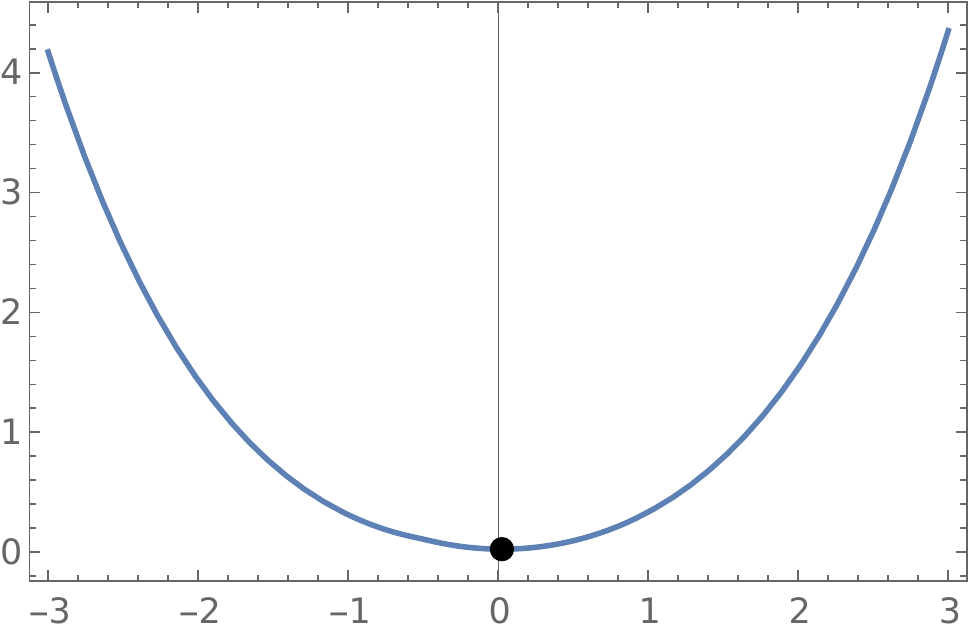} 
    \caption*{$A$: $\b=3$, $m_s^2=0.6$} 
    %\label{a} 
    %\vspace{1ex}
  \end{subfigure}%
  %\hspace{-10em}
  \begin{subfigure}[b]{0.5\linewidth}
    \centering
    \includegraphics[width=0.95\linewidth]{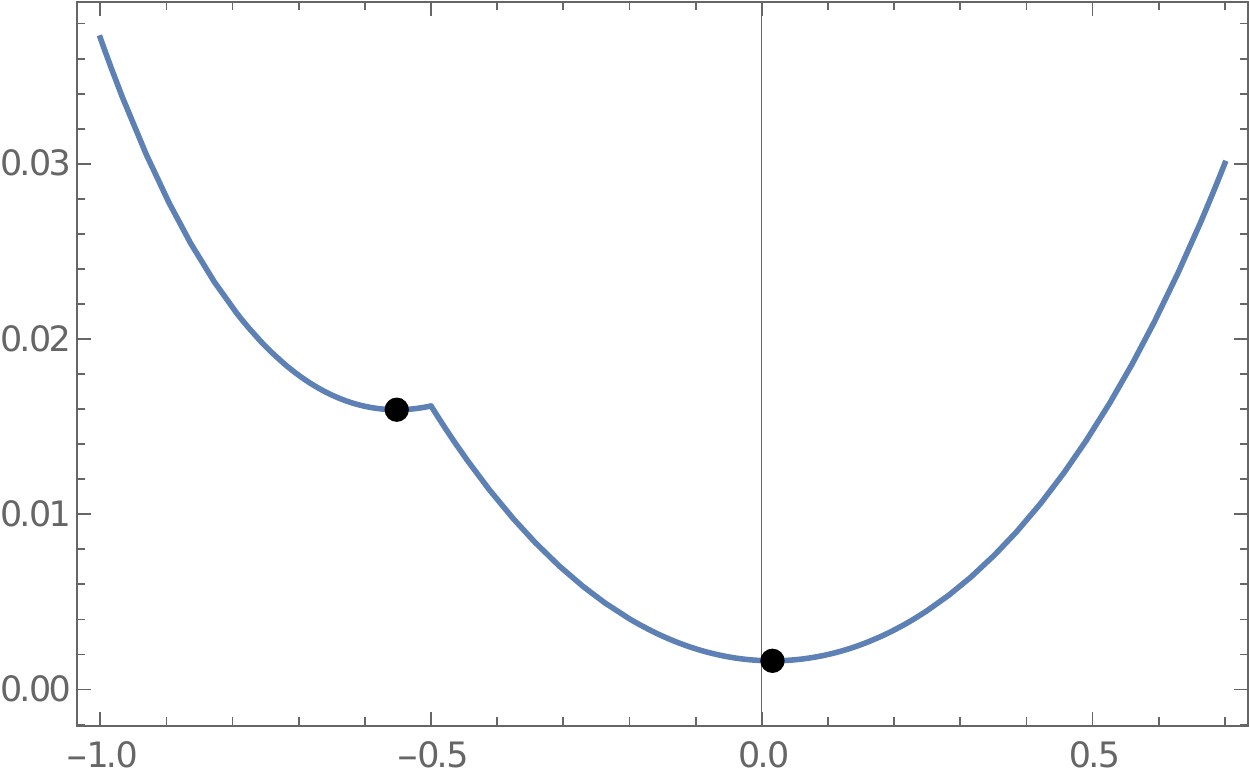} 
    \caption*{$B$: $\b=5$, $m_s^2=0.1$} 
    %\label{b} 
   % \vspace{1ex}
  \end{subfigure} 
  \begin{subfigure}[b]{0.48\linewidth}
    \centering
    \includegraphics[width=0.95\linewidth]{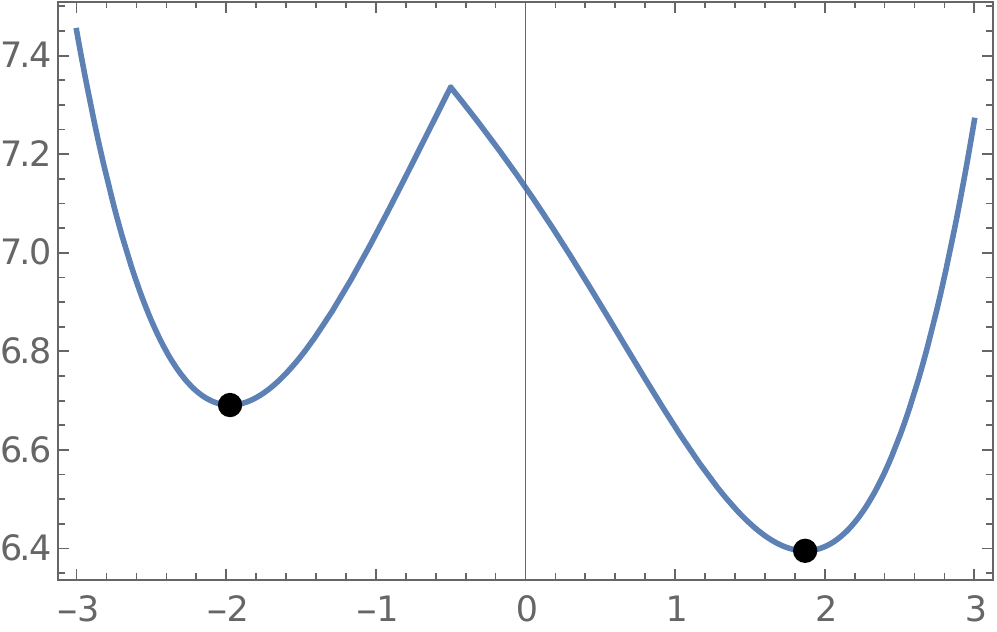} 
    \caption*{$B$: $\b=0.5$, $m_s^2=0.1$} 
    %\label{b} 
    %\vspace{1ex}
  \end{subfigure} 
  \begin{subfigure}[b]{0.5\linewidth}
    \centering
    \includegraphics[width=0.95\linewidth]{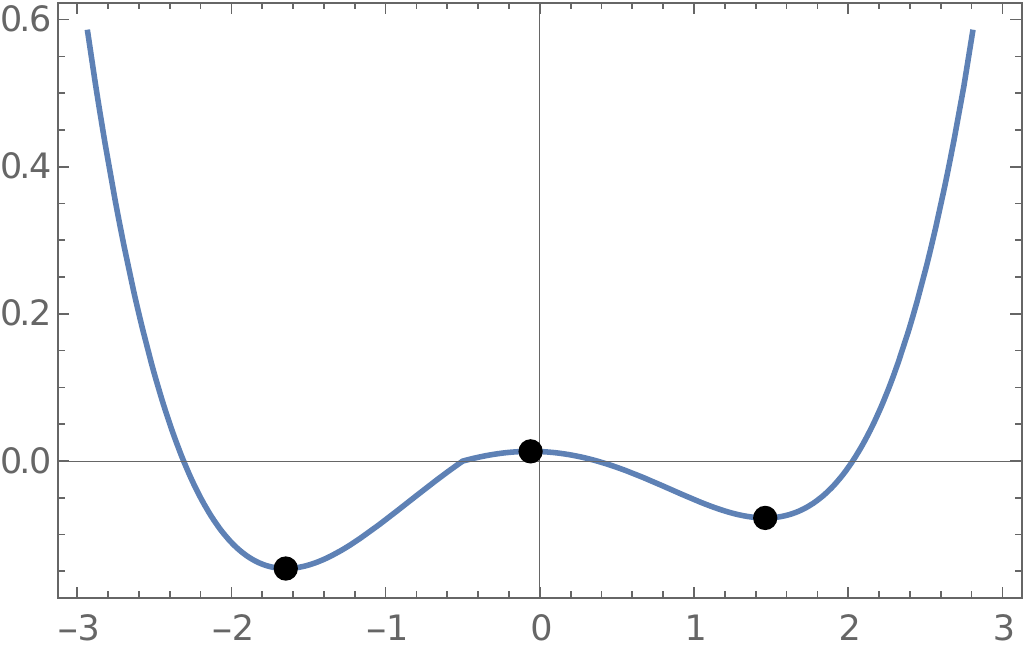} 
    \caption*{\small{$C$:$\b=3.5$, $m_s^2=-0.15$}} 
    %\label{c} 
  \end{subfigure}%%
  %\hspace{-5em}
 % \label{fig7} 
\end{minipage}
  \caption{(a): Regions in the $\b$-$m_s^2$ plane for $m_f=0.1$, $g=0.2$ and representative potentials (right). Black dots on the potential plots are the extrema. (b): Shows the phase boundaries for $m_f=0.1$ and various values of $g$.}
  \label{YAdS2tbms} 
  \end{center} 
\end{figure}

\noindent
2. At finite temperatures, $\b-m^2_s$ plots are shown in Figures  \ref{YAdS2tbms1r} and \ref{YAdS2tbmsr} corresponding to a point in region $I$ and $III$ of Figure \ref{YAdS20tmfgr} respectively. In these phase plots, the region $A$ has only one minimum. Regions $B$ and $C$ have two minima. Note that at zero temperature (large $\b$), a ($m_f$,$g$) point (for $m_s^2=0.2$) corresponding to $I$ and $III$ lies in $A$ and in $C$ respectively.

\noindent
3. The dashed lines separating regions $B$ and $C$ in Figures  \ref{YAdS2tbms1r} and \ref{YAdS2tbmsr} are  phase boundaries where the two minima exchange dominance. Potentials corresponding to representative points in these regions are shown in the extreme right plots of Figures \ref{YAdS2tbms1} and \ref{YAdS2tbms}. The relevant boundaries for $m_f=0.1$ and various values of $g$ are shown in Figure \ref{YAdS2tbmsc}. The asymptotic valuses of $m_s^2$ at zero temperature, $m_{sA}^{2*}$, $m_{sB}^{2*}$ for the solid and dashed boundary curves respectively are given within the Figure \ref{YAdS2tbmsc}. The corresponding values in Figure \ref{YAdS2tbms1r} are
$m_{sA}^{2*}=0.4$, $m_{sB}^{2*}=0.2$.

\subsubsection{AdS$_4$}\label{Yads4sec}

Expanding the zero temperature traces about $d=3$ with $\e=3-d$, from \cite{Kakkar:2022hub} we have

\beqa\label{expads4s}
\frac{\m^{-\e}}{{\cal V}^{}_{d+1}}\mbox{tr} \frac{1}{-\square+M_s^{2}}=\frac{(2+M_s^2)}{16\pi^2}\left[-\frac{2}{\e}-1+\gamma-\log(4\pi)+\psi^{(0)}\left(\n-\f{1}{2}\right)+\psi^{(0)}\left(\n+\f{3}{2}\right)+\log(\m^2)\right]\non
\eeqa

and from (\ref{trzerotf4})

\beqa\label{expads4f}
\f{\m^{-\e}}{{\cal V}_{d+1}}\mbox{tr} \frac{1}{\slashed{D}+M_f}=\frac{M_f(M_f^2-1)}{4\pi^2}\left[-\frac{2}{\e}-1+\gamma-\log(2\pi)+\psi^{(0)}\left(|M_f|-1\right)+\psi^{(0)}\left(|M_f|+2\right)+\log(\m^2)\right]\non
\eeqa

In four dimensions one to four point functions in $\phi_{cl}$ are UV divergent. Here again we use the renormalization scheme as in the previous section for AdS$_2$. The counterterms have the following form

\[\phi_{cl}~\d\l_1+\f{1}{2}\phi_{cl}^2~\d m^2_s+\f{1}{3!}\phi_{cl}^3~\d \l_3+\f{1}{4!}\phi_{cl}^4~\d \l_4
\]

We use the with renormalization conditions at $\phi_{cl}=0$

\beqa\label{rcads4f1}
\f{\pa V^0_{eff}}{\pa \phi_{cl}}=0~~~~~~~~~~\f{\pa^2 V^0_{eff}}{\pa \phi_{cl}^2}=m_s^2~~~~~~~~~~\f{\pa^3 V^0_{eff}}{\pa \phi_{cl}^3}=\l_3~~~~~~~~~~\f{\pa^4 V^0_{eff}}{\pa \phi_{cl}^4}=\l_4.
\eeqa

The expression for the effective potential $V^0_{eff}$ utilizing the above renormalization condition is quite long and can be found in  
equation (\ref{vztads4f}) of appendix \ref{Yads4app}. In the large $\phi_{cl}$ limit, the leading contributions from the scalar and the fermion to the potential come from the integrals in (\ref{vztads4f}), 

\beqa
\f{1}{2}\int^{M_s^2}_{m_s^2} \left(\frac{2+M_s^2}{16 \pi^2}\right) \left(\psi^{(0)} \left(\nu(\phi_{cl})-\frac{1}{2}\right)+\psi^{(0)} \left(\nu(\phi_{cl})+\frac{3}{2}\right)\right)dM_s^2 \sim \f{\l_4^2}{128\pi^2}\phi^4_{cl}\log(\phi_{cl})
\eeqa

\beqa
-\f{1}{4 \pi^2}\int^{M_f}_{m_f}M_f\left(M_f^2-1 \right) \left(\psi^{(0)} \left(|M_f|-1\right)+\psi^{(0)} \left(|M_f|+2\right)\right)dM_f\sim -\f{g^4}{8\pi^2}\phi^4_{cl}\log(\phi_{cl})
\eeqa
   
These asymptotic behavior imply that the potential remains bounded for $\l_4>4g^2$. Note that this constraint is same as that in flat space \cite{Weinberg:1973am}. It is taken care of in the following numerical analysis where $\l_4=0.8$ so that $g<0.447$. $\l_3$ is set to zero.

The full potential including the finite temperature piece is 
\beqa
V_{eff}&=&V^0_{eff}-\f{3}{2\pi\b}\sum^{\infty}_{n=1}\frac{1}{n}\frac{e^{-n\beta\left(\frac{3}{2}+\sqrt{\frac{9}{4}+M_s^2}\right)}}{|1-e^{-n\beta}|^3}+\f{6}{\pi\b}\sum^{\infty}_{n=1}\frac{(-1)^n}{n}\frac{e^{-n\beta\left(\frac{3}{2}+|M_f|\right)}}{|1-e^{-n\beta}|^3}~.
\eeqa

\begin{figure}[H] 
\begin{center} 
  \begin{minipage}{0.28\textwidth}%   
   \begin{subfigure}[b]{0.98\linewidth}
    \centering
    \includegraphics[width=0.98\linewidth]{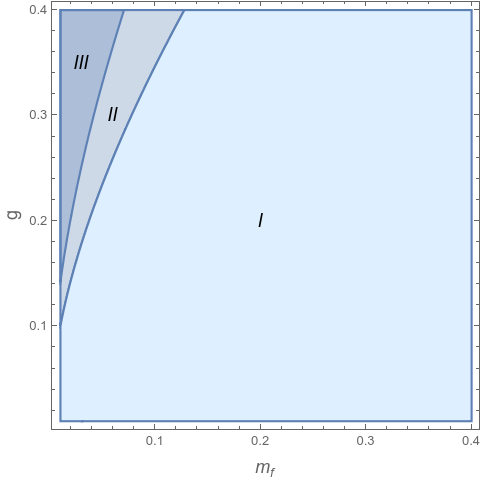} 
    \caption{} 
    \label{YAdS40tmfgr} 
    \vspace{1ex}
  \end{subfigure}
  \end{minipage}%%%
  \begin{minipage}{0.28\textwidth}%   
  \begin{subfigure}[b]{0.98\linewidth}
    \centering
    \includegraphics[width=0.98\linewidth]{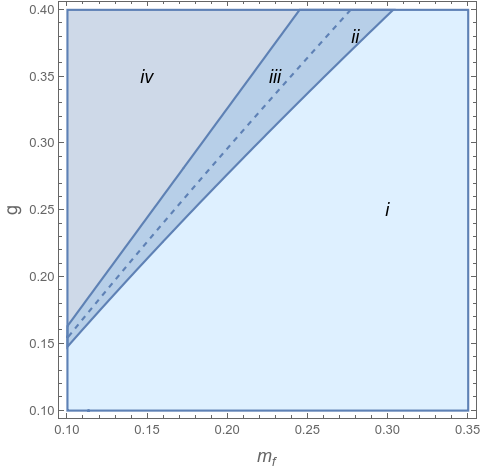} 
   \caption{} 
    \label{YAdS40tmfg1r} 
  \end{subfigure}   
  \end{minipage}%%%
  \begin{minipage}{0.44\textwidth}%
  \begin{subfigure}[b]{0.5\linewidth}
    \centering
    \includegraphics[width=0.95\linewidth]{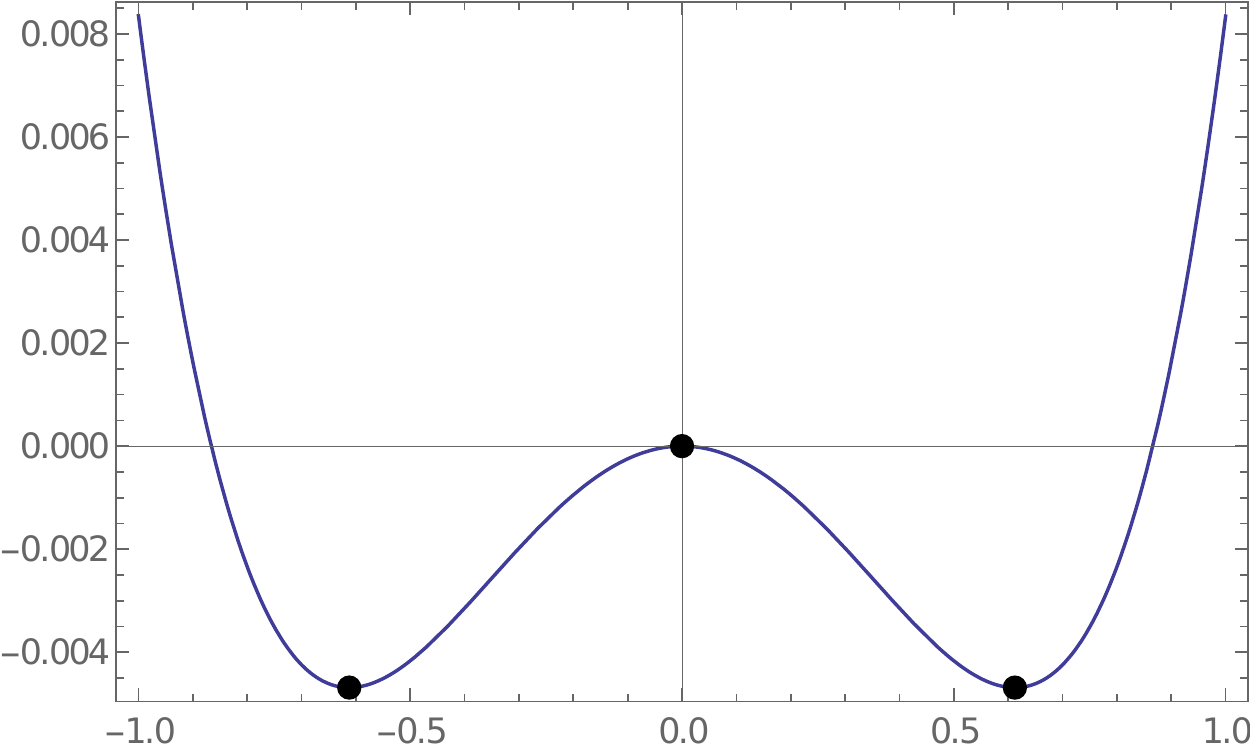} 
    \caption*{$i$: $m_f=0.3$, $g=0.25$} 
    %\label{a} 
    %\vspace{1ex}
  \end{subfigure}%
  %\hspace{-10em}
  \begin{subfigure}[b]{0.5\linewidth}
    \centering
    \includegraphics[width=0.95\linewidth]{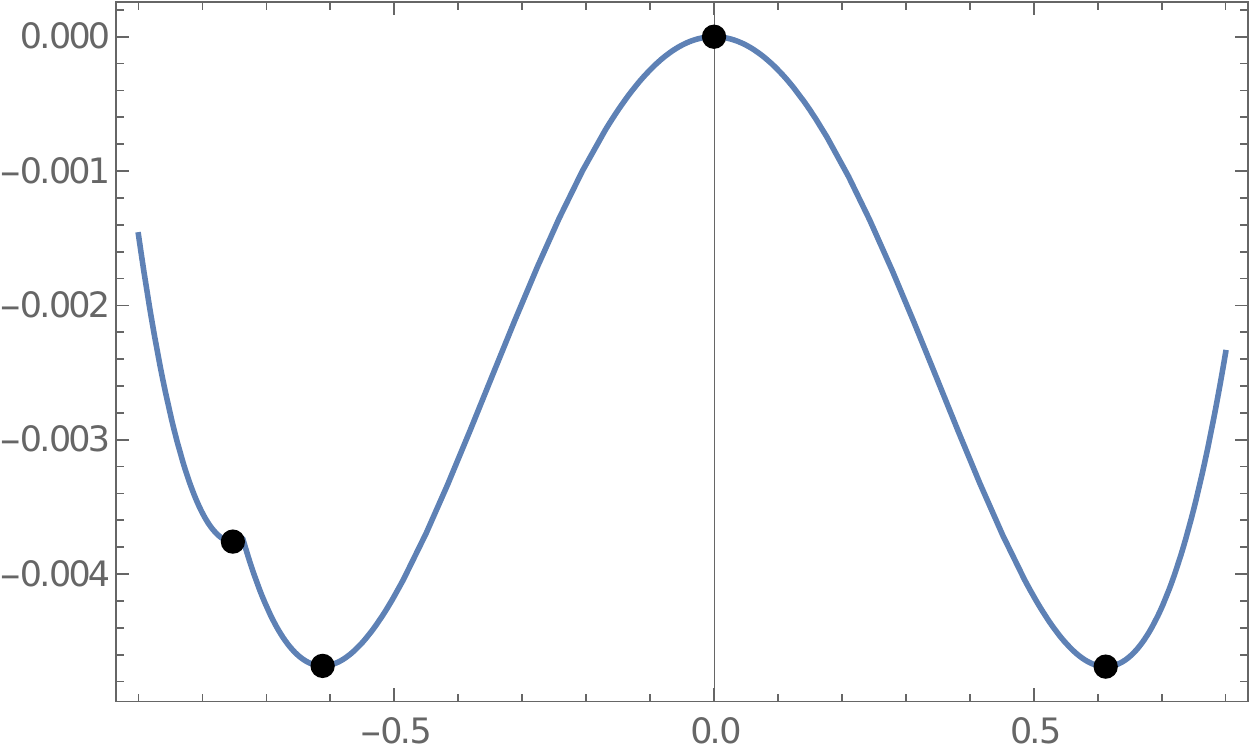} 
    \caption*{$ii$: $m_f=0.28$, $g=0.38$} 
    %\label{b} 
   % \vspace{1ex}
  \end{subfigure} 
  \begin{subfigure}[b]{0.48\linewidth}
    \centering
    \includegraphics[width=0.95\linewidth]{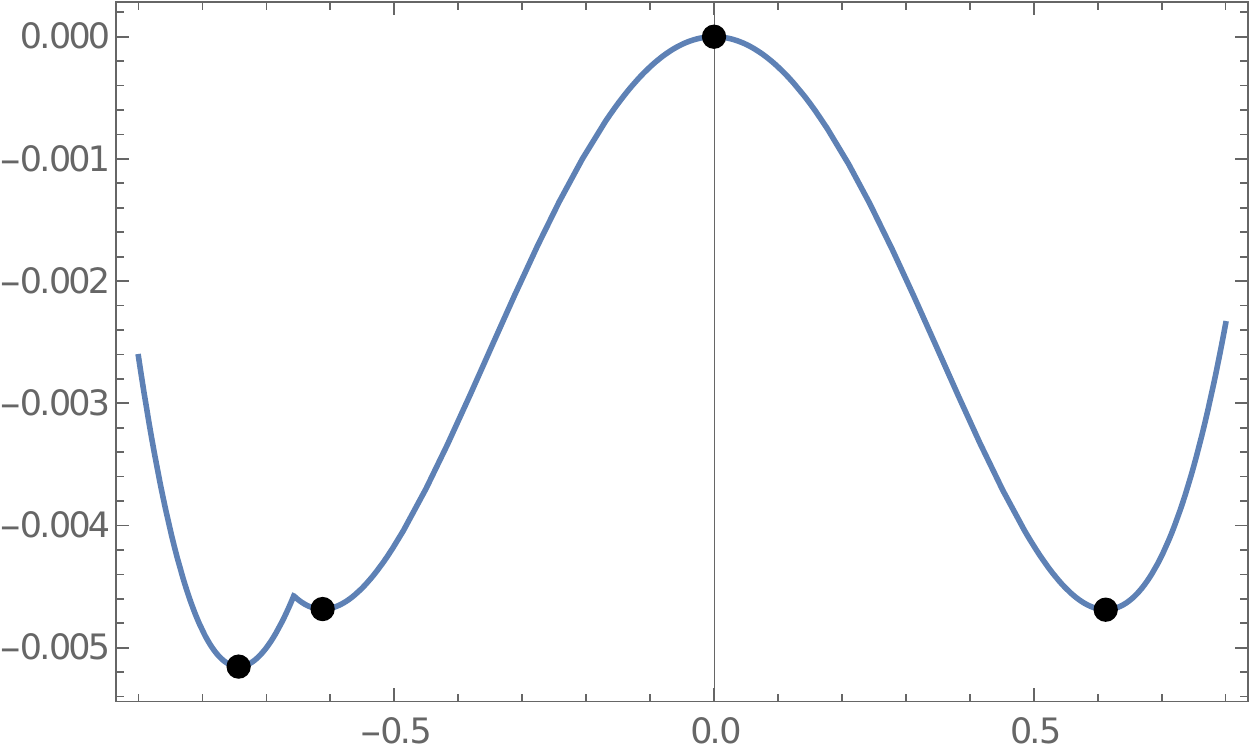} 
    \caption*{$iii$: $m_f=0.23$, $g=0.35$} 
    %\label{b} 
    %\vspace{1ex}
  \end{subfigure} 
  \begin{subfigure}[b]{0.5\linewidth}
    \centering
    \includegraphics[width=0.95\linewidth]{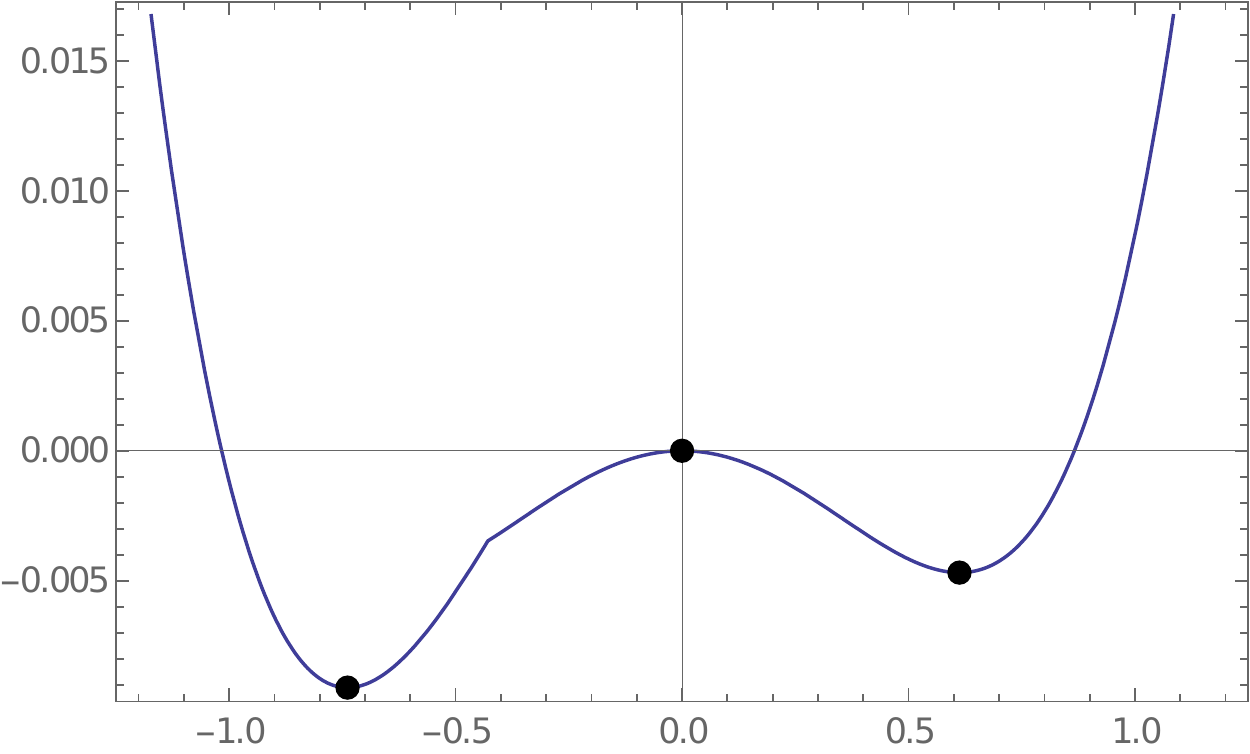} 
    \caption*{\small{$iv$: $m_f$=0.15, $g=0.35$}} 
    %\label{c} 
  \end{subfigure}%%
  %\hspace{-5em}
 % \label{fig7} 
\end{minipage}
  \caption{Phases at zero temperature in the $m_f$-$g$ plane for (a) $m_s^2=0.05$  and (b) $m_s^2=-0.05$. Representative potentials $i-iv$ corresponding to regions in (b) are on extreme right. Black dots on the potential plots are the extrema. }
  \label{YAdS40tmfg} 
  \end{center} 
\end{figure}

At zero temperature Figure \ref{YAdS40tmfg} shows the phases on the $m_f$-$g$ plane for positive and negative values of $m_s^2$ and related potential plots. In Figure \ref{YAdS40tmfgr} there is a transition across the boundary separating $II$ and $III$. This is same as in the case of AdS$_2$ . The potential plots corresponding to regions in Figure \ref{YAdS40tmfgr} are similar to those in Figure \ref{YAdS2tbms1} for the case of AdS$_2$.  

In Figure \ref{YAdS40tmfg1r}, we have plotted the regions $i$-$iv$ corresponding to the nature of the potentials (shown on extreme right) on the left of the origin $\phi_{cl}=0$. The minimum right of the origin and a maximum at the origin (as per the renormalization condition (\ref{rcads4f1})) always exist for negative values of $m_s^2$. The dashed line is the phase boundary. In regions $ii$ and $iii$ there are two minima left of the origin. 

\begin{figure}[H] 
\begin{center} 
  \begin{minipage}{0.28\textwidth}%   
   \begin{subfigure}[b]{0.98\linewidth}
    \centering
    \includegraphics[width=0.98\linewidth]{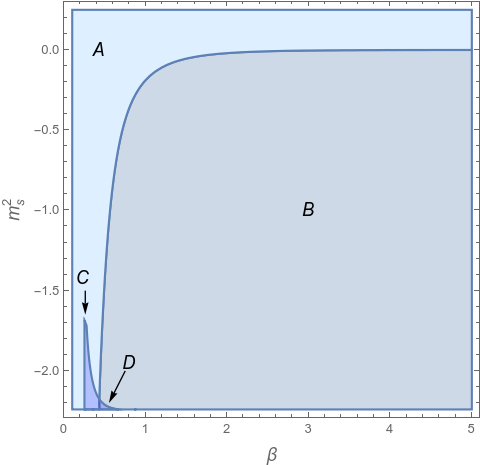} 
    \caption{} 
    \label{YAdS4tbmsr} 
    \vspace{1ex}
  \end{subfigure}
  \end{minipage}%%%
  \begin{minipage}{0.28\textwidth}%   
  \begin{subfigure}[b]{0.98\linewidth}
    \centering
    \includegraphics[width=0.98\linewidth]{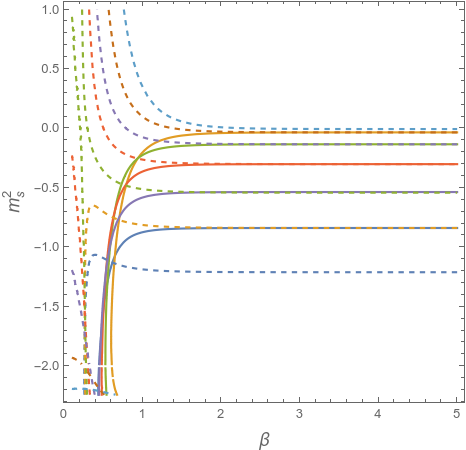} 
   \caption{} 
    \label{YAdS4tbmscr} 
  \end{subfigure}   
  \end{minipage}%%%
  \begin{minipage}{0.44\textwidth}%
  \begin{subfigure}[b]{0.5\linewidth}
    \centering
    \includegraphics[width=0.95\linewidth]{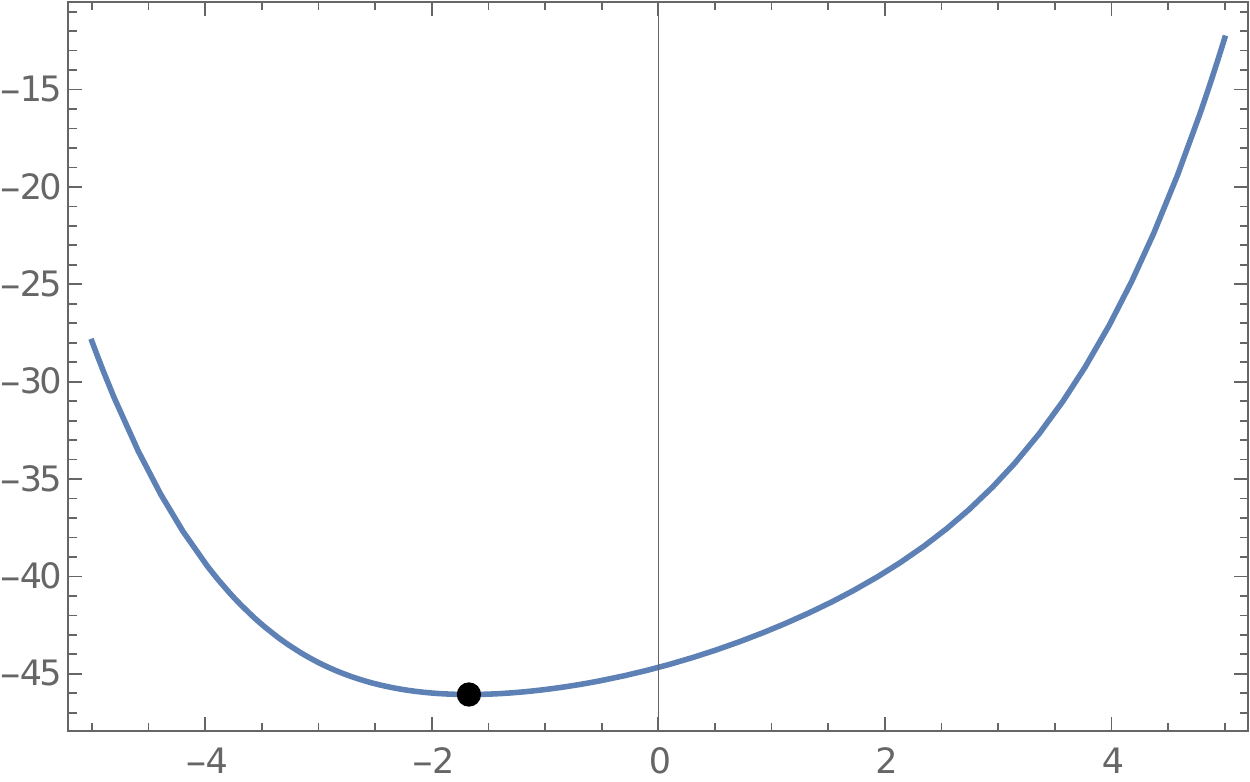} 
    \caption*{$A$: $\b=0.43$, $m^2_s=0$} 
    %\label{a} 
    %\vspace{1ex}
  \end{subfigure}%
  %\hspace{-10em}
  \begin{subfigure}[b]{0.5\linewidth}
    \centering
    \includegraphics[width=0.95\linewidth]{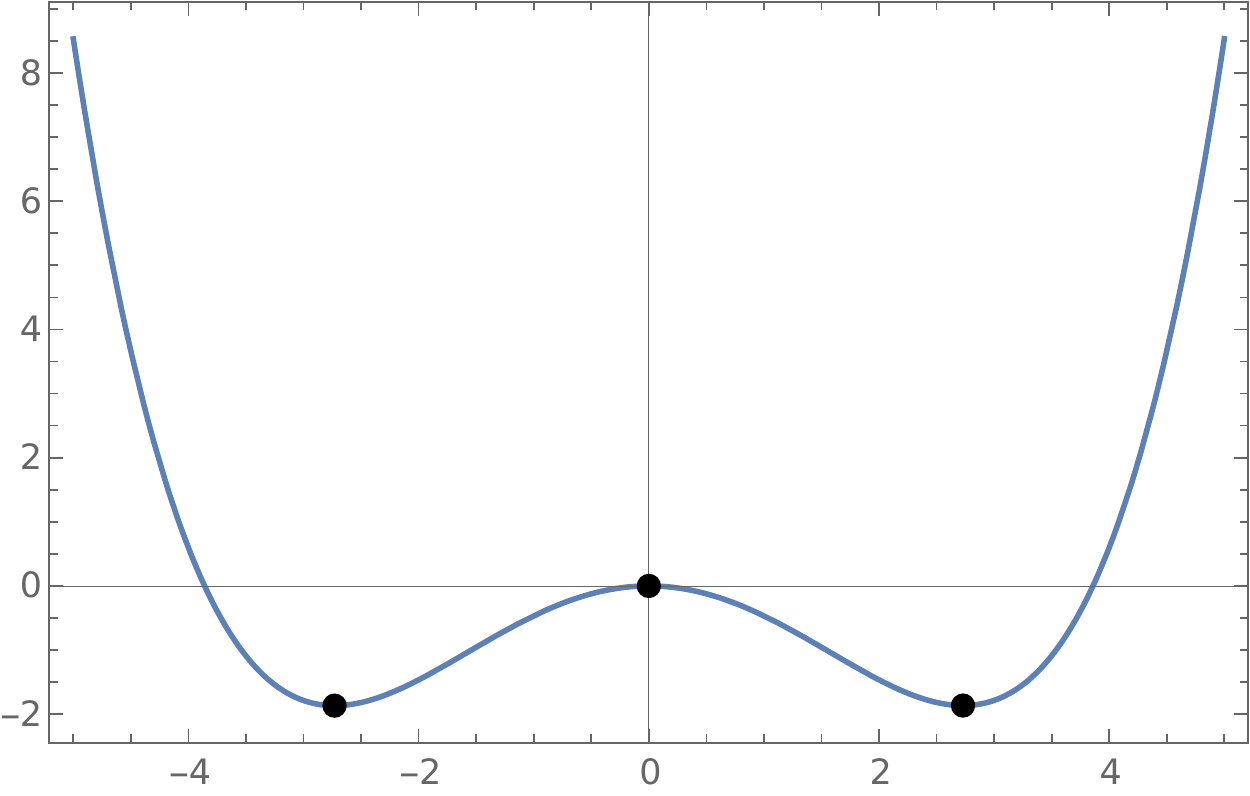} 
    \caption*{$B$: $\b=3$, $m^2_s=-1$} 
    %\label{b} 
   % \vspace{1ex}
  \end{subfigure} 
  \begin{subfigure}[b]{0.48\linewidth}
    \centering
    \includegraphics[width=0.98\linewidth]{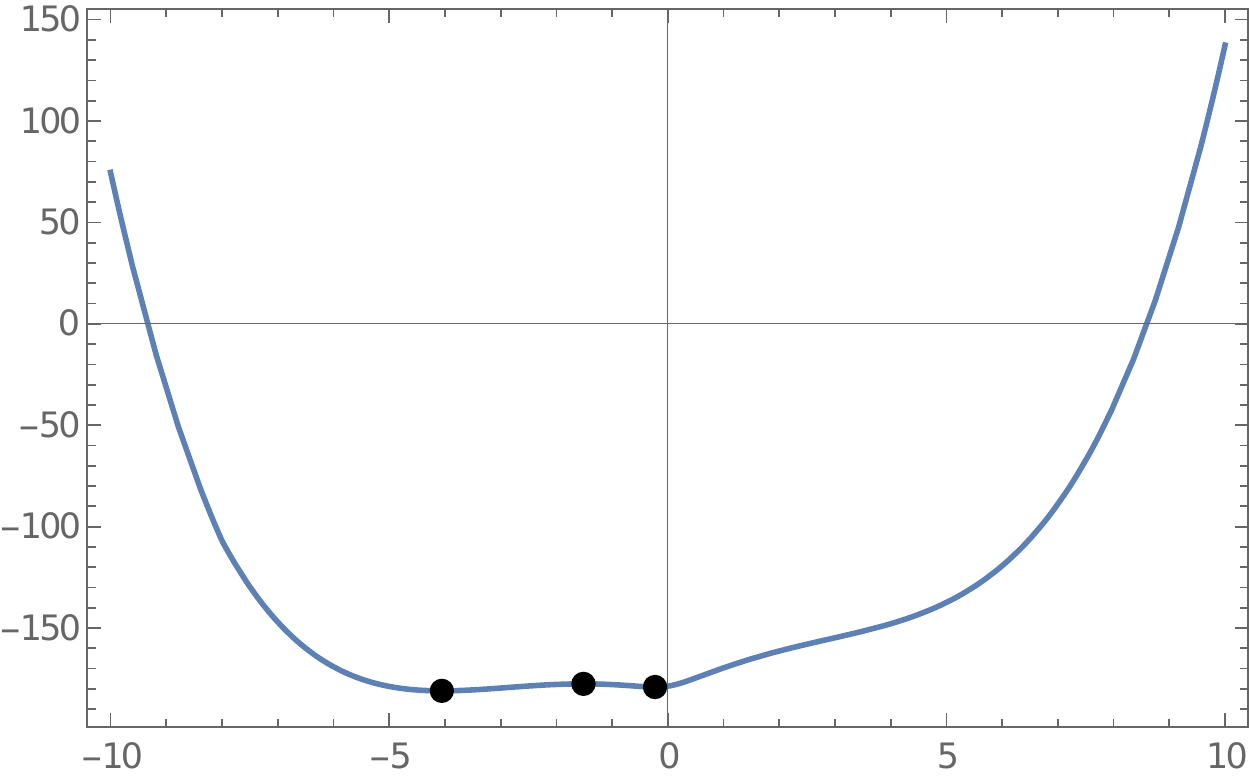} 
    \caption*{$C$:$\b=0.32$,$m^2_s=-2.18$} 
    %\label{b} 
    %\vspace{1ex}
  \end{subfigure} 
  \begin{subfigure}[b]{0.5\linewidth}
    \centering
    \includegraphics[width=0.95\linewidth]{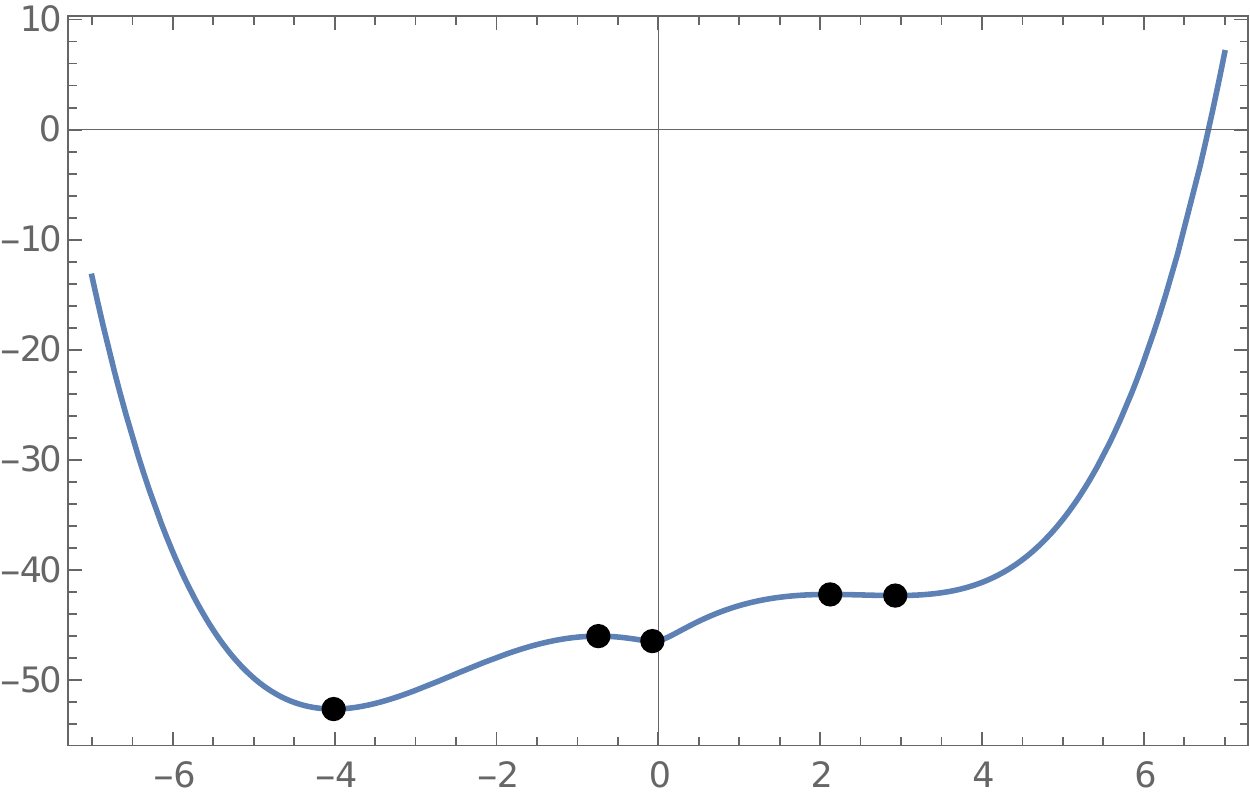} 
    \caption*{\small{$D$:$\b=0.44$, $m_s^2=-2.24$}} 
    %\label{c} 
  \end{subfigure}%%
  %\hspace{-5em}
 % \label{fig7} 
\end{minipage}
  \caption{(a): Phases for $m_f=0.8$ and $g=0.1$ in the $\b$-$m^2_s$ plane and representative potentials $A$-$D$ on extreme right. Black dots on the potential plots are the extrema. (b): Contours for the extrema at various values of $\phi_{cl}$. The solid lines range from $\phi_{cl}=2.5$ to $0.5$ and the dashed lines are for $\phi_{cl}=-3$ to $-0.2$. }
  \label{YAdS4tbms} 
  \end{center} 
\end{figure}

The presence of the cusp at $m_f+g\phi_{cl}=0$ poses a complication in the analysis of phases at finite temperature. We thus set $m_f=0.8$, $g=0.1$ so that the cusp falls on the left of the interesting region. This $(m_f,g)$ point corresponds to a point in region $I$ at zero temperature in Figure \ref{YAdS40tmfgr}. The phase plots and the corresponding potentials at finite temperature are given in Figure \ref{YAdS4tbms}. We see the following features.

\noindent
1. There is always a root of the saddle point equation at the left of the origin which is a minimum and this is the only one for all positive $m^2_s$. This is region $A$ in Figure \ref{YAdS4tbmsr}. 

\noindent
2. For $m^2_s<0$ which at low temperatures in region $B$ we have two additional extrema, a maximum
and a minimum. These extrema vanish at high temperatures as in theories of scalars \cite{Kakkar:2022hub} where the the $O(N)$ symmetry gets restored.

\noindent
3. Another set of new extrema appear for lower values of masses. This is in region $D$ in Figure \ref{YAdS4tbmsr}. This feature is also similar to that of the scalars. The differences in this case is that the potential is asymmetric about $\phi_{cl}=0$ and there is a cusp (which are not shown in the potential plots). The additional extrema in region $D$ survive in region $C$ which vanish as one moves into region $A$.

\noindent 
4. Figure \ref{YAdS4tbmscr} shows the contours for extrema at various values of $\phi_{cl}$. Two contours intersect when two extrema exist simultaneously. The boundaries of regions $C$ and $D$ of  Figure \ref{YAdS4tbmsr} envelopes the regions of the contour intersection giving rise to additional extrema.

\noindent
5. We find that the extreme left minimum remains the global minimum for all values of masses and temperatures.  

\subsection{Gross-Neveu model}\label{gnmodel}

In this section we study the phases of massive Gross-Neveu model. The Euclidean theory is given by

\beqa\label{gn1}
{\cal L}=\bar{\psi}^i\left(\slashed{D}+m_f\right)\psi^i+\f{g}{2N}\left(\bar{\psi}^i\psi^i\right)^2~,~~~~~~~~i=1,\cdots,N
\eeqa

The large $N$ computation proceeds similar to that in flat space. We thus review the essential ingredients. 
It is convenient to introduce an auxiliary field $\s(x)$ and define

\beqa\label{gn2}
{\cal L}^{\prime}&=&{\cal L}-\f{N}{2g}\left(\s(x)-\f{g}{N} \bar{\psi}^i\psi^i\right)^2\non
&=& \bar{\psi}^i\left(\slashed{D}+m_f+\s(x)\right)\psi^i-\f{N}{2g}\s^2(x)
\eeqa

Imposing the $\s$ equation of motion $\s=(g/N)\bar{\psi}^i\psi^i$ one recovers the original theory (\ref{gn1}) with four fermion interaction.
In (\ref{gn2}) the fermions being quadratic, can now be integrated over

\beqa
Z&=&\int {\cal D}\s{\cal D}\bar{\psi}^i{\cal D}\psi^i\exp\left(-\int d^{d+1}x\sqrt{g}{\cal L}^{\prime}\right)\non
&=& \int {\cal D}\s~\left[\det\left(\slashed{D}+m_f+\s(x)\right)\right]^N\exp\left(\f{N}{2g}\s^2(x)\right).
\eeqa

Writing $\s(x)=\s_{cl}+\d\s(x)$, one arrives at the following effective potential as a function of $\s_{cl}$ at the leading order in $1/N$

\beqa\label{gnp}
\f{V_{eff}}{N}&=&-\f{\s_{cl}^2}{2g}-\mbox{tr}\log\left(\slashed{D}+m_f+\s_{cl}\right)
\eeqa

The saddle point equation is

\beqa\label{gns}
\f{1}{N}\f{\pa V_{eff}}{\pa \s_{cl}}=-\f{\s_{cl}}{g}-\mbox{tr}\left[\f{1}{\slashed{D}+m_f+\s_{cl}}\right]
\eeqa

We now analyze the potential (\ref{gnp}) for $m_f,g>0$.   
The extrema of the potential is determined by the real roots of the saddle point equation (\ref{gns}). The changes in the phase plots at finite temperature are discussed further.

\subsubsection{AdS$_3$}

For AdS$_3$, the potential and the saddle point equations at zero temperature are respectively

\beqa\label{gnpads3}
\f{V^0_{eff}}{N}=-\f{\s_{cl}^2}{2g}+\f{1}{2\pi}\left(\f{1}{3}|m_f+\s_{cl}|^3-\f{1}{4}|m_f+\s_{cl}|\right)
\eeqa

\beqa\label{gnsads3}
\f{1}{N}\f{\pa V^0_{eff}}{\pa \s_{cl}}=-\f{\s_{cl}}{g}+\f{1}{2\pi}\mbox{sgn}(m_f+\s_{cl})\left((m_f+\s_{cl})^2-\f{1}{4}\right)
=0.
\eeqa

We summarize below the essential features of the potential. The saddle point equation being quadratic, has a maximum of two roots.
\\

\begin{figure}[H] 
\begin{center} 
  \begin{minipage}{0.45\textwidth}%   
   \begin{subfigure}[b]{0.8\linewidth}
    \centering
    \includegraphics[width=1.1\linewidth]{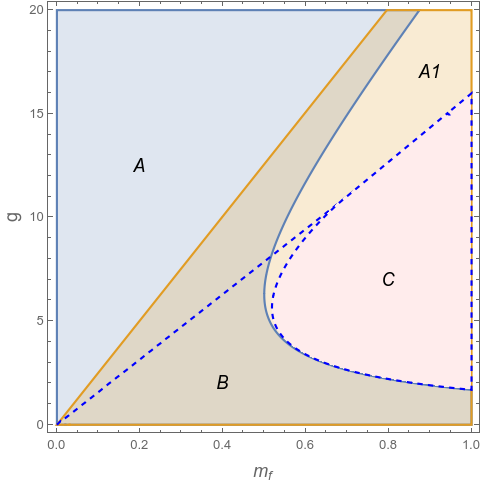} 
    %\caption{Phases on the $\b-m^2$ plane} 
    \label{OnAds4region} 
    \vspace{1ex}
  \end{subfigure}
  \end{minipage}%%%
  \begin{minipage}{0.55\textwidth}%
  \begin{subfigure}[b]{0.5\linewidth}
    \centering
    \includegraphics[width=0.95\linewidth]{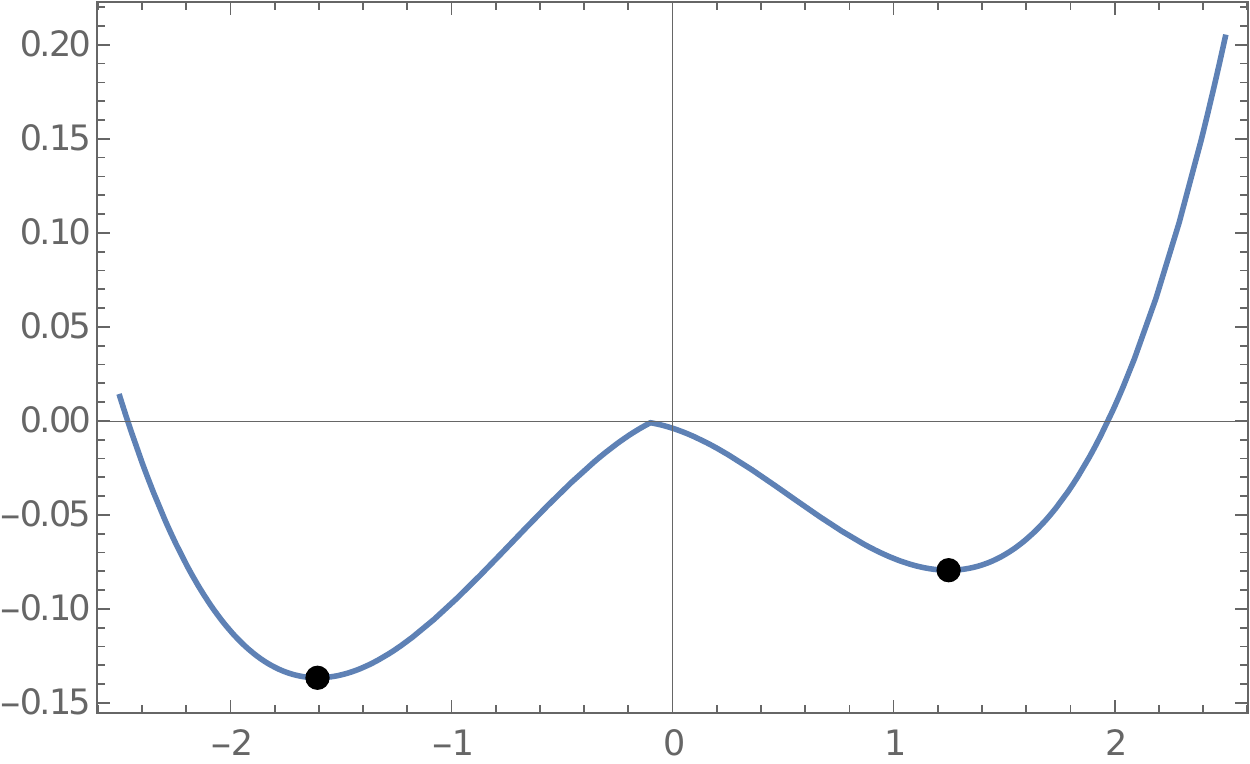} 
    \caption*{$A$: $m_f=0.1$, $g=5$} 
    %\label{a} 
    \vspace{1ex}
  \end{subfigure}%
  %\hspace{-10em}
  \begin{subfigure}[b]{0.5\linewidth}
    \centering
    \includegraphics[width=0.95\linewidth]{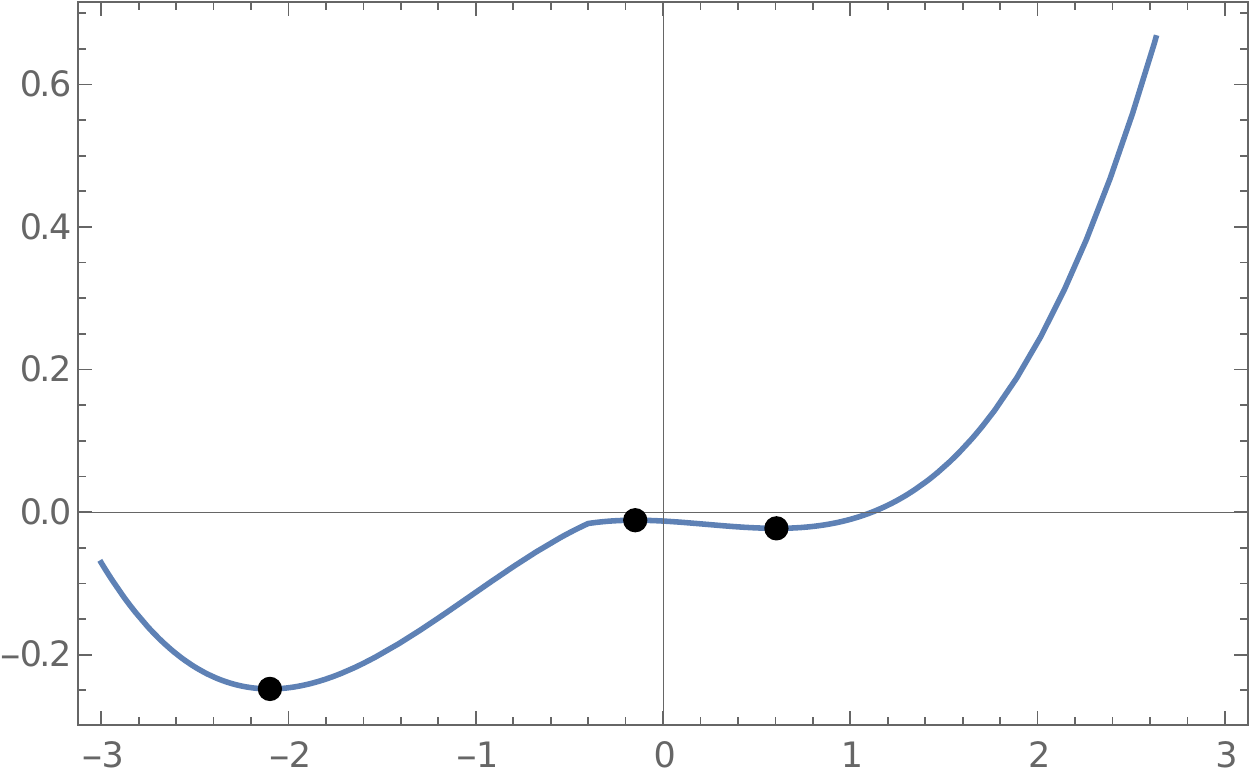} 
    \caption*{$B$: $m_f=0.4$, $g=5$} 
    %\label{b} 
    \vspace{1ex}
  \end{subfigure} 
  \begin{subfigure}[b]{0.5\linewidth}
    \centering
    \includegraphics[width=0.95\linewidth]{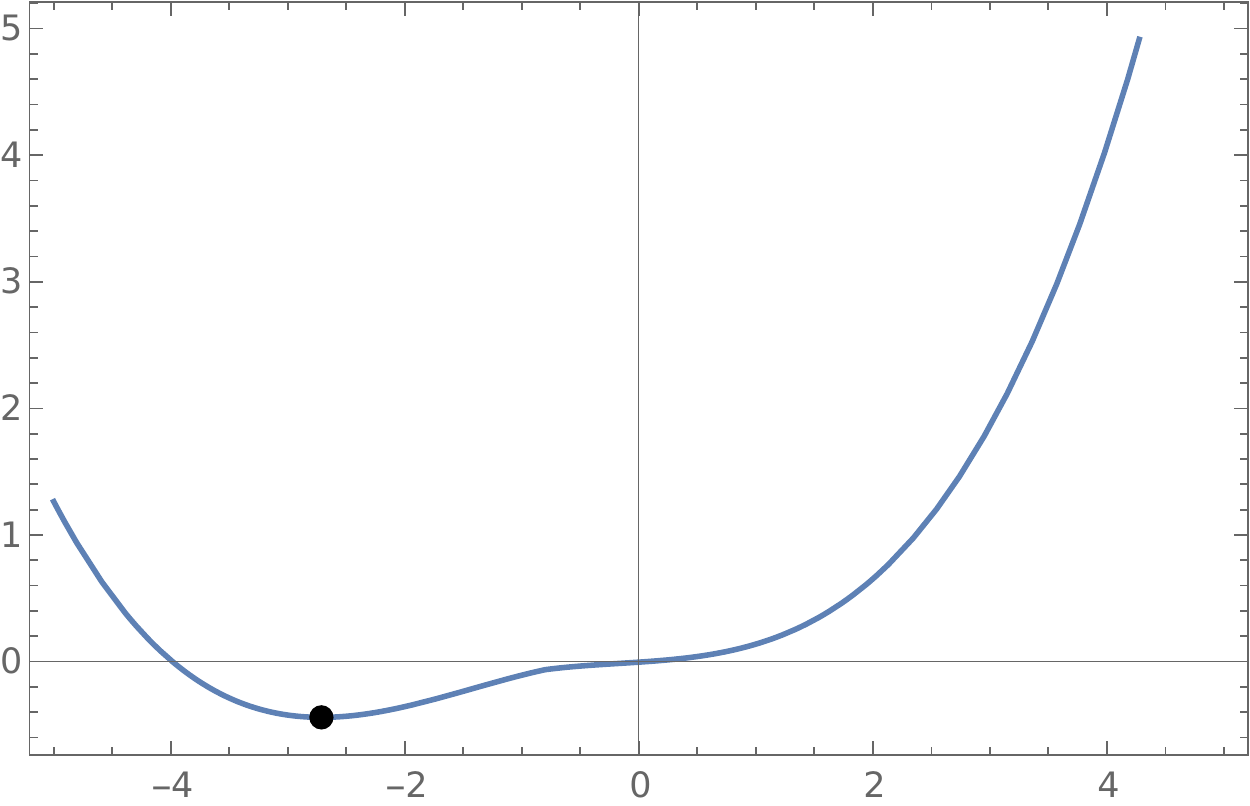} 
    \caption*{$C$: $m_f=0.8$, $g=5$} 
    %\label{c} 
  \end{subfigure}%%
  %\hspace{-5em}
  \begin{subfigure}[b]{0.5\linewidth}
    \centering
    \includegraphics[width=0.95\linewidth]{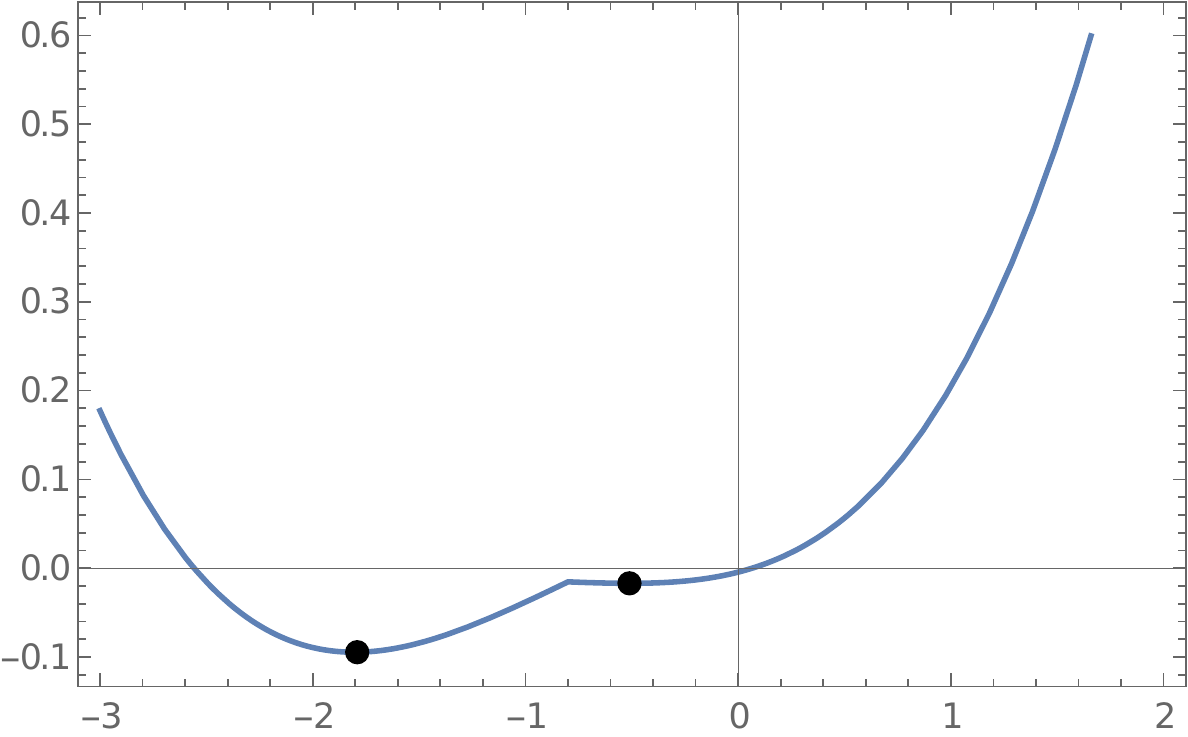} 
   \caption*{$A1$: $m_f=0.8$, $g=15$, $\b=4$} 
    %\label{d} 
  \end{subfigure} 
 % \label{fig7} 
\end{minipage}
  \caption{Regions (left) corresponding to various representative potentials (right). The solid lines correspond to boundaries at zero temperature and the dashed lines for $\b=4$ and $n=10$. Black dots on the potential plots are the extrema. }
  \label{GNAds3} 
  \end{center} 
\end{figure}

\noindent
1. For $\mbox{sgn}(m_f+\s_{cl})=+1$ i.e. for  $\s_{cl}>-m_f$, there are two roots when $m_f>g/(8\pi)$ corresponding to a minimum and a maximum of the potential. On the $(m_f-g)$ plane this region is $B$ in Figure \ref{GNAds3}. Otherwise in region $A$ there is only one minimum. No roots exist when $\pi^2/g^2-2\pi m_f/g+1/4<0$ (region $C$). The solid lines in Figure \ref{GNAds3} are the boundaries of these regions at zero temperature. At the boundary separating regions $B$ and $C$ the maximum and the minimum in region $B$ coincide.
 
\noindent 
2. For $\mbox{sgn}(m_f+\s_{cl})=-1$ i.e. for  $\s_{cl}<-m_f$, there exists one root corresponding to a minimum of the potential for all values of $m_f$ and $g$.

\noindent
3. The minimum on the left always remains lower than that of the right, whenever the right minimum exists. In other words, there is no first order phase transition. This has been checked numerically.

\noindent
4. For $m_f=0$ the potential is symmetric under $\s_{cl}\rightarrow -\s_{cl}$. This originates from the discrete chiral symmetry for $m_f=0$, $\psi\rightarrow \g^5\psi$ exactly as in flat space case. However the deviation from the flat space appears in the last term on the r.h.s in equation (\ref{gnpads3}) as also in the Yukawa theory in AdS$_3$ discussed earlier. This term results in the potential being non-differentiable at $m_f+\s_{cl}=0$. Putting back factors of the AdS radius $L$ it is easy to see that the term can be ignored as compared to the first two terms in r.h.s of equation (\ref{gnpads3}) as $L \rightarrow \infty$. Thus the maximum that appears in the flat space case at $\s_{cl}=0$ for $m_f=0$ is replaced by a cusp.
\\

At finite temperature,

\beqa\label{gnptads3}
\f{V_{eff}}{N}=\f{V^0_{eff}}{N}-\f{4}{\pi\b}\sum^{\infty}_{n=1}\frac{(-1)^n}{n}\frac{e^{-n\beta\left(1+|m_f+\s_{cl}|\right)}}{|1-e^{-n\beta}|^2}.
\eeqa

The corresponding boundary that separates regions $A$ and $B$  at zero temperature is now given by

\beqa
\left.\f{1}{N}\f{\pa V_{eff}}{\pa \s_{cl}}\right|_{\s_{cl}+m_f=0+}=\f{m_f}{g}-\f{1}{8\pi}+\f{4}{\pi}\sum^{\infty}_{n=1}\frac{(-1)^ne^{-n\b}}{|1-e^{-n\beta}|^2}=0.
\eeqa

This boundary is shown as the dashed straight line in Figure  \ref{GNAds3}. The finite temperature contribution decreases the slope and as a result the region $A$ now expands to include the region $A1$. The region where no right extremum exists shrinks at finite temperature to the region enclosed by the dashed curve. It is known that in flat space,  at high temperatures (for tree-level, $m_f=0$) the discrete chiral symmetry is restored \cite{Klimenko:1987gi,Rosenstein:1988dj}. However we find that in AdS$_3$ it remains broken for all temperatures.

\subsubsection{AdS$_2$}

The zero temperature trace is UV divergent and can be regularized using dimensional regularization. We expand about $d=1$ with $\e=1-d$. The expression for the zero temperature trace is given in (\ref{expads2f}). In terms of Feynman diagrams, the UV divergences arise from the two-point and the one-point diagrams. We thus use the following structure of the counterterms

\[\s_{cl}~\d\l_1+\f{1}{2}\s_{cl}^2~\d \left(\f{1}{g}\right)
\]

with renormalization conditions

\beqa\label{rcads2g1}
\left. \f{1}{N}\f{\pa V^0_{eff}}{\pa \s_{cl}}\right|_{\s_{cl}=0}=0~~~~~~~~~~\left. \f{1}{N}\f{\pa^2 V^0_{eff}}{\pa \s_{cl}^2}\right|_{\s_{cl}=0}=-\f{1}{g}.
\eeqa

These give the following zero temperature effective potential

\beqa\label{vztads2g}
V^0_{eff}&=& -\f{\s_{cl}^2}{2g} +\f{1}{2 \pi} \int^{M_f}_{m_f} M_f \left(\psi^{(0)}\left(|M_f|\right)+\psi^{(0)}\left(|M_f|+1\right)\right) dM_f-\s_{cl}\left[\f{m_f}{2\pi} \left(\psi^{(0)}\left(m_f\right)+\psi^{(0)}\left(m_f+1\right)\right)\right] \non
&-& \f{\s^2_{cl}}{4\pi} \left[\left(\psi^{(0)}\left(m_f\right)+\psi^{(0)}\left(m_f+1\right)\right)+m_f\left(\psi^{(1)}\left(m_f\right)+\psi^{(1)}\left(m_f+1\right)\right)\right]
\eeqa

To see the large $\s_{cl}$ behaviour we expand the integrand in (\ref{vztads2g}). $\psi^{(0)}(\s_{cl})$ for large $\s_{cl}$ behaves as $\log(\s_{cl})$. Thus after integration the leading positive term is $\s_{cl}^2\log(\s_{cl})$. This is behaviour of the large $N$ correction in flat space in two dimensions. In three dimensions the leading term is $\s_{cl}^3$. Since $\log(\s_{cl})$ grows slowly than $\s_{cl}$, we need to choose somewhat larger values of $g$ in the numerics so as to get the extrema of the potential within a small range of $\s_{cl}$. A further point to note is that since we have set the renormalization condition at $\s_{cl}=0$, $m_f$ cannot be taken to be zero due to infrared divergences. One needs to set the renormalization conditions (\ref{rcads2g1}) at some nonzero value, $\s_{cl}=\s_0$. In this case, because of the the discrete chiral and $\s\rightarrow -\s$ symmetries for $m_f=0$, the potential is symmetric about $\s_{cl}=0$ with two minima as in flat space. The discrete chiral symmetry (of the massless theory) is spontaneously broken.

Returning back to the massive theory, the finite temperature effective potential is 

\beqa
\f{V_{eff}}{N}&=&\f{V^0_{eff}}{N} - \f{2}{\b}\sum^{\infty}_{n=1}\frac{(-1)^n}{n}\frac{e^{-n\beta\left(\frac{1}{2}+|M_f|\right)}}{|1-e^{-n\beta}|}~.
\eeqa

We now summarize the essential features of the potential below.
\\

\begin{figure}[H] 
\begin{center} 
  \begin{minipage}{0.45\textwidth}%   
   \begin{subfigure}[b]{0.85\linewidth}
    \centering
    \includegraphics[width=1.1\linewidth]{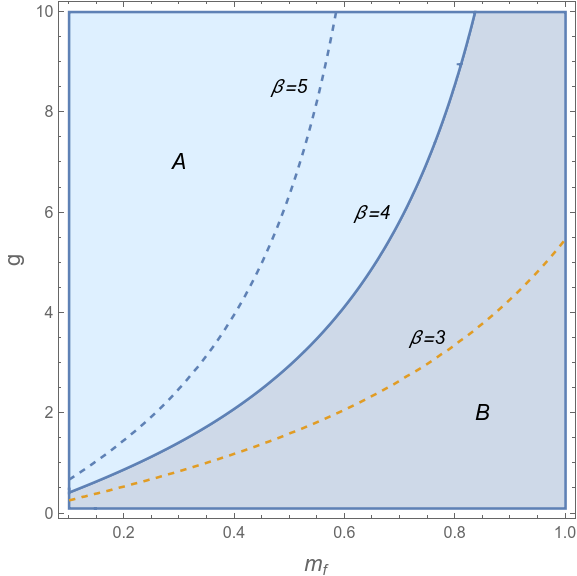} 
    \caption{} 
    \label{GNAdS2r} 
    \vspace{1ex}
  \end{subfigure}
  \end{minipage}%%%
  \begin{minipage}{0.55\textwidth}%
  \begin{subfigure}[b]{0.5\linewidth}
    \centering
    \includegraphics[width=0.95\linewidth]{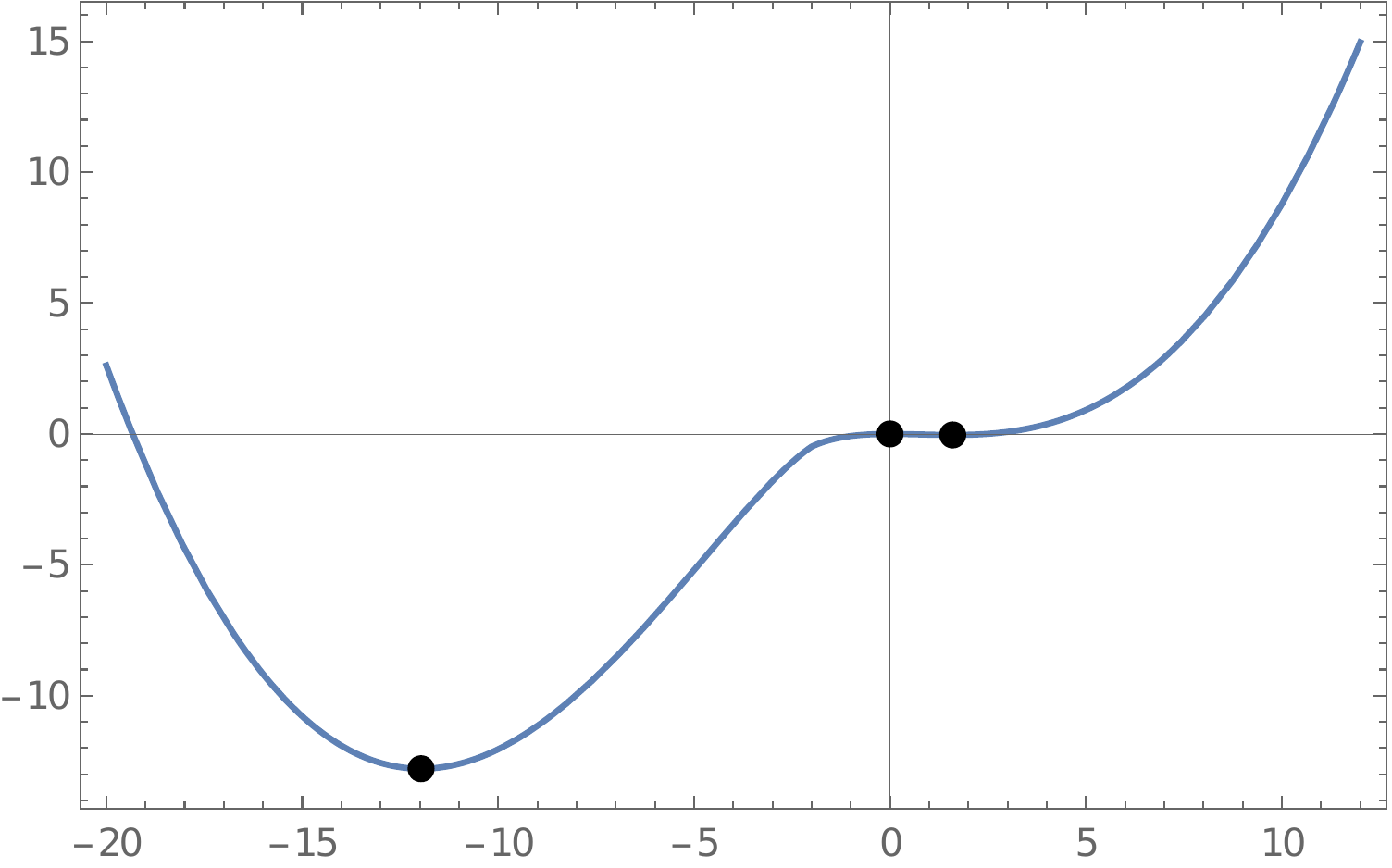} 
    \caption*{$T=0$, $m_f=2$, $g=10$} 
    %\label{a} 
    \vspace{1ex}
  \end{subfigure}%
  %\hspace{-10em}
  \begin{subfigure}[b]{0.5\linewidth}
    \centering
    \includegraphics[width=0.95\linewidth]{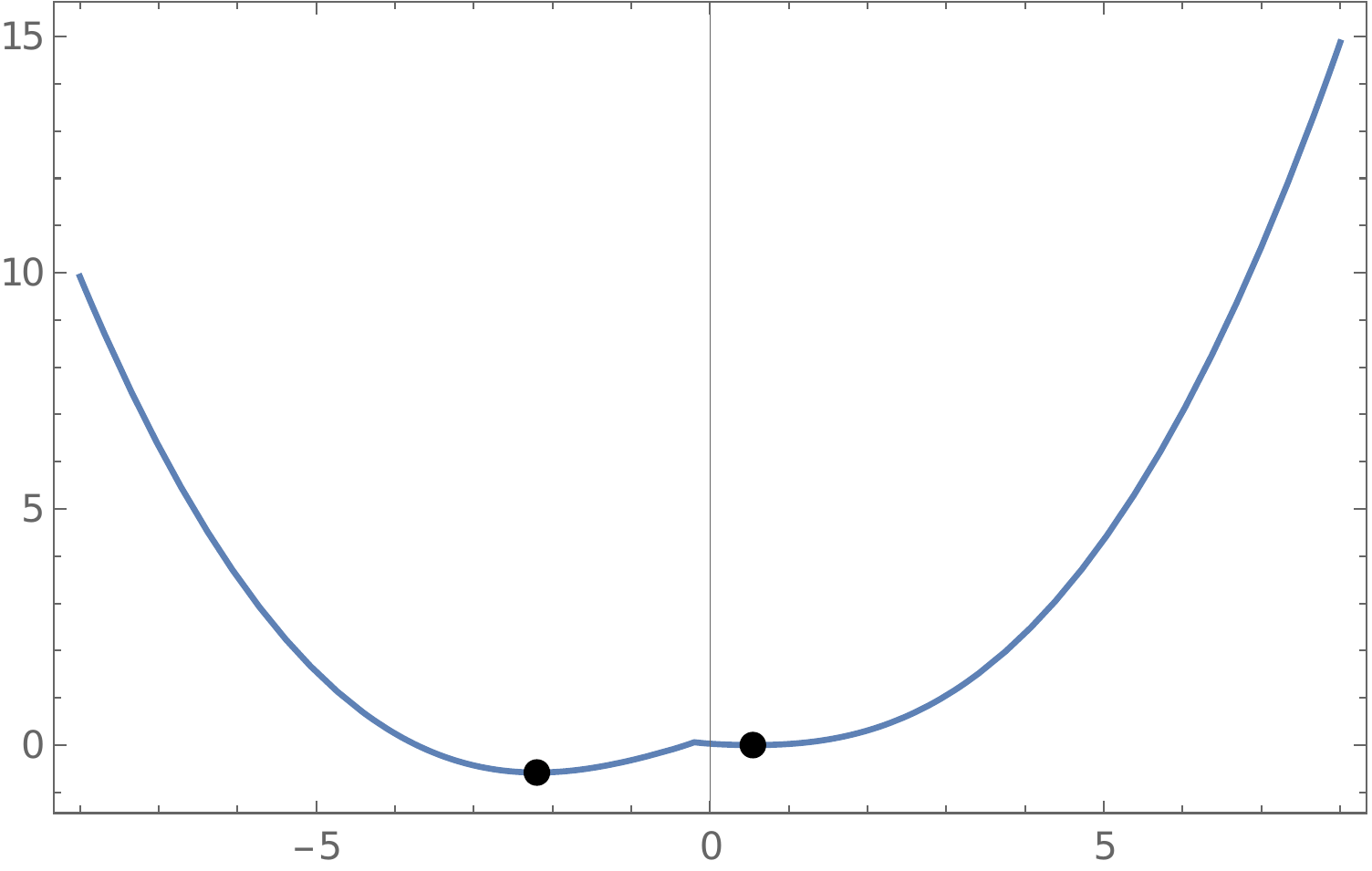} 
    \caption*{$A$: $m_f=0.2$, $g=8$} 
    %\label{b} 
    \vspace{1ex}
  \end{subfigure} 
  \begin{subfigure}[b]{0.5\linewidth}
    \centering
    \includegraphics[width=0.95\linewidth]{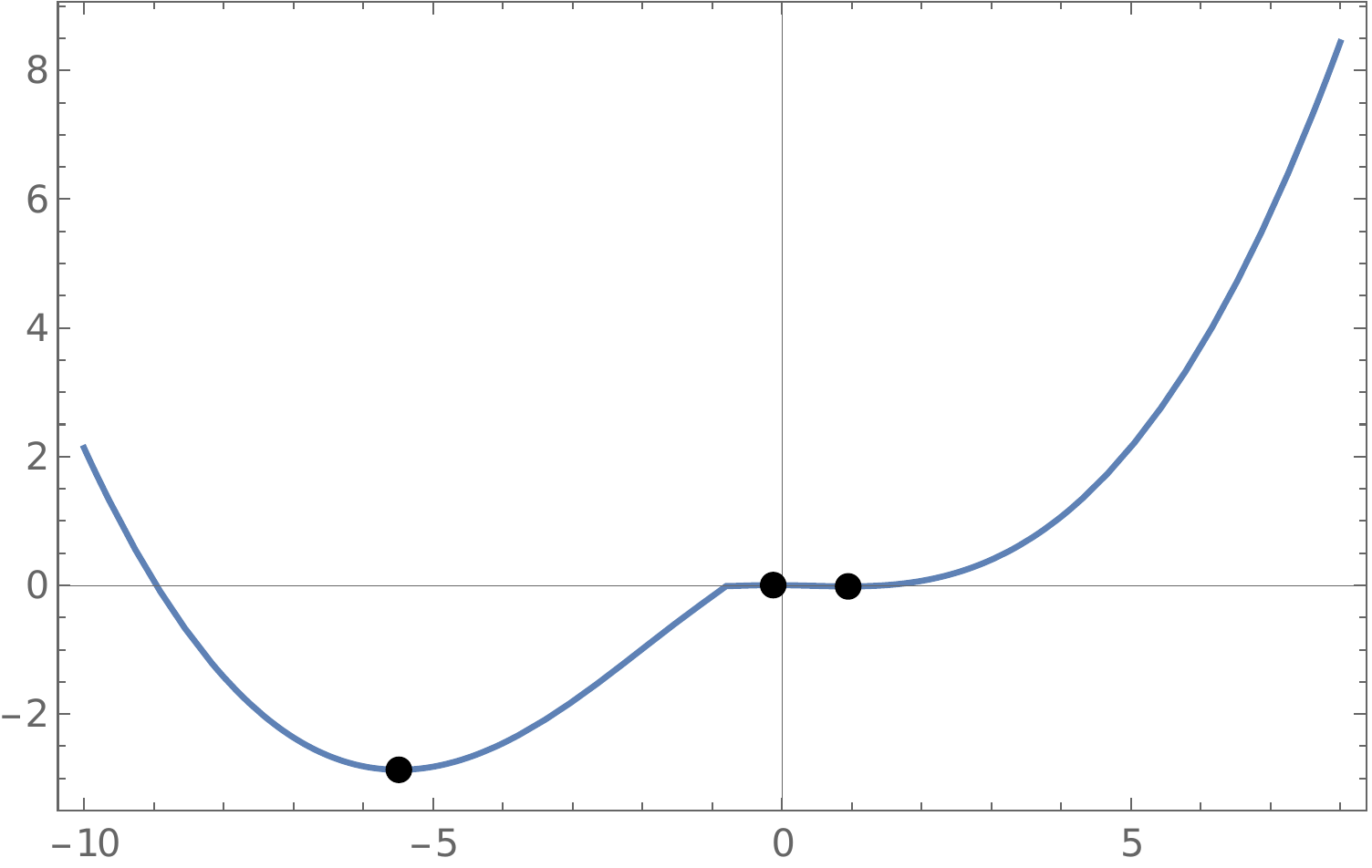} 
    \caption*{$B$: $m_f=0.8$, $g=8$} 
    %\label{c} 
  \end{subfigure}%%
  %\hspace{-5em}
  \begin{subfigure}[b]{0.5\linewidth}
    \centering
    \includegraphics[width=0.95\linewidth]{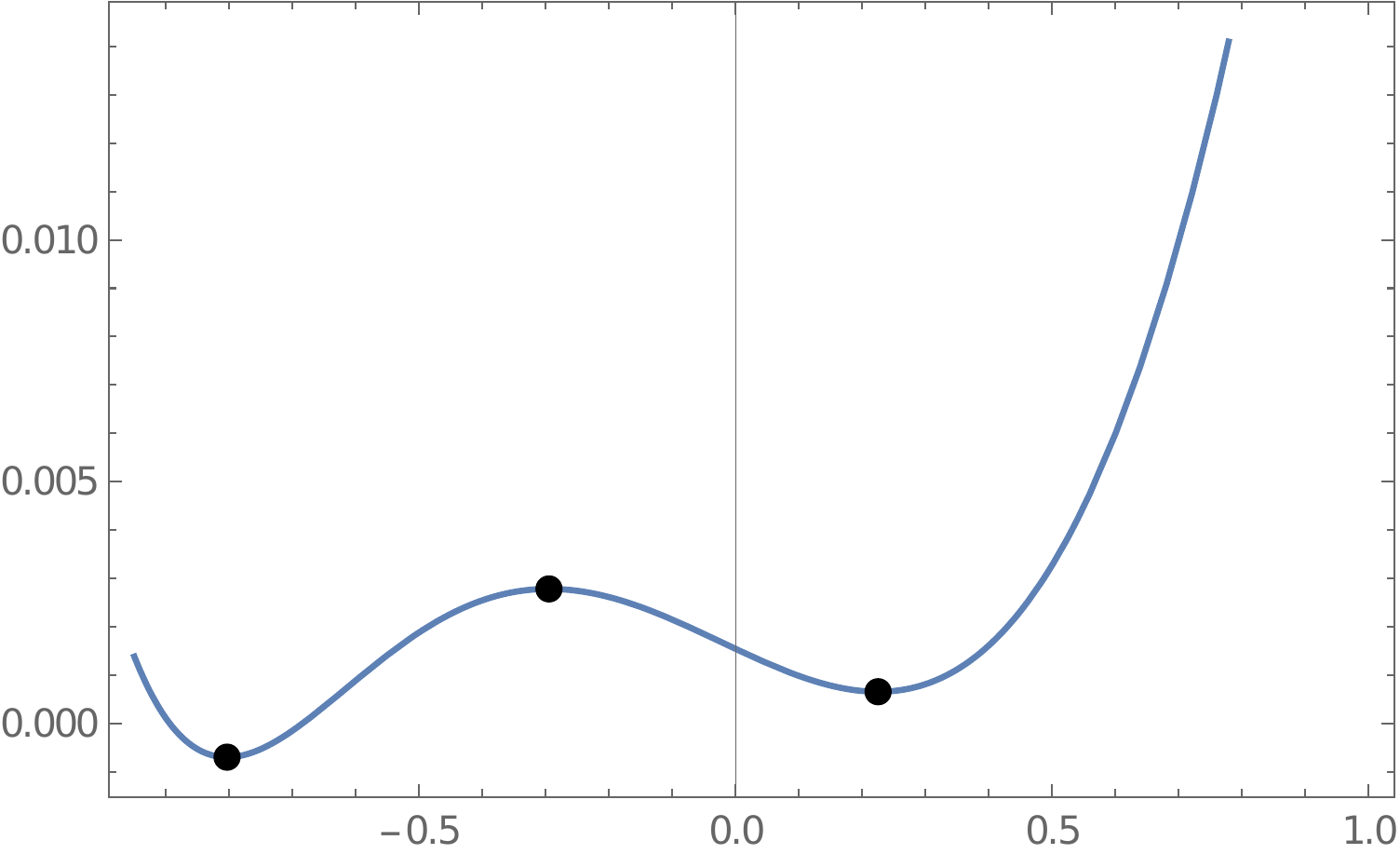} 
   \caption*{$R$: $m_f=0.95$, $g=50$, $\b=4$} 
    %\label{d} 
  \end{subfigure} 
 % \label{fig7} 
\end{minipage}
  \caption{Regions in $(m_f,g)$ space (left) and some potential plots (right). The solid boundary separates the shaded regions $A$ and $B$ for $\b=4$. The corresponding boundaries for $\b=3,5$ are shown as dashed lines. Black dots on the potential plots are the extrema. }
  \label{GNAdS2} 
  \end{center} 
\end{figure}

\noindent
1. At zero temperature there are two minima for all values of $(m_f,g)$ coming from the two branches of the potential $\s_{cl}+m_f>0$ and $\s_{cl}+m_f<0$. The left minimum is lower than the right in the $(m_f,g)$ plane. In the left branch we only have the minimum. The renormalization condition (\ref{rcads2g1}) imply that at $\s_{cl}=0$ there is a maximum for all positive values of $g$. This is unlike the case of AdS$_3$ where there are additional $B$ and $C$ regions. Note that the maximum always comes right of the cusp because $m_f>0$ and the cusp is at $\s_{cl}+m_f=0$. A plot of the potential at $T=0$ is shown in Figure \ref{GNAdS2} (right).

\noindent
2. The two minima remain at finite temperatures and the left minimum remains lower than the right. Figure \ref{GNAdS2r} shows two regions $A$ and $B$. In region $A$ the maximum ceases to exist unlike that in region $B$. The dashed lines in the figure are the corresponding boundaries for $\b=3,5$. Representative potential plots are shown in the right of Figure \ref{GNAdS2r}. For high values of $g$ a new minimum appears from the right branch $(m_f+\s_{cl}>0)$ shown in the figure labelled as $R$.

\subsubsection{AdS$_4$}

In four dimensions, since the four-fermion theory is not renormalizable, we need to add additional scalar interactions. To see this, we expand the zero temperature fermion trace about $d=3$ with $\e=3-d$ as in (\ref{expads4f}). One to four point functions in $\s_{cl}$ are UV divergent. The counterterms have the following form

\[\s_{cl}~\d\l_1+\f{1}{2}\s_{cl}^2~\d\left(\f{1}{g}\right)+\f{1}{3!}\s_{cl}^3~\d \l_3+\f{1}{4!}\s_{cl}^4~\d \l_4.\]

We use the renormalization scheme as in the previous section for AdS$_2$, with the following renormalization conditions at $\s_{cl}=0$

\beqa\label{rcads4g1}
\f{1}{N}\f{\pa V^0_{eff}}{\pa \s_{cl}}=0~~~~~~~~~~\f{1}{N}\f{\pa^2 V^0_{eff}}{\pa \s_{cl}^2}=-\f{1}{g}~~~~~~~~~~\f{1}{N}\f{\pa^3 V^0_{eff}}{\pa \s_{cl}^3}=\l_3~~~~~~~~~~\f{1}{N}\f{\pa^4 V^0_{eff}}{\pa \s_{cl}^4}=\l_4~.
\eeqa

These give the following zero temperature effective potential

\beqa\label{vztads4g}
V^0_{eff}&=& -\f{\s_{cl}^2}{2g} +\l_3\f{\s_{cl}^3}{3!}+\l_4\f{\s_{cl}^4}{4!}-\int_{m_f}^{M_f} \f{M_f(M^2_f - 1)}{4 \pi^2}\left(\psi^{(0)}\left(|M_f|-1\right)+\psi^{(0)}\left(|M_f|+2\right)\right)dM_f\non
&+& \f{\s^{4}_{cl}}{4!}\left[\f{3}{2 \pi^2}\left(\psi^{(0)}\left(m_f -1\right)+\psi^{(0)}\left(m_f +2\right)\right)+ \f{9 m_f}{2 \pi^2}\left(\psi^{(1)}\left(m_f -1\right)+\psi^{(1)}\left(m_f +2\right)\right)\right.\non
&+& \left. \f{3}{4\pi^2}\left(3m^2_f-1\right)\left(\psi^{(2)}\left(m_f -1\right)+\psi^{(2)}\left(m_f +2\right)\right)+ \f{\left(m^3_f-m_f\right)}{4 \pi^2}\left(\psi^{(3)}\left(m_f -1\right)+\psi^{(3)}\left(m_f +2\right)\right)\right]\non
&+& \f{\s^{3}_{cl}}{3!}\left[ \f{3 m_f}{2 \pi^2}\left(\psi^{(0)}\left(m_f -1\right)+\psi^{(0)}\left(m_f +2\right)\right)+ \f{ \left( 3m^2_f -1\right)}{2 \pi^2}\left(\psi^{(1)}\left(m_f -1\right)+\psi^{(1)}\left(m_f +2\right)\right)\right.\non
&+& \left. \f{\left(m^3_f-m_f\right)}{4 \pi^2}\left(\psi^{(2)}\left(m_f -1\right)+\psi^{(2)}\left(m_f +2\right)\right)\right]\non
&+& \f{\s^{2}_{cl}}{2!}\left[ \f{ \left(3 m^2_f-1\right)}{4 \pi^2}\left(\psi^{(0)}\left(m_f -1\right)+\psi^{(0)}\left(m_f +2\right)\right)+\f{\left( m^3_f -m_f\right)}{4 \pi^2}\left(\psi^{(1)}\left(m_f -1\right)+\psi^{(1)}\left(m_f +2\right)\right) \right] \non
&+&  \s_{cl}\left[\f{\left( m^3_f -m_f\right)}{4 \pi^2}\left(\psi^{(0)}\left(m_f -1\right)+\psi^{(0)}\left(m_f +2\right)\right)\right]
\eeqa

Note however that the the fermion trace in the first line of (\ref{vztads4g}) gives the following leading behavior for large values of $\s_{cl}$

\beqa
-\int_{m_f}^{M_f} \f{M_f(M^2_f - 1)}{4 \pi^2}\left(\psi^{(0)}\left(|M_f|-1\right)+\psi^{(0)}\left(|M_f|+2\right)\right)dM_f\sim - \s_{cl}^4\log(\s_{cl})
\eeqa

which makes the potential unbounded below. We thus also add the kinetic term for the $\s$ field and write the modified theory as 

\beqa\label{gny4}
{\cal L}^{\prime}&=& \bar{\psi}^i\left(\slashed{D}+m_f+g\s\right)\psi^i+\f{1}{2}(\pa_{\m}\s)^2+\f{1}{2}m_s^2\s^2+\l_3\f{\s^3}{3!}+\l_4\f{\s^4}{4!}
\eeqa

which is known as the Gross-Neveu-Yukawa model \cite{Zinn-Justin:1991ksq}. To study the large $N$ behavior we re-scale $\s\rightarrow \sqrt{N}\s$, $g\rightarrow g/\sqrt{N}$, $\l_3\rightarrow \l_3/\sqrt{N}$, $\l_4\rightarrow \l_4/N$. Then writing $\s=\s_{cl}+\d\s$ and integrating over the  fluctuations $\d\s$ gives

\beqa
\f{V^0_{eff}}{N}&=&\f{1}{2}m_s^2\s_{cl}^2+\f{\l_3}{3!}\s_{cl}^3+\f{\l_4}{4!}\s_{cl}^4 -\frac{1}{2{\cal V}^{}_{d+1}}\int_{M_s^2}^{\infty}\mbox{tr}\left[ \frac{1}{-\square^{}_{E}+M_s^2}\right]dM_s^2
-\frac{1}{{\cal V}^{}_{d+1}}\int_0^{M_f}\mbox{tr}\left[ \frac{1}{\slashed{D}+M_f}\right]dM_f\non &+& \mbox{counterterms}\nonumber
\eeqa 

with $M_s^2=m_s^2+\l_3\s_{cl}+\l_4\s_{cl}^2/2$ and $M_f=m_f+g\s_{cl}$, which is essentially same as the Yukawa model studied in section \ref{Yads4sec}.

\section{Discussion}\label{summaryf}

In this paper we have explored the phases of Yukawa theories and the Gross-Neveu models in AdS$_{d+1}$ spaces for $d=1,2,3$ at both zero and finite temperature. The analysis leads to qualitative features which at certain places deviate from those in flat space. This is in line with the expectation that the infrared behaviour of the theories is different from those in flat space. The paper begins with the description of a method, based on the premise that the generalized eigenfunctions of the Laplacian operator in Euclidean AdS obey the desired periodicities under the required thermal identification, for computing the one-loop partition function $Z^{(1)}$ for fermions. Our method, as described in this paper, reproduces the partition function results already known in literature. This computation is also shown to generalize to thermal AdS spaces of arbitrary dimensions. 

It was observed from our previous analysis of scalars in \cite{Kakkar:2022hub} that, while the zero temperature contribution to the trace is proportional to the divergent volume of $AdS_{d+1}$, the finite temperature contribution is not. We have thus used the regularized volumes as has been discussed in \cite{Kakkar:2022hub}. The ultraviolet behavior of the theories being the same as that of flat space, we used dimensional regularization and renormalized perturbation theory to renormalize the divergent effective potentials. An alternate scheme, which is similar to the minimal subtraction scheme, has also been discussed for the Yukawa theories in appendix \ref{alternateads2Y}. 

We first performed the analysis for the Yukawa theories wherein we studied the phases of these field theories in the corresponding parameter spaces first at zero temperature and then at finite temperature. For all cases at zero temperature we found a phase boundary where two minima exchange dominance. This feature was also observed at finite temperature for AdS$_2$ and AdS$_3$. We next studied the massive Gross-Neveu models. The Gross-Neveu Yukawa model was studied for $d=3$. In the large $N$ limit with proper re-scaling of the parameters by $N$, the potential in this case reduces to that of the Yukawa model in AdS$_4$. For the Gross-Neveu models for $d=1,2$, the phases at zero temperature in  the corresponding parameter space and the subsequent changes in them at finite temperature were discussed. We found that in the large $N$ limit, unlike in flat space where the discrete chiral symmetry gets restored beyond a certain temperature \cite{Jacobs:1974ys}-\cite{Rosenstein:1988dj}, the discrete chiral symmetry appearing in the $m_f = 0$ limit remains broken at all temperatures both for two and three dimensions. 

This study of the one-loop partition functions paves way for future research in various directions. The analysis can be extended to other theories involving fermions such as those with continuous chiral symmetries and with vector fields in thermal AdS spaces. Work along these lines is in progress. Till now our analysis has been limited to spaces where the background space is a fixed thermal AdS. A more interesting and involved setup is to consider asymptotically AdS black hole geometry. At high temperatures it is known that the black hole is a more stable background through the Hawking-Page transition when one considers the gravitational fluctuations. Another possible promising direction is to compute non-perturbative effective potentials in the flat space limit                                        using techniques implemented in recent studies on flat space S-Matrix as noted in the opening paragraph of the introduction.

\vspace{2.5mm}
\noindent
{\bf Acknowledgements :}
\\ Astha Kakkar acknowledges the support of Department of Science and Technology (DST), 
Ministry of Science and Technology, Government of India, for the DST INSPIRE Fellowship with the INSPIRE
Fellowship Registration Number: IF180721. S.S. thanks the University Grants Commission (UGC), and DST, New
Delhi, India, for providing special assistance and infrastructural support to the Department of Physics,
Vidyasagar University, through the SAP and FIST program respectively.

\appendix

\section{Some details of zero temperature computation}\label{appaf}

%\subsection{Representation of Gamma Matrices}\label{gammarep}

\subsection{Derivation of the measure $\m(\l)$} \label{measuref}

Consider the integrals \cite{Grad}

\beqa\label{in1}
\int_0^{\infty}dy ~y^{\a} K_{i\l-\frac{1}{2}}(y)K_{-i\lp-\frac{1}{2}}(y)&=&\frac{2^{\a-2}}{\G(1+\a)}\left[\G\left(\frac{\a}{2}+\frac{i}{2}(\l-\lp)\right)\G\left(\frac{\a+2}{2}-\frac{i}{2}(\l-\lp)\right)\right.\non
&&\left.\G\left(\frac{\a+1}{2}+\frac{i}{2}(\l+\lp)\right)
\G\left(\frac{\a+1}{2}-\frac{i}{2}(\l+\lp)\right)\right]
\eeqa

\beqa\label{in2}
\int_0^{\infty}dy ~y^{\a} K_{i\l+\frac{1}{2}}(y)K_{-i\lp+\frac{1}{2}}(y)&=&\frac{2^{\a-2}}{\G(1+\a)}\left[\G\left(\frac{\a}{2}-\frac{i}{2}(\l-\lp)\right)\G\left(\frac{\a+2}{2}+\frac{i}{2}(\l-\lp)\right)\right.\non
&&\left.\G\left(\frac{\a+1}{2}+\frac{i}{2}(\l+\lp)\right)
\G\left(\frac{\a+1}{2}-\frac{i}{2}(\l+\lp)\right)\right]
\eeqa

Adding equations (\ref{in1}) and  (\ref{in2}), the r.h.s is

\beqa\label{addint}
\frac{2^{\a-2}}{\G(\a)}\G\left(\frac{\a+1}{2}+\frac{i}{2}(\l+\lp)\right)
\G\left(\frac{\a+1}{2}-\frac{i}{2}(\l+\lp)\right)\G\left(\frac{\a}{2}+\frac{i}{2}(\l-\lp)\right)\G\left(\frac{\a}{2}-\frac{i}{2}(\l-\lp)\right)
\eeqa

We are interested in the limit $\a \rightarrow 0 $. In this limit, for $\l\ne \lp$, the r.h.s $\rightarrow 0$ since

\beqa\label{gl}
\G(x) \sim \frac{1}{x}~~~ \mbox{as} ~~~x \rightarrow 0.
\eeqa

For $\l\rightarrow \lp$ we write the r.h.s (\ref{addint}) using (\ref{gl})

\beqa\label{measuremu}
\G\left(\frac{1}{2}+i\l\right)\G\left(\frac{1}{2}-i\l\right)\left[\frac{\a}{(\l-\lp)^2+\a^2}\right]&\sim& \pi \d(\l-\lp)\G\left(\frac{1}{2}+i\l\right)\G\left(\frac{1}{2}-i\l\right)\non &=& \d(\l-\lp)/\m(\l)
\eeqa

\subsection{$\l$ contour integral in (\ref{trzerotf3})}\label{contourf}

The integrand:
\beqa
f(\l)=\f{1}{\l^2+M_f^2}\frac{\G\left(\frac{d+1}{2}+i\l\right)\G\left(\frac{d+1}{2}-i\l\right)}{\G\left(\frac{1}{2}+i\l\right)\G\left(\frac{1}{2}-i\l\right)}
\eeqa

In the upper half of the complex $\l$-plane the contributions from poles are as follows.

\begin{enumerate}

\item $\l=iM_f$  with $M_f>0$ pole from $\frac{1}{\l^2+M_f^2}$

\beqa\label{res1}
[2\pi i \mbox{Res}f(\l)]_1=\frac{1}{M_f}\G\left(\frac{d+1}{2}+M_f\right)\G\left(\frac{d+1}{2}-M_f\right)\cos(\pi M_f)
\eeqa

\item The gamma function $\G\left(\frac{d+1}{2}+i\l\right)$ has poles for

\[
\frac{d+1}{2}+i\l=-n ~~~~~~n=0,1,2,\cdots~~~~~~
\mbox{or}~~~~~ \l=i\left(n+\frac{d+1}{2}\right).
\]

Using $\underset{z=in}{\mbox{Res}\G(iz)}=(-i)(-1)^n/n!$

\beqa\label{res2}
[2\pi i \mbox{Res}f(\l)]_2&=&2\pi\sum_{n=0}^{\infty}\left[\frac{1}{M_f^2-((d+1)/2+n)^2}\right]\frac{(-1)^n}{n!}\frac{\G\left(n+(d+1)\right)}{\G\left(1/2+\left(n+\frac{d+1}{2}\right)\right)\G\left(1/2-\left(n+\f{d+1}{2}\right)\right)}\non
&=& -\frac{1}{M_f}\G\left(\frac{d+1}{2}+M_f\right)\G\left(\frac{d+1}{2}-M_f\right)\sin(\pi M_f)\cot(\pi(d+1)/2)
\eeqa

\end{enumerate}   

Similar computation can be done for $M_f<0$.

Thus

\beqa
(\ref{res1})+(\ref{res2})=\frac{1}{M_f}\frac{\pi\G\left((d+1)/2+|M_f|\right)}{\G\left(1/2-d/2+|M_f|\right)\sin(\pi(d+1)/2)}
\eeqa

Putting in the other factors in (\ref{trzerotf3}) and using the identity

\[\G\left(1/2-d/2\right)\G\left(1/2+d/2\right)=\frac{\pi}{\sin(\pi(d+1)/2)}\]

gives the final expression (\ref{trzerotf4}).

\subsection{An alternate computation of trace (\ref{trzerotf1})} \label{alternatef} 

In this appendix we perform an alternate computation of the trace (\ref{trzerotf1}). The computation here differs from that in the main text in the order of $k$ and $\l$ integrals. Here we first perform the $\l$ integral.   
We begin from equation (\ref{trzerotf1}) for the case of odd $d$. 
\beqa
\tr\left[\frac{1}{\slashed{D}+M_f}\right]&=&\int d^{d+1}x\sqrt{g}\int \f{d^dk}{(2\pi)^d}\f{1}{k^d}\int_{-\infty}^{\infty}\f{d\l ~\m(\l)}{i\l+M_f}\psi^{\dagger}_{\vec{k},\l}(\vec{x},y)\psi_{\vec{k},\l}(\vec{x},y)\label{trzerotf1a}\\
&=& 2^{\frac{d-1}{2}} \int d^{d+1}x\sqrt{g}\int \f{d^dk}{(2\pi)^d}\f{1}{k^d}\int_{-\infty}^{\infty}\f{d\l ~\m(\l)M_f}{\l^2+M_f^2}\times\non
&\times& (ky)^{d+1}\left[K_{-i\l-\frac{1}{2}}(ky)K_{i\l-\frac{1}{2}}(ky)+K_{-i\l+\frac{1}{2}}(ky)K_{i\l+\frac{1}{2}}(ky)\right].\label{trzerotf2da}
\eeqa

Noting that $\m(\l)$ is an even function, the following has been used in going from (\ref{trzerotf1a}) to (\ref{trzerotf2da})

\beqa
\int_{-\infty}^{\infty}\f{d\l ~\m(\l)}{i\l+M_f}&=&\int_{0}^{\infty}\f{d\l ~\m(\l)}{i\l+M_f}+\int_{0}^{\infty}\f{d\l ~\m(\l)}{-i\l+M_f}\non
&=&M_f\int_{0}^{\infty}\f{d\l ~2\m(\l)}{\l^2+M_f^2}=M_f\int_{-\infty}^{\infty}\f{d\l ~\m(\l)}{\l^2+M_f^2}.
\eeqa

We next use the following relation
\beqa\label{brel}
K_{i\l\pm\frac{1}{2}}(ky)= \frac{\pi}{2}\frac{I_{-i\l\mp\frac{1}{2}}(ky)-I_{i\l\pm\frac{1}{2}}(ky)}{i \sinh(\pi(\l+\frac{i}{2}))}
= \frac{\pi}{2}\frac{I_{i\l\pm\frac{1}{2}}(ky)-I_{-i\l\mp\frac{1}{2}}(ky)}{ \cosh(\pi\l)}.
\eeqa

Therefore
\beqa
K_{-i\l\pm\frac{1}{2}}(ky)K_{i\l\pm\frac{1}{2}}(ky)
= \frac{\pi}{2}\frac{K_{-i\l\pm\frac{1}{2}}(ky)I_{i\l\pm\frac{1}{2}}(ky)-K_{i\l\mp\frac{1}{2}}(ky)I_{-i\l\mp\frac{1}{2}}(ky)}{ \cosh(\pi\l)}
\eeqa

where we have used the even parity property of the Bessel function of the second kind in the second term of r.h.s.
Next consider the following identities \cite{Grad}
\beqa 
\int^{\infty}_{0} J_{\mu+\nu}(2 z \sinh t) \cosh[(\mu-\nu)t]dt&=&\frac{1}{2}[I_{\nu}(z)K_{\mu}(z)+I_{\mu}(z)K_{\nu}(z)]\label{id1}\\
\int^{\infty}_{0}J_{\mu+\nu}(2 z \sinh t) \sinh[(\mu-\nu)t]dt&=&\frac{1}{2}[I_{\nu}(z)K_{\mu}(z)-I_{\mu}(z)K_{\nu}(z)]\label{id2}.
\eeqa

Adding equations (\ref{id1}) and (\ref{id2})

\beqa
\int^{\infty}_{0}J_{\mu+\nu}(2 z \sinh t) \exp{((\mu-\nu)t)}dt=I_{\nu}(z)K_{\mu}(z).
\eeqa

Out of the four terms from (\ref{brel}), we get for the following term
\beqa
\frac{\pi}{2}\frac{1}{\cosh(\pi\l)}[K_{-i\l-\frac{1}{2}}(ky)I_{i\l-\frac{1}{2}}(ky)]=\frac{\pi}{2}\frac{1}{\cosh(\pi\l)}\int^{\infty}_{0}J_{-1}(2z\sinh t)\exp[-2 i \l t]
\eeqa

Using the property for the Bessel functions $J_{-\mu}(z)=(-1)^{\mu}J_{\mu}(z)$ for $\mu\in \mathbb{Z}
$, the $\l$ integral in the trace (\ref{trzerotf2da}) becomes

\beqa
-\int^{\infty}_{-\infty} \frac{d \l \mu(\l) M_f}{\l^{2}+M_f^{2}}\frac{\pi}{2} (\cosh \pi \l)^{-1} \int^{\infty}_{0} J^{}_{1}(2 k y \sinh t) \exp[-2 i \l t] dt
\eeqa

and performing the $\l$ integral in the lower half plane with the pole at $\l=-i M_f$ with $M_f>0$

\beqa
-\int^{\infty}_{0} J^{}_{1}(2 k y \sinh t) \exp[-2 M_f t] dt=-K^{}_{\frac{1}{2}-M_f}(ky)I^{}_{\frac{1}{2}+M_f}(ky)
\eeqa

Next performing the $x$ and the $k$ integrals in (\ref{trzerotf2da}) and considering also the case for $M_f<0$ gives

\beqa
&&-2^{(d-1)/2} V^{}_{d+1}  \left(\frac{2 \pi^{d/2}}{\Gamma(d/2) (2 \pi)^{d}}\right)\times \non
&&\mbox{sgn}(M_f) \f{1}{2}\left[2^{d-1}\Gamma\left(1+\frac{d}{2}\right) \Gamma\left(\left(\frac{1+d}{2}\right)+|M_f|\right) {}_{2}F^{}_{1R}\left(\frac{2+d}{2},\frac{1+d}{2}+|M_f|,\frac{3}{2}+|M_f|,1\right)\right]\non
\eeqa

The above expression can be simplified using the following relations
\beqa
^{}_{2}F^{}_{1R}(a,b,c;1)=\frac{\Gamma(c-b-a)}{\Gamma(c-a)\Gamma(c-b)}~~~~~\mbox{and}~~~~~\Gamma(1-d/2)=\frac{2^{d}\sqrt{\pi}\Gamma(1-d)}{\Gamma(\frac{1}{2}-\frac{d}{2})}
\eeqa

This gives the first contribution to the trace as

\beqa
2^{(d-1)/2}\times\f{1}{2}\times\mbox{sgn}(M_f) \frac{V^{}_{d+1}}{(4\pi)^{(d+1)/2}}\frac{\Gamma(\frac{d+1}{2}+|M_f|)\Gamma(\frac{1}{2}-\frac{d}{2})}{\Gamma(\frac{1}{2}-\frac{d}{2}+|M_f|)}
\eeqa

We use similar transformations for the other three terms from (\ref{brel}) as well. We perform the $\lambda$ integral in the appropriate half of the plane to get four equal contributions from the four terms. This gives the final expression for trace  given by the r.h.s of (\ref{trzerotf4}). The factor of $2^{\frac{d-1}{2}}$ in the equation (\ref{trzerotf2da}) comes from the number of components of $\psi_{\pm}(ky)$ for odd $d$. For even $d$ this factor needs to be replaced by $2^{\frac{d-2}{2}}$. The result in this case is

\beqa
\tr\left[\frac{1}{\slashed{D}+M}\right]&=& \mbox{sgn}(M_f)\frac{{\cal V}_{d+1}2^{\frac{d}{2}}}{(4\pi)^{(d+1)/2}}\frac{\G\left(\frac{d+1}{2}+|M_f|\right)\G\left(\frac{1}{2}-\f{d}{2}\right)}{\G\left(\frac{1}{2}-\f{d}{2}+|M_f|\right)}
\eeqa

\section{Effective potentials}\label{potentialf}
\subsection{Yukawa theory on AdS$_4$}\label{Yads4app}

Effective potential for Yukawa theory on AdS$_4$ at zero temperature resulting from the renormalization conditions (\ref{rcads4f1})
is

\beqa\label{vztads4f}
V^0_{eff}&=& \frac{1}{2}m_s^2\phi_{cl}^2+\frac{\l_3}{3!}\phi_{cl}^3+ \frac{\l_4}{4!}\phi_{cl}^4+\f{1}{2}\int^{M_s^2}_{m_s^2} \left(\frac{2+M_s^2}{16 \pi^2}\right) \left(\psi^{(0)} \left(\nu(\phi_{cl})-\frac{1}{2}\right)+\psi^{(0)} \left(\nu(\phi_{cl})+\frac{3}{2}\right)\right)dM_s^2 \non
&-&\f{1}{4 \pi^2}\int^{M_f}_{m_f}M_f\left(M_f^2-1 \right) \left(\psi^{(0)} \left(|M_f|-1\right)+\psi^{(0)} \left(|M_f|+2\right)\right)dM_f\non
&+& \phi_{cl} \left[\f{g\left(m_f^3-m_f\right)}{4 \pi^2 } \left(\psi^{(0)} \left( m_f-1\right)+\psi^{(0)} \left(m_f+2\right)\right)\right.\non
&-&\left.\f{\l_3 \left(m_s^2+2\right)}{32 \pi^2 } \left(\psi^{(0)} \left(\sqrt{m_s^2+\frac{9}{4}}-\frac{1}{2}\right)+\psi^{(0)} \left(\sqrt{m_s^2+\frac{9}{4}}+\frac{3}{2}\right)\right)\right]\non
&+& \frac{1}{2} \phi_{cl}^2 \Biggl[\frac{g^2 \left(3 m_f^2-1\right)}{4 \pi^2} \left(\psi^{(0)} (m_f-1)+\psi^{(0)} (m_f+2)\right)+\f{g^2 \left(m_f^3-m_f\right)}{4 \pi^2 } (\psi^{(1)}(m_f-1)+\psi^{(1)}(m_f+2))\bigr.\non
&-&\frac{\l_3}{32 \pi^2 }\left(\l_3 \left(\psi^{(0)} \left(\sqrt{m_s^2+\frac{9}{4}}-\frac{1}{2}\right)+\psi^{(0)} \left(\sqrt{m_s^2+\frac{9}{4}}+\frac{3}{2}\right)\right)\right.\non
&+&\left.\left.\l_3 (m_s^2+2) \left(\f{\psi ^{(1)}\left(\sqrt{m_s^2+\frac{9}{4}}-\frac{1}{2}\right)}{2 \sqrt{m_s^2+\frac{9}{4}}}+\f{\psi ^{(1)}\left(\sqrt{m_s^2+\frac{9}{4}}+\frac{3}{2}\right)}{2 \sqrt{m_s^2+\frac{9}{4}}}\right)\right)\right.\non
&-&\left.\f{\l_4}{32 \pi ^2 }\left((m_s^2+2) \left(\psi^{(0)} \left(\sqrt{m_s^2+\frac{9}{4}}-\frac{1}{2}\right)+\psi^{(0)} \left(\sqrt{m_s^2+\frac{9}{4}}+\frac{3}{2}\right)\right)\right)\right]\non
&+&\frac{1}{3!}\phi_{cl}^3\Biggl[\frac{3 g^3 m_f (\psi ^{(0)}(m_f-1)+\psi ^{(0)}(m_f+2))}{2 \pi ^2}+\frac{g^3 \left(3 m_f^2-1\right) (\psi ^{(1)}(m_f-1)+\psi ^{(1)}(m_f+2))}{2 \pi ^2}\Biggr.\non
&+&\frac{g^3 \left(m_f^3-m_f\right) (\psi ^{(2)}(m_f-1)+\psi ^{(2)}(m_f+2))}{4 \pi ^2}\non
&-&\frac{\l_4}{16 \pi ^2} \left(\l_3 \left(\psi ^{(0)}\left(\sqrt{m_s^2+\frac{9}{4}}-\frac{1}{2}\right)+\psi ^{(0)}\left(\sqrt{m_s^2+\frac{9}{4}}+\frac{3}{2}\right)\right)\right.\non
&+&\left.\l_3 (m_s^2+2)\left( \f{(\psi ^{(1)}\left(\sqrt{m_s^2+\frac{9}{4}}-\frac{1}{2}\right)}{2 \sqrt{m_s^2+\frac{9}{4}}}+\f{\psi ^{(1)}\left(\sqrt{m_s^2+\frac{9}{4}}+\frac{3}{2}\right)}{2 \sqrt{m_s^2+\frac{9}{4}}}\right)\right)\non
&-&\f{\l_3}{32 \pi^2}\left(\l_4 \left(\psi ^{(0)}\left(\sqrt{m_s^2+\frac{9}{4}}-\frac{1}{2}\right)+\psi ^{(0)}\left(\sqrt{m_s^2+\frac{9}{4}}+\frac{3}{2}\right)\right)\right.\non
&+&2\l_3^2 \left(\frac{\psi ^{(1)}\left(\sqrt{m_s^2+\frac{9}{4}}-\frac{1}{2}\right)}{2 \sqrt{m_s^2+\frac{9}{4}}}+\frac{\psi ^{(1)}\left(\sqrt{m_s^2+\frac{9}{4}}+\frac{3}{2}\right)}{2 \sqrt{m_s^2+\frac{9}{4}}}\right)\non
&+&(m_s^2+2)\left(\frac{\l_4 \psi ^{(1)}\left(\sqrt{m_s^2+\frac{9}{4}}-\frac{1}{2}\right)}{2 \sqrt{m_s^2+\frac{9}{4}}}-\frac{\l_3^2 \psi ^{(1)}\left(\sqrt{m_s^2+\frac{9}{4}}-\frac{1}{2}\right)}{4 \left(m_s^2+\frac{9}{4}\right)^{3/2}}-\frac{\l_3^2 \psi ^{(1)}\left(\sqrt{m_s^2+\frac{9}{4}}+\frac{3}{2}\right)}{4 \left(m_s^2+\frac{9}{4}\right)^{3/2}}\right.\non
&+&\left.\left.\left.\frac{\l_4 \psi ^{(1)}\left(\sqrt{m_s^2+\frac{9}{4}}+\frac{3}{2}\right)}{2 \sqrt{m_s^2+\frac{9}{4}}}+\frac{\l_3^2 \psi ^{(2)}\left(\sqrt{m_s^2+\frac{9}{4}}-\frac{1}{2}\right)}{4 \left(m_s^2+\frac{9}{4}\right)}+\frac{\l_3^2 \psi ^{(2)}\left(\sqrt{m_s^2+\frac{9}{4}}+\frac{3}{2}\right)}{4 \left(m_s^2+\frac{9}{4}\right)}\right)\right)\right]\non
&+&\frac{1}{4!}\phi_{cl}^4\Biggl[\frac{3 g^4 (\psi ^{(0)}(m_f-1)+\psi ^{(0)}(m_f+2))}{2 \pi ^2}+\frac{9 g^4 m_f (\psi ^{(1)}(m_f-1)+\psi ^{(1)}(m_f+2))}{2 \pi ^2}\Biggr.\non
&+&\frac{3 g^4 \left(3 m_f^2-1\right) (\psi ^{(2)}(m_f-1)+\psi ^{(2)}(m_f+2))}{4 \pi ^2}+\frac{g^4 m_f \left(m_f^2-1\right) (\psi ^{(3)}(m_f-1)+\psi ^{(3)}(m_f+2))}{4 \pi ^2}\non
&-&\frac{3}{32 \pi ^2}\l_4\left(\l_4 \left(\psi ^{(0)}\left(\sqrt{m_s^2+\frac{9}{4}}-\frac{1}{2}\right)+\psi ^{(0)}\left(\sqrt{m_s^2+\frac{9}{4}}+\frac{3}{2}\right)\right)\right.\non
&+&2 \l_3^2 \left(\frac{\psi ^{(1)}\left(\sqrt{m_s^2+\frac{9}{4}}-\frac{1}{2}\right)}{2 \sqrt{m_s^2+\frac{9}{4}}}+\frac{\psi ^{(1)}\left(\sqrt{m_s^2+\frac{9}{4}}+\frac{3}{2}\right)}{2 \sqrt{m_s^2+\frac{9}{4}}}\right)\non
&+&(m_s^2+2)\left(\frac{\l_4 \psi ^{(1)}\left(\sqrt{m_s^2+\frac{9}{4}}-\frac{1}{2}\right)}{2 \sqrt{m_s^2+\frac{9}{4}}}-\frac{\l_3^2 \psi ^{(1)}\left(\sqrt{m_s^2+\frac{9}{4}}-\frac{1}{2}\right)}{4 \left(m_s^2+\frac{9}{4}\right)^{3/2}}+\frac{\l_4 \psi ^{(1)}\left(\sqrt{m_s^2+\frac{9}{4}}+\frac{3}{2}\right)}{2 \sqrt{m_s^2+\frac{9}{4}}}\right.\non
&-&\left.\left.\frac{\l_3^2 \psi ^{(1)}\left(\sqrt{m_s^2+\frac{9}{4}}+\frac{3}{2}\right)}{4 \left(m_s^2+\frac{9}{4}\right)^{3/2}}+\frac{\l_3^2 \psi ^{(2)}\left(\sqrt{m_s^2+\frac{9}{4}}-\frac{1}{2}\right)}{4 \left(m_s^2+\frac{9}{4}\right)}+\frac{\l_3^2 \psi ^{(2)}\left(\sqrt{m_s^2+\frac{9}{4}}+\frac{3}{2}\right)}{4 \left(m_s^2+\frac{9}{4}\right)}\right)\right)\non
&-&\frac{1}{32 \pi ^2}\l_3\left(3 \l_4 \left(\frac{\l_3 \psi ^{(1)}\left(\sqrt{m_s^2+\frac{9}{4}}-\frac{1}{2}\right)}{2 \sqrt{m_s^2+\frac{9}{4}}}+\f{\l_3 \psi ^{(1)}\left(\sqrt{m_s^2+\frac{9}{4}}+\frac{3}{2}\right)}{2 \sqrt{m_s^2+\frac{9}{4}}}\right)\right.\non
&+&3\l_3\left(\frac{\l_4 \psi ^{(1)}\left(\sqrt{m_s^2+\frac{9}{4}}-\frac{1}{2}\right)}{2 \sqrt{m_s^2+\frac{9}{4}}}-\frac{\l_3^2 \psi ^{(1)}\left(\sqrt{m_s^2+\frac{9}{4}}-\frac{1}{2}\right)}{4 \left(m_s^2+\frac{9}{4}\right)^{3/2}}-\frac{\l_3^2 \psi ^{(1)}\left(\sqrt{m_s^2+\frac{9}{4}}+\frac{3}{2}\right)}{4 \left(m_s^2+\frac{9}{4}\right)^{3/2}}\right.\non
&+&\left.\frac{\l_4 \psi ^{(1)}\left(\sqrt{m_s^2+\frac{9}{4}}+\frac{3}{2}\right)}{2 \sqrt{m_s^2+\frac{9}{4}}}+\frac{\l_3^2 \psi ^{(2)}\left(\sqrt{m_s^2+\frac{9}{4}}-\frac{1}{2}\right)}{4 \left(m_s^2+\frac{9}{4}\right)}+\frac{\l_3^2 \psi ^{(2)}\left(\sqrt{m_s^2+\frac{9}{4}}+\frac{3}{2}\right)}{4 \left(m_s^2+\frac{9}{4}\right)}\right)\non
&+&(m_s^2+2)\left(\frac{3 \l_3^3 \psi ^{(1)}\left(\sqrt{m_s^2+\frac{9}{4}}-\frac{1}{2}\right)}{8 \left(m_s^2+\frac{9}{4}\right)^{5/2}}-\frac{3 \l_3 \l_4 \psi ^{(1)}\left(\sqrt{m_s^2+\frac{9}{4}}-\frac{1}{2}\right)}{4 \left(m_s^2+\frac{9}{4}\right)^{3/2}}+\frac{3 \l_3^3 \psi ^{(1)}\left(\sqrt{m_s^2+\frac{9}{4}}+\frac{3}{2}\right)}{8 \left(m_s^2+\frac{9}{4}\right)^{5/2}}\right.\non
&-&\frac{3 \l_3 \l_4 \psi ^{(1)}\left(\sqrt{m_s^2+\frac{9}{4}}+\frac{3}{2}\right)}{4 \left(m_s^2+\frac{9}{4}\right)^{3/2}}-\frac{3 \l_3^3 \psi ^{(2)}\left(\sqrt{m_s^2+\frac{9}{4}}-\frac{1}{2}\right)}{8 \left(m_s^2+\frac{9}{4}\right)^2}+\frac{3 \l_3 \l_4 \psi ^{(2)}\left(\sqrt{m_s^2+\frac{9}{4}}-\frac{1}{2}\right)}{4 \left(m_s^2+\frac{9}{4}\right)}\non
&-&\frac{3 \l_3^3 \psi ^{(2)}\left(\sqrt{m_s^2+\frac{9}{4}}+\frac{3}{2}\right)}{8 \left(m_s^2+\frac{9}{4}\right)^2}+\frac{3 \l_3 \l_4 \psi ^{(2)}\left(\sqrt{m_s^2+\frac{9}{4}}+\frac{3}{2}\right)}{4 \left(m_s^2+\frac{9}{4}\right)}+\frac{\l_3^3 \psi ^{(3)}\left(\sqrt{m_s^2+\frac{9}{4}}-\frac{1}{2}\right)}{8 \left(m_s^2+\frac{9}{4}\right)^{3/2}}\non
&+&\left.\left.\left.\frac{\l_3^3 \psi ^{(3)}\left(\sqrt{m_s^2+\frac{9}{4}}+\frac{3}{2}\right)}{8 \left(m_s^2+\frac{9}{4}\right)^{3/2}}\right)\right)\right].
\eeqa

\subsection{A different renormalization scheme}\label{alternateads2Y}

In this section we discuss an alternate renormalization scheme for the zero temperature partition function in the Yukawa theories. For simplicity we set the renormalized parameters $\l_1=\l_3=0$ in (\ref{generalveff}) and rename $\l_4=\l$.
\\

\noindent
{\bf AdS$_2$}

We see that the coefficient of the UV divergent $1/\e$ terms in (\ref{expads2s}) and in (\ref{expads2f}) adds up to $(M_s^2-2M_f^2)$ after the integral in (\ref{generalveff}), where $M_s^2$ and $M_f$ are defined in (\ref{defmsmf}). We can absorb these divergences into a single counterterm defined as

\[
(M_s^2-2M_f^2)~\d
\] 

and using the renormalization condition

\beqa\label{rcads2f1a}
\left.\f{\pa V^0_{eff}}{\pa \phi_{cl}}\right|_{\phi_{cl}=\phi_0}=0.
\eeqa

$\phi_0>0$ is some undetermined constant and $V^0_{eff}$ is the zero temperature effective potential.  
The renormalization scheme used here is similar to that of the Minimal Subtraction (MS) scheme. We have made a slight departure from the MS scheme, by imposing (\ref{rcads2f1a}) because it is easier to track the extremum (at $\phi_0$) in the numerics. This results in addition of finite terms proportional to $(M_s^2-2M_f^2)$.

The final expressions for the zero temperature effective potential is

\beqa
&&V^0_{eff}=\f{1}{2}m_s^2\phi_{cl}^2-m_s^2\f{\phi_0}{\chi}(M_s^2-2M_f^2)+\f{\l}{4!}\phi_{cl}^4-\f{\l}{3!}\f{\phi_0^3}{\chi}(M_s^2-2M_f^2)
\non&-&\f{1}{4\pi}\left[\int_0^{M_s^2}\psi^{(0)}\left(\nu(\phi_{cl})+\f{1}{2}\right)dM_s^2-\l\f{\phi_0}{\chi}\psi^{(0)}\left(\nu(\phi_0)+\f{1}{2}\right)(M_s^2-2M_f^2)\right]\non
&+&\f{1}{2\pi}\left[\int_0^{M_f}M_f\left[\psi^{(0)}\left(|M_f|\right)+\psi^{(0)}\left(1+|M_f|\right)\right]dM_f-\left.g\f{M_f}{\chi}\left[\psi^{(0)}\left(M_f\right)+\psi^{(0)}\left(1+M_f\right)\right]\right|_{\phi_{cl}=\phi_0}(M_s^2-2M_f^2)\right]\non
&+&\f{1}{8 \pi}\left[(M_s^2\log(4\pi)-2M_f^2\log(2\pi))-\left.\f{1}{\chi}(\l\phi_0\log(4\pi)-4M_fg\log(2\pi))\right|_{\phi_{cl}=\phi_0}(M_s^2-2M_f^2)\right]
\eeqa 

where $\chi=\l\phi_0-4g M_f(\phi_0)$. In the above expression, pairs of terms written in a single line cancel for the l.h.s of (\ref{rcads2f1a}). 
\\

\noindent
{\bf AdS$_4$}

These UV divergences in (\ref{generalveff}) can be cancelled by the following counterterms 

\beqa
(M_s^2+2M_f^2)~\d_1+(M_s^4-4M_f^4)~\d_2.
\eeqa 

along with the renormalization conditions,

\beqa\label{rcads2f2a}
\left.\f{\pa V^0_{eff}}{\pa \phi_{cl}}\right|_{\phi_{cl}=\phi_0}=0~~~~~~~~~~\left.\f{\pa^2 V^0_{eff}}{\pa \phi_{cl}^2}\right|_{\phi_{cl}=\phi_0}=m_s^2.
\eeqa

These conditions which imply that the extremum at $\phi_0$ is a minimum or maximum depending on the sign of $m_s^2$ give the following counterterms.

\beqa
\d_1=\f{1}{16\pi^2}\left[\frac{y_1z_2-y_2z_1}{x_1y_2-x_2y_1}\right]~~~~~~;~~~~~~\d_2=\f{1}{64\pi^2}\left[\frac{x_1z_2-x_2z_1}{x_2y_1-x_1y_2}\right]
\eeqa 

where
\beqa
x_1=\f{1}{16\pi^2}(\l\phi_0+4gM_f)~~~~~~&;&~~~~~~x_2=\f{\l}{16\pi^2}+\f{g^2}{4\pi^2}\\
y_1=\f{1}{64\pi^2}(2\l M_s^2\phi_0-16gM_f^3)~~~~~~&;&~~~~~~y_2=\f{1}{64\pi^2}(2\l^2\phi_0^2+2\l M_s^2-48g^2M_f^2)
\eeqa

\beqa
z_1&=&m_s^2\phi_0+\f{\l}{3!}\phi_0^3 + \frac{\l\phi_0(2+M_s^2)}{32 \pi^2}\left[-\f{2}{\epsilon}-1+\g-\log(4\pi/\m^2) +\psi^{(0)}(\n+3/2)+\psi^{(0)}(\n-1/2)\right]\non
&-&\frac{gM_f(M_f^2-1)}{4 \pi^2}\left[-\f{2}{\epsilon}-1+\g-\log(2\pi/\m^2) +\psi^{(0)}(M_f+2)+\psi^{(0)}(M_f-1)\right]
\eeqa

\beqa
z_2&=&\l\phi_0^2+ \frac{\l(2+M_s^2+\l\phi_0^2)}{32 \pi^2}\left[-\f{2}{\epsilon}-1+\g-\log(4\pi/\m^2) +\psi^{(0)}(\n+3/2)+\psi^{(0)}(\n-1/2)\right]\non
&+& \frac{\l\phi_0(2+M_s^2)}{32 \pi^2}\left[\psi^{(1)}(\n+3/2)+\psi^{(1)}(\n-1/2)\right]\left(\frac{\pa \n}{\pa\phi}\right)\non
&-&\frac{g^2(3M_f^2-1)}{4 \pi^2}\left[-\f{2}{\epsilon}-1+\g-\log(2\pi/\m^2) +\psi^{(0)}(M_f+2)+\psi^{(0)}(M_f-1)\right]\non
&-&\frac{g^2M_f(M_f^2-1)}{4 \pi^2}\left[\psi^{(1)}(M_f+2)+\psi^{(1)}(M_f-1)\right]
\eeqa

$x_1,x_2,y_1,y_2,z_1,z_2$ are evaluated at $\phi_{cl}=\phi_0>0$.

\end{document}